\documentclass[longauth]{aa}

\usepackage[utf8]{inputenc}
\usepackage[T1]{fontenc}

\usepackage{float}
\usepackage{subcaption}
\usepackage{graphicx}
\usepackage{txfonts}
\usepackage[export]{adjustbox}% http://ctan.org/pkg/adjustbox

\usepackage[bottom]{footmisc}

\PassOptionsToPackage{hyphens}{url}\usepackage{hyperref}
\usepackage{mathtools}
\usepackage{amsmath}

\newcommand*\diff{\mathop{}\!\mathrm{d}}

\usepackage{multirow}
\usepackage{makecell}

\usepackage{stackengine}
\newcommand\xrowht[2][0]{\addstackgap[.5\dimexpr#2\relax]{\vphantom{#1}}}

\usepackage{float}

\newcommand{\ie}{i.e.}
\newcommand{\eg}{e.g.}

\newcommand{\Pth}{\ensuremath{P_\text{th}}}
\newcommand{\Guv}{\ensuremath{G_0}} %\text{UV}}}
\newcommand{\AV}{\ensuremath{A_V^{\text{tot}}}}

\newcommand{\Av}{\AV{}}
\newcommand{\Gnaught}{\Guv{}}

\newcommand{\ch}[1]{\ensuremath{\mathrm{#1}}}
\newcommand{\HH}{\ch{H_2}}

\usepackage{xifthen}
\newcommand{\ifequals}[3]{\ifthenelse{\equal{#1}{#2}}{#3}{}}
\newcommand{\case}[2]{#1 #2} % Dummy, so \renewcommand has something to overwrite...
\newenvironment{switch}[1]{\renewcommand{\case}{\ifequals{#1}}}{}

\newcommand{\latexline}[3]{%
  \latexmol{#1}\,\transp{#2}{#3}%
}%

\newcommand{\latexmol}[1]{%
    \begin{switch}{#1}%
        \case{12cn}{\ch{^{12}CN}}%
        \case{12co}{\ch{^{12}CO}}%
        \case{12cs}{\ch{^{12}CS}}%
        \case{13co}{\ch{^{13}CO}}%
        \case{32so}{\ch{^{32}SO}}%
        \case{34so}{\ch{^{34}SO}}%
        \case{c3h2}{\ch{C_3H_2}}%
        \case{c17o}{\ch{C^{17}O}}%
        \case{c18o}{\ch{C^{18}O}}%
        \case{cch}{\ch{CCH}}%
        \case{ch3oh}{\ch{CH_3OH}}%
        \case{hcn}{\ch{HCN}}%
        \case{hcop}{\ch{HCO^+}}%
        \case{hnc}{\ch{HNC}}%
        \case{n2hp}{\ch{N_2H^+}}%
        \case{sio}{\ch{SiO}}%
        \case{h2}{\ch{H_2}}%
        \case{h2co}{\ch{H_2CO}}%
        \case{c2h}{\ch{C_2H}}%
        \case{13c18o}{\ch{^{13}C^{18}O}}%
    \end{switch}%
}%

\newcommand{\trans}[2]{\ch{#1-#2}}%
\newcommand{\transp}[2]{\ch{(\trans{#1}{#2})}}%

\newcommand{\ci}{\ch{[CI]}}%
\newcommand{\cp}{\ch{[CII]}}%

\newcommand{\unit}[1]{\ensuremath{\mathrm{#1}}}

\newcommand{\kms}{\unit{km\,s^{-1}}}

\newcommand{\microm}{\unit{\text{\textmu{}}m}}
\newcommand{\pccm}{\unit{cm^{-3}}}

\usepackage{letltxmacro}
\LetLtxMacro{\originaleqref}{\eqref}

\renewcommand{\eqref}{Eq.~\ref}
\newcommand{\eqrefp}[1]{(Eq.~\ref{#1})}

\DeclareMathOperator*{\argmin}{arg\,min}
\DeclareMathOperator*{\argmax}{arg\,max}

\DeclareMathOperator{\asinh}{asinh}

\newcommand{\R}{\mathbb{R}}

\newcommand{\noiseGeneral}{\mathcal{A}}

\newcommand{\multnoise}{\varepsilon_{n\ell}^{(m)}}

\newcommand{\addnoise}{\varepsilon_{n\ell}^{(a)}}

\newcommand{\obsfull}{Y}
\newcommand{\obsvect}[1]{\mathbf{y}_{#1}}
\newcommand{\obselt}{y_{n\ell}}

\newcommand{\obsvectsubset}[1]{\mathbf{y}_{#1}^{(s)}}
\newcommand{\obsfullsubset}{Y^{(s)}}
\newcommand{\paramfull}{\Theta}

\newcommand{\paramvect}[1]{\boldsymbol{\theta}_{#1}}
\newcommand{\paramelt}[1]{\theta_{#1}}

\newcommand{\truef}{\mathbf{f}}

\newcommand{\truefell}{f_\ell}

\newcommand{\fvalidity}{\mathcal{C}}

\newcommand{\boldmu}{\boldsymbol{\mu}} % mean vector
\newcommand{\Unif}[1]{\text{Unif} \left(#1 \right)}

\newcommand{\lognormal}{\ensuremath{\mathrm{log}\,\mathcal{N}}}
\newcommand{\logd}{\ensuremath{\log_{10}}}

\newcommand{\h}[1]{\ensuremath{h\left(#1\right)}}
\newcommand{\condh}[2]{\ensuremath{h\left(#1 \,\vert\, #2\right)}}

\newcommand{\mi}[2]{\ensuremath{I\left(#1,\,#2\right)}}
\newcommand{\miEst}[2]{\ensuremath{\widehat{I}_N \left(#1,\,#2\right)}}

\begin{document}

\title{Quantifying the informativity of emission lines to infer physical conditions in giant molecular clouds}
\subtitle{I. Application to model predictions}

\author{%
  %% Paper team
  Lucas Einig\inst{\ref{IRAM},\ref{GIPSA-Lab}}\thanks{Equal contribution.} %
  \and Pierre Palud\inst{\ref{LERMA/MEUDON},\ref{CRISTAL},\ref{APC}}\footnotemark[1] %
  \and Antoine Roueff\inst{\ref{Toulon}} %
  \and J\'er\^ome Pety\inst{\ref{IRAM},\ref{LERMA/PARIS}} %
  \and Emeric Bron\inst{\ref{LERMA/MEUDON}} %
  \and Franck Le Petit\inst{\ref{LERMA/MEUDON}} %
  \and Maryvonne Gerin\inst{\ref{LERMA/PARIS}} %
  %% Supervisors by alphabetical order
  \and Jocelyn Chanussot\inst{\ref{GIPSA-Lab}} %
  \and Pierre Chainais\inst{\ref{CRISTAL}} %
  \and Pierre-Antoine Thouvenin\inst{\ref{CRISTAL}} %
  \and David Languignon\inst{\ref{LERMA/MEUDON}} %
  %% Non-permanent researchers by alphabetical order
  \and Ivana Be\v{s}li\'c\inst{\ref{LERMA/PARIS}} %
  \and Simon Coud\'e\inst{\ref{WORCESTER},\ref{CfA}}%
  \and Helena Mazurek\inst{\ref{LERMA/PARIS}} %
  \and Jan H. Orkisz\inst{\ref{IRAM}} %
  \and Miriam G. Santa-Maria\inst{\ref{CSIC},\ref{UF}} %
  \and L\'eontine S\'egal\inst{\ref{IRAM},\ref{Toulon}} %
  \and Antoine Zakardjian\inst{\ref{IRAP}}
  %% Permanent by alphabetical order
  \and S\'ebastien Bardeau\inst{\ref{IRAM}} %
  \and Karine Demyk\inst{\ref{IRAP}} %
  \and Victor de Souza Magalh\~aes\inst{\ref{NRAO}} %
  \and Javier R. Goicoechea\inst{\ref{CSIC}} %
  \and Pierre Gratier \inst{\ref{LAB}} %
  \and Viviana V. Guzm\'an\inst{\ref{Catholica}} %
  \and Annie Hughes\inst{\ref{IRAP}} %
  \and François Levrier\inst{\ref{LPENS}} %
  \and Jacques Le Bourlot\inst{\ref{LERMA/MEUDON}} %
  \and Dariusz C. Lis\inst{\ref{JPL}} %
  \and Harvey S. Liszt\inst{\ref{NRAO}} %
  \and Nicolas Peretto\inst{\ref{UC}} %
  \and Evelyne Roueff\inst{\ref{LERMA/PARIS}}
  \and Albrecht Sievers\inst{\ref{IRAM}} %
}

\institute{%
  IRAM, 300 rue de la Piscine, 38406 Saint-Martin-d'H\`eres,  France. \email{einig@iram.fr}\label{IRAM} %
  \and LERMA, Observatoire de Paris, PSL Research University, CNRS, Sorbonne Universit\'es, 92190 Meudon, France. \email{pierre.palud@obspm.fr}\label{LERMA/MEUDON} %
  \and Univ. Grenoble Alpes, Inria, CNRS, Grenoble INP, GIPSA-Lab, Grenoble, 38000, France.\label{GIPSA-Lab} %
  \and Univ. Lille, CNRS, Centrale Lille, UMR 9189 - CRIStAL, 59651 Villeneuve d'Ascq, France.\label{CRISTAL} %
  \and Université Paris Cité, CNRS, Astroparticule et Cosmologie, F-75013 Paris, France.\label{APC} %
  \and Univ. Toulon, Aix Marseille Univ., CNRS, IM2NP, Toulon, France.\label{Toulon} %
  \and LERMA, Observatoire de Paris, PSL Research University, CNRS, Sorbonne Universit\'es, 75014 Paris, France.\label{LERMA/PARIS} %
  \and Department of Earth, Environment, and Physics, Worcester State University, Worcester, MA 01602, USA.\label{WORCESTER} %
  \and Center for Astrophysics | Harvard \& Smithsonian, 60 Garden Street, Cambridge, MA 02138, USA.\label{CfA} %
  \and Instituto de Física Fundamental (CSIC), Calle Serrano 121, 28006, Madrid, Spain.\label{CSIC} %
  \and Department of Astronomy, University of Florida, P.O. Box 112055, Gainesville, FL 32611, USA.\label{UF} %
  \and Institut de Recherche en Astrophysique et Planétologie (IRAP), Université Paul Sabatier, Toulouse cedex 4, France.\label{IRAP} %
  \and National Radio Astronomy Observatory, 520 Edgemont Road, Charlottesville, VA, 22903, USA.\label{NRAO} %
  \and Laboratoire d'Astrophysique de Bordeaux, Univ. Bordeaux, CNRS,  B18N, Allée Geoffroy Saint-Hilaire, 33615 Pessac, France.\label{LAB} %
  \and Instituto de Astrofísica, Pontificia Universidad Católica de Chile, Av. Vicuña Mackenna 4860, 7820436 Macul, Santiago, Chile.\label{Catholica} %
  \and Laboratoire de Physique de l'\'Ecole normale supérieure, ENS, Université PSL, CNRS, Sorbonne Université, Université de Paris, Sorbonne Paris Cité, Paris, France.\label{LPENS} %
  \and Jet Propulsion Laboratory, California Institute of Technology,  4800 Oak Grove Drive, Pasadena, CA 91109, USA.\label{JPL}
  \and School of Physics and Astronomy, Cardiff University, Queen's buildings, Cardiff CF24 3AA, UK.\label{UC} %
} %

% \date{Received September 15, 1996; accepted March 16, 1997}

\abstract%
% context heading (optional) {} leave it empty if necessary
{
    Observations of ionic, atomic, or molecular lines are performed to improve our understanding of the interstellar medium (ISM).
    However, the potential of a line to constrain the physical conditions of the ISM is difficult to assess quantitatively, because of the complexity of the ISM physics.
    The situation is even more complex when trying to assess which combinations of lines are the most useful.
    Therefore, observation campaigns usually try to observe as many lines as possible for as much time as possible.
}
{
    We search for a quantitative statistical criterion to evaluate the full constraining power of a (or combination of) tracer(s) with respect to physical conditions.
    Our goal with such a criterion is twofold.
    First, we want to improve our understanding of the statistical relationships between ISM tracers and physical conditions.
    Secondly, by exploiting this criterion, we aim to propose a method that helps observers to motivate their observation proposals, \eg, by choosing to observe the lines with the highest constraining power given limited resources and observation time.
}
{
    We propose an approach based on information theory, in particular the concepts of conditional differential entropy and mutual information.
    The best (combination of) tracer(s) is obtained by comparing the mutual information between a physical parameter and different sets of lines.
    The presented analysis is independent of the choice of the estimation algorithm (\eg, neural network or $\chi^2$ minimization).
    We apply this method to simulations of radio molecular lines emitted by a photodissociation region similar to the Horsehead Nebula.
    In this simulated data, we consider the noise properties of a state-of-the-art single dish telescope such as the IRAM~30m telescope.
    We search for the best lines to constrain the visual extinction \AV{} or the far UV illumination \Guv{}.
    We run this search for different gas regimes, namely translucent gas, filamentary gas, and dense cores.
}
{
    The most informative lines change with the physical regime (\eg, cloud extinction).
    However, the determination of the optimal combination of lines to constrain a physical parameter such as the visual extinction depends not only on the radiative transfer of the lines and chemistry of the associated species, but also on the achieved mean signal-to-noise ratio.
    Short integration time of the CO isotopologue $J=1-0$ lines already yields much information on the total column density for a large range of (\AV{}, \Guv{}) space.
    The best set of lines to constrain the visual extinction does not necessarily combine the most informative individual lines.
    Precise constraints on the radiation field are more difficult to achieve with molecular lines.
    They require spectral lines emitted at the cloud surface (\eg, \cp{} and \ci{} lines).
}
{
  This approach allows one to better explore the knowledge provided by ISM codes, and to guide future observation campaigns.
}


\maketitle{} %

%%%%%%%%%%%%%%%%%%%%%%%%%%%%%%%%%%%%%%%%%%%%%%%%%%%%%%%%%%%%%%%%%%%%%%%%%%%

\section{Introduction}%
\label{sec:introduction}

The effect of the feedback of a newborn star on its parent molecular cloud is to this day poorly understood.
The newborn star overall dissipates the parent cloud, leading to a decrease in its star forming capability.
However, it also causes a local compression of the gas, which may trigger a gravitational collapse.
Both spatially resolved observations of star forming regions and refined numerical models are needed to better understand the physical phenomena involved.
A difficulty for interstellar medium (ISM) studies is that observing many lines in the infrared or millimeter domains is expensive and can require several successive observations with different instrument settings.
It appears that using statistical arguments to determine the most relevant
tracer to observe in order to estimate a given physical parameter (\eg, the
cloud visual extinction, the gas volume density, the thermal pressure) received only limited attention from the ISM community.
This work provides a general approach based on information theory to compare the information provided by different tracers and sets of tracers.

This paper is the first of a series of two on applications of information theory concepts to ISM studies.
This paper has two goals.
First, it aims to show that tools from information theory can be exploited to visualize and better understand the complex statistical relationships between physical conditions and noisy observations.
Second, it aims to provide a tool to guide future observations in choosing the best lines to observe, and for how long, to accurately estimate physical parameters such as the gas column density (or visual extinction), the intensity of the incident UV field, and the thermal pressure.
The results of such a study heavily depend on the signal-to-noise ratio (S/N) for each line, \ie{} on the instrument properties, on the integration time and on the observed environment.
To achieve these two goals, we define a general method and apply it to data simulated with a fast, accurate emulation of the Meudon PDR code~\citep{lepetitModelAtomicMolecular2006,paludNeuralNetworkbasedEmulation2023} and a realistic noise model.
The proposed approach is applicable to any ISM model combined with any noise model.
The next paper will use real data from the \mbox{ORION-B} Large Program~(co-PIs: J.\,Pety \& M.\,Gerin,~\citealt{petyAnatomyOrionGiant2017}), with a focus on photodissociation regions (PDRs).

Selecting the most informative lines to estimate a physical parameter (\eg, visual extinction or gas volume density) is an instance of a machine learning problem called feature selection~\citep[chapter 25]{shalev-shwartzUnderstandingMachineLearning2014}.
A straightforward and common approach is to evaluate Pearson's correlation coefficient between individual lines and individual physical parameters of interest.
The lines with the highest correlation with a given physical parameter would then be selected.
This method is common in ISM studies, see, \eg,~\citet{petyAnatomyOrionGiant2017}.
However, it suffers from three main drawbacks.
First, it is restricted to one-to-one relationships, while one might be interested in selecting multiple lines to predict multiple physical parameters at once.
Second, it is restricted to linear relationships, and cannot fully capture nonlinear dependencies between lines and physical parameters.
Third, by considering tracers individually, it neglects their complementarity -- \ie, the possibility for a group of lines to be more informative than any single emission line from the group -- while such complementarities are already known and studied with line ratios or line combinations.
For instance,~\citep{kaufmanFarInfraredSubmillimeter1999} studies line combinations and ratios in order to disentangle several physical parameters whose estimates would be degenerate with a single tracer.

The canonical coefficient analysis~\citep{hardleCanonicalCorrelationAnalysis2007} enables considering correlations between multiple lines and multiple physical parameters.
It alleviates the one-to-one relationship restriction and enables to account for many-to-many relationships and thus to include line complementarities.
This approach provides multiple correlation coefficients in the many-to-many case.
The difficulty with this method is that ranking lines based on multiple correlation coefficients is not trivial.
As shown in the following, these coefficients can be combined into one number which is  interpretable if both observed lines and physical parameters are normally distributed.

Predictor-dependent methods can address the linear and Gaussian limitations.
Such methods rely on a regression model, \eg, random forests or neural networks.
The greedy selection algorithm~\citep[sect.~25.1]{shalev-shwartzUnderstandingMachineLearning2014} would iteratively select tracers to reduce the error of a type of regression model.
Similarly, the greedy elimination method would iteratively remove tracers.
For instance,~\citet{bronTracersIonizationFraction2021} applied numerous random forest regressions to predict ionization fraction using only one tracer at a time.
Then, they defined the best tracers as those leading to the minimum sum of residual squares.
Other statistical methods exploit specificities of a predictor class to explain the predictions of a model and remove unused features.
For instance,~\citep{gratierQuantitativeInferenceColumn2021} used feature importance from random forests to assess the predictive power of individual lines or on the H$_2$ column density.
However, the tracer subsets obtained with these approaches heavily depend on the considered type of regression model.

Finally, explainable AI methods such as SHAP values~\citep{lundbergUnifiedApproachInterpreting2017} can be used to understand a numerical model and identify its most important features.
This kind of approach was already applied in ISM studies, for instance in~\citet{heylUnderstandingMolecularAbundances2023} and~\citet{ramosFastNeuralEmulator2024}.
However, this class of methods only addresses deterministic methods, and is thus not able to handle noisy observations.
Besides, it is limited to one-to-one relationships and scales poorly with the number of features.
Some fast variants exist such as Kernel SHAP~\citep{lundbergUnifiedApproachInterpreting2017}, but require the features to be independent, which is strongly violated with ISM lines.

In this work, we propose to exploit entropy and mutual information~\citep[sect.~8.6]{coverElementsInformationTheory2006}.
Mutual information has already been exploited in astrophysics tasks (see, \eg,~\citealt{pandeyHowMuchGalaxy2017}), although not in the ISM community to the best of our knowledge.
It does not depend on the choice of a regression model, handles at once multiple lines and multiple physical parameters, does not assume any distribution for lines or physical parameters, and accounts for nonlinearities and line complementarities.
The methodology proposed in this work can be adapted to other problems with the associated Python package called \textsc{InfoVar}%
\footnote{\url{https://pypi.org/project/infovar/}}%
, which stands for ``informative variables''.
The results in this paper are produced using a dedicated Python package%
\footnote{\url{https://github.com/einigl/iram-30m-emir-obs-info}}%
, which is based on \textsc{InfoVar} and designed for the generation and the statistical analysis of synthetic line observations.
All the scripts used to generate these results are freely available%
\footnote{\url{https://github.com/einigl/informative-obs-paper}}%
.

Section~\ref{sec:entropy} reviews the three information theory quantitative criteria our method builds upon, namely entropy, conditional entropy and mutual information.
Section~\ref{sec:optimization_pb} formalizes the line selection problem and introduces an approximate solution that accounts for numerical uncertainties.
Section~\ref{sec:dataset} sets up an application of the proposed method to PDRs with the Meudon PDR code on IRAM's EMIR instrument.
Section~\ref{sec:global-results} presents and analyzes global results of this application.
Section~\ref{sec:results_line_selection} applies the line selection method to different environments.
Section~\ref{sec:conclu} provides some concluding remarks.

%%%%%%%%%%%%%%%%%%%%%%%%%%%%%%%%%%%%%%%%%%%%%%%%%%%%%%%%%%%%%%%%%%%%%%%%%%%

\section{Information theory toolkit}%
\label{sec:entropy}

This section reviews the information theory concepts that the proposed approach builds upon.
We first define the considered physical model.
Secondly, Shannon and differential entropies are introduced.
Entropy is the building block of mutual information, which allows us to compare how informative subsets of lines are.
Table~\ref{tab:summary-info-theory} summarizes the information theory quantities to be introduced in sections~\ref{subsec:entropy_diff} to~\ref{sec:mi}.

In a nutshell, the physical parameters and the line intensities are considered as dependent random variables.
The entropy of physical parameters characterizes their distribution uncertainty before any measurements.
The mutual information between a physical parameter and a set of line intensities quantifies the information gain on the physical parameter when observing line intensities.
A high value of mutual information for a given line thus indicates that an observation would constrain well the inferred value of the physical parameter.

\begin{table*}
    \centering
    \caption{
        Overview of the information theory quantities used in this work.
    }%
    \label{tab:summary-info-theory}
    \begin{tabular}{ccccc}
        \hline
        \hline\\[-1mm]
        Quantity
        & Notation
        & Domain
        & Relationship with other quantities
        & Interpretation\\[2mm]
        \hline\\[-1mm]
        Differential entropy
        & $\h{\paramfull}$
        & $]-\infty,+\infty[$
        & --
        & uncertainty on $\paramfull$ before any measurement\\[2mm]
        Conditional diff. entropy
        & $\condh{\paramfull}{\obsfull}$
        & $]-\infty,+\infty[$
        & $\condh{\paramfull}{\obsfull} = \h{\paramfull, \obsfull} - \h{\obsfull}$
        & remaining uncertainty on $\paramfull$ when $\obsfull$ is known \\[2mm]
        Mutual information
        & $\mi{\paramfull}{\obsfull}$
        & $[0,+\infty[$
        & $\mi{\paramfull}{\obsfull} = \h{\paramfull} - \condh{\paramfull}{\obsfull}$
        & statistical dependence between $\paramfull$ and $\obsfull$ \\[2mm]
        \hline
    \end{tabular}
\end{table*}

\subsection{Physical model}%
\label{subsec:obs_model_and_proba}

A physical model links physical conditions $\paramvect{}$ with observables $\obsvect{}$ by combining an ISM model $\truef$ and a simulator of observation $\noiseGeneral$ that includes all sources of noise.
In this work, we use it to generate a realistic set of $(\paramvect{}, \obsvect{})$ pairs, called sets of physical models.
We consider an ISM model $\truef$ that predicts the true value $\truef(\paramvect{}) = (\truefell(\paramvect{}) )_{\ell=1}^L$ of $L$ observables from a limited number of $D \lesssim 10$ physical parameters $\paramvect{} = (\paramelt{d})_{d=1}^D$.
For instance, in its version 7 released in 2024, the Meudon PDR code~\citep{lepetitModelAtomicMolecular2006} computes the integrated intensity of $5\ 375$ emission lines from the thermal pressure (or gas volume density), the intensity of the incident UV radiative field, the cloud visual extinction, the cosmic ray ionization rate, grain distribution properties, etc.
The model $\truef$ is assumed to simulate accurately the physics of the ISM.
This means that for a given set of physical conditions $\paramvect{}$ and a line of index $1 \leq \ell \leq L$, the predicted value $\truefell(\paramvect{})$ is considered to be the one a telescope would measure in the absence of noise.
In the remainder of this work, the considered observables $\obsvect{}$ are integrated intensities of emission lines associated with ionic, atomic or molecular quantum transitions.
However, the approach we propose could be applied with any kind of observable, such as line ratios, raw line profiles or other summary values such as the line width or maximum value.

The noise, as well as other observational effects, are included through the observation simulator $\noiseGeneral$.
Observed integrated intensities $\obsvect{} = (y_\ell)_{\ell=1}^L$ can thus be associated with physical conditions~$\paramvect{}$ using
\begin{align}
    y_\ell
    =
    \noiseGeneral \left( \truefell(\paramvect{}) \right)
    .
    \label{eq:obs_model_abstract}
\end{align}
This observation simulator can include, for instance, additive Gaussian noise for thermal effects or photon counting error, or multiplicative lognormal noise for calibration error.
To model the uncertainties due to the noise, we resort to random variables denoted $\paramfull$ and $\obsfull$ for physical conditions and observations, respectively.
For instance, for a subset $s$ of $K \in \{1,\dots,L\}$ lines, the simulator of observations in~\eqref{eq:obs_model_abstract} defines a probability distribution on observation $\obsfullsubset$ for a physical condition $\paramfull = \paramvect{}$.
This random variable is fully described with a probability density function (PDF) $\pi\left( \cdot \vert \paramvect{} \right)$, that is a function such that for any physical condition vector $\paramvect{} \in \R^D$ and observation $\obsvectsubset{} \in \R^K$, $\pi\left(\obsvectsubset{} \vert \paramvect{} \right) \geq 0$ and $\int \pi\left(\obsvectsubset{} \vert \paramvect{} \right) \, \diff \obsvectsubset{} = 1$.
Common probability distributions on multivariate random variables include the uniform distribution Unif$(\mathcal{C})$ on a set $\mathcal{C}$ and the normal distribution $\mathcal{N}(\boldmu, \Sigma)$ with $\boldmu$ the mean of the distribution and $\Sigma$ its covariance matrix -- also called Gaussian distribution.
This paper will also resort to the lognormal distribution which corresponds to the exponential of a normally distributed random variable.
In other words, if a random variable follows a lognormal distribution $\lognormal(\boldmu, \Sigma)$, then its log follows a Gaussian distribution of parameters $\boldmu$ and $\Sigma$.

This work aims at determining the subset of $K$ lines that best constrains the physical parameters $\paramfull$.
We expect the most informative lines to differ depending on the type of physical regime.
For instance, a line that can quickly become optically thick may be most informative on the visual extinction \AV{} in translucent or filamentary conditions, before it saturates.
We thus define different types of regime, characterized by different priors $\pi(\paramvect{})$, and determine the most informative subset of $K$ emission lines in each of these regimes.

\subsection{Two-dimensional illustrative example}%
\label{subsec:example}

We now introduce a simple synthetic example that will illustrate the information theory concepts defined below.
We use the simplest case where a physical process, controlled by a physical parameter
$\paramfull$, yields one value of $\obsfull$ per value of $\paramfull$.
Sources of uncertainty such as the presence of noise or hidden control variables can however blur the relationship between $\paramfull$ and $\obsfull$.
This implies that inferring the physical parameters from the observed quantity yields uncertain values.
By representing $\paramfull$ and $\obsfull$ as dependent random variables, the concepts of information theory allow us to quantify the uncertainty on the physical parameter $\paramfull$ before and after measuring $\obsfull$.

The distribution chosen to represent the couple $(\paramfull, \obsfull)$ is a two-dimensional lognormal distribution.
Its parameters correspond to the mean vector and covariance matrix in the logarithmic scale.
They are set to obtain unit expectations, a standard deviation such that a $1\sigma$ error corresponds to a factor 1.3, and a $\rho = 0.9$ correlation coefficient in linear scale.
Appendix~\ref{subsec:example_detailed} gathers details on the associated computations.

\begin{figure}
    \centering
    \includegraphics[width=0.9\linewidth, trim={2mm 0 0 0}, clip]{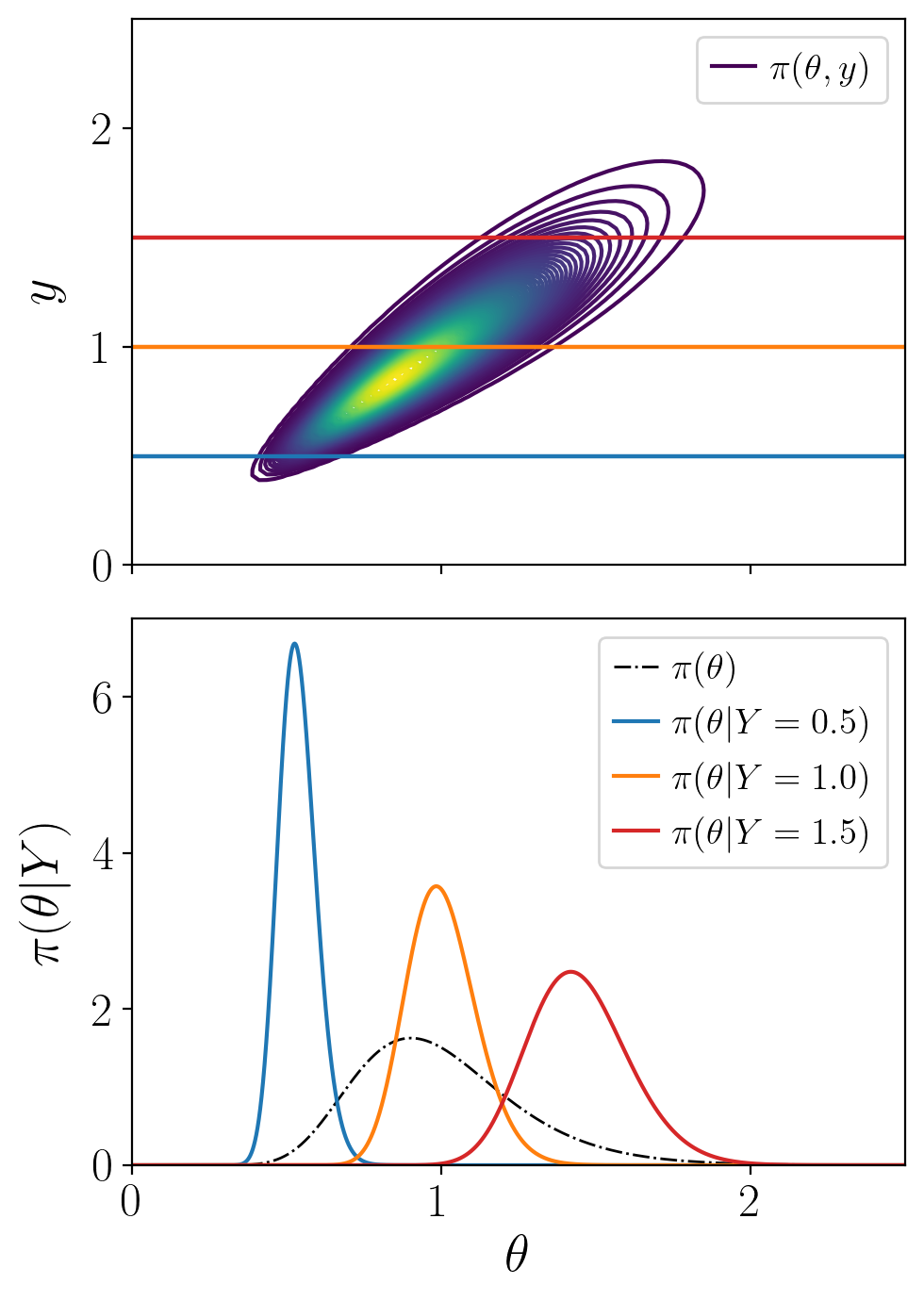}
    \caption{
        A simple synthetic example of a joint distribution on the couple $(\paramfull{}, \obsfull)$.
        Top: contour levels of the PDF of the joint distribution with lognormal marginals and a clear correlation.
        Three observed values are indicated with horizontal lines.
        Bottom: comparison of the distribution on $\paramfull{}$ before any observation (prior, in dashed black) and for the three $y$ values (conditional distributions, in colors).
    }%
    \label{fig:gaussian-example}
\end{figure}

The top panel of Fig.~\ref{fig:gaussian-example} shows the PDF of the joint distribution $\pi(\theta, y)$.
The bottom panel compares the prior distribution $\pi(\theta)$ (\ie, the distribution of the physical parameter before any observation) with three conditional distributions $\pi(\theta \vert y)$ (\ie, each distribution of the physical parameter values consistent with one observed value $Y=y$).
Each represented conditional distribution is tighter and has lighter tails than the prior distribution, which indicates that observing $\obsfull$ reduces the uncertainty on $\paramfull$.
Besides, among the three considered observed values of $\obsfull$, the lower ones lead to the tightest conditional distribution, and thus to lower uncertainty on $\paramfull$.
The information theory concepts to be introduced in the next sections quantify this notion of uncertainty.

\subsection{Entropy for discrete random variables}%
\label{subsec:entropy_finite}

The notion of entropy was first introduced by Boltzmann and Gibbs in the 1870s as a measure of the disorder of a system.
It plays a key role in the second law of thermodynamics, which establishes the irreversibility of the macroscopic evolution of an isolated particle system despite the reversibility of microscopic processes.
In a large system where particles can only be in a finite set $\mathcal{X}$ of $\Omega \geq 1$ states, the state of one particle can be modeled as a discrete random variable $X$.
This random variable is fully described with a probability mass function $\pi$, \ie, a function such that for any state $x \in \mathcal{X}$, $\pi(x) \geq 0$ and $\sum_{x \in \mathcal{X}} \pi(x) = 1$.
In this setting, $\pi(x)$ is the probability for a particle to be in the state $x$.
The entropy is then defined as~\citep{wehrl1978general}
\begin{align}
    \label{eq:boltzmann-entropy}
    S = - k_B \sum_{x \in \mathcal{X}} \left[\ln \pi(x) \right] \; \pi(x)
    ,
\end{align}
with $k_B$ the Boltzmann constant.

In information theory, the entropy refers to that introduced in~\citet{shannonMathematicalTheoryCommunication1948}.
Informally, it measures the uncertainty or lack of information in a probability distribution.
The entropy of a discrete random variable $X$ is defined by~\citep[chapter 2]{coverElementsInformationTheory2006}
\begin{align}
    \label{eq:entropy_finite}
    H(X)
    =
    \mathbb{E}_{X} \left[
        - \log_2 \pi(X)
    \right]
    =
    - \sum_{x \in \mathcal{X}}
    \left[ \log_2 \pi(x) \right] \; \pi(x)
    .
\end{align}
The two definitions are equivalent up to the considered units.
The base-2 logarithm in~\eqref{eq:entropy_finite} leads to entropy values in bits.

The entropy is bounded and always positive.
The entropy equals exactly $0$ when $\pi(x) = 1$ for a single state $x \in \mathcal{X}$ and $0$ for all the others.
In this first case, the probability distribution does not contain any uncertainty.
For a particle system, this case corresponds to all particles being in the same state $x$.
Conversely, both definitions are maximized with the uniform distribution, \ie, when for all states $x \in \mathcal{X}$, $\pi(x) = 1/ \Omega$.
In this second case, the uncertainty is indeed maximum, in the sense that none of the states is favored.
This uniform distribution limit corresponds to a macroscopic thermodynamic equilibrium, where~\eqref{eq:boltzmann-entropy} reduces to the well known formula (often called the Boltzmann equation) $S = k_B \ln \Omega$ or, equivalently,~\eqref{eq:entropy_finite} reduces to $H(X) = \log_2 \Omega$.

Shannon used the entropy to prove that there exists a code that can compress the data for storage and transmission.
Shannon not only proposed the algorithm, but also quantified the optimal performances that can be reached.
In this context, Shannon entropy in base 2 corresponds to the average minimum length of a binary message to encode an information.
A fundamental property of entropy, namely the additivity of independent sources of information, states that, for any couple of independent random variables $X_1,\,X_2$,
\mbox{$H(X_1,X_2) = H(X_1) + H(X_2)$}.
In other words, the minimum length of a message containing two uncorrelated parts is the sum of the lengths required to encode each of the parts.
More generally, the uncertainty of a couple of independent random variables is the sum of their individual uncertainties.

\subsection{Differential entropy for continuous random variables}%
\label{subsec:entropy_diff}

As introduced in Sect.~\ref{subsec:obs_model_and_proba}, this work relies on continuous random variables, namely subsets of lines $\obsfullsubset \in \R^K$ and physical parameters $\paramfull \in \R^D$, \eg, visual extinction or incident UV radiative field intensity.
For continuous random variables, the information theory notion of entropy is generalized with the so-called differential entropy~\citep[chapter 8]{coverElementsInformationTheory2006}:
\begin{align}
    \label{eq:diff_entropy}
    \h{\paramfull}
    =
    \mathbb{E}_{\paramfull} \left[
        - \log_2 \pi(\paramfull)
    \right]
    =
    - \int
    \left[ \log_2\pi(\paramvect{}) \right]
    \; \pi(\paramvect{})
    \, \diff \paramvect{}
    ,
\end{align}
with $\pi(\paramvect{})$ the PDF of $\paramfull$.
The differential entropy $\h{\paramfull}$ is the limit of the discrete
entropy $H$ of a quantized variable $\paramfull^\Delta$, where $\Delta$
corresponds to quantization step~\citep[theorem
8.3.1]{coverElementsInformationTheory2006}
\begin{align}
    \h{\paramfull} = \lim_{\Delta \to 0} H \left( \paramfull^\Delta \right) + \log_2 \Delta
    .
\end{align}
Unlike the finite case, the differential entropy can take negative values, as $\log_2 \Delta < 0$ when $\Delta < 1$.
Table~\ref{tab:entropy_formulae} lists the differential entropy formulae of a few common parametric distributions.
For instance, the entropy of a Gaussian distribution only depends on its variance and not on its mean.
The entropy of a uniform distribution on a compact set is the logarithm of the set volume.

For the example from Sect.~\ref{subsec:example},
using the lognormal formula from Table~\ref{tab:entropy_formulae}, the
uncertainty on $\paramfull$ before any observation is $\h{\paramfull} =
0.07$\,bits.
This corresponds to the uncertainty contained in a uniform distribution on an interval of size \mbox{$2^{0.07} = 1.05$}, or in a univariate Gaussian distribution of standard deviation \mbox{$\sigma = 0.25$}.

The entropy can also be computed for couples of random variables.
For instance, when considering the problem of inferring $\paramfull$ from $\obsfullsubset$, we can now introduce the differential entropy on the couple $\left(\paramfull,\obsfullsubset \right)$ that is defined as
\begin{align}
    \label{eq:joint_diff_entropy}
    \h{\paramfull, \obsfullsubset}
    & = \mathbb{E}_{\paramfull{}, \obsfullsubset} \left[ - \log_2 \pi \left(\paramfull{}, \,\obsfullsubset\right)\right] \\
    & =
    - \int
    \left[ \log_2 \pi \left(\paramvect{}, \obsvectsubset{}\right) \right]
    \; \pi \left(\paramvect{}, \obsvectsubset{} \right)
    \, \diff \paramvect{} \, \diff \obsvectsubset{},
\end{align}
where $\pi\left(\paramvect{}, \obsvectsubset{} \right)$ is the joint PDF of the couple $\left(\paramfull{}, \obsfullsubset{} \right)$.

\begin{table}[t]
  \centering
    \caption{Differential entropy for a few common distributions.}
    \label{tab:entropy_formulae}
    \renewcommand{\arraystretch}{1.3}
    \addtolength{\tabcolsep}{-0.5mm}
    \begin{tabular}{c|c|c}
        \hline
        \hline
        \multicolumn{2}{c|}{Distribution on $\paramfull$} & Differential entropy $\h{\paramfull}$ \\
        \hline
        \xrowht[()]{20pt}
        \multirow{6}{*}{\begin{tabular}{@{}c@{}}
            General \\
            $\paramfull \in \R^D$ \\
            $D \geq 1$
        \end{tabular}}
        & \begin{tabular}{@{}c@{}}
            $\mathcal{N}(\boldmu, \Sigma)$ \\
            $\boldmu \in \R^D$, $\Sigma  \in \R^{D\times D}$
        \end{tabular}
        & \(\displaystyle \frac{1}{2} \log_2\left[ (2\pi e)^D \, \vert \Sigma \vert \right] \)\\
        %
        % \xrowht[()]{20pt}
        \cline{2-3}
        \xrowht[()]{20pt}
        &  \begin{tabular}{@{}c@{}}
            $\Unif{\fvalidity}$ \\
            $\fvalidity \subset \R^D$
        \end{tabular}
        & $\log_2\text{Vol} \, \fvalidity$ \\
        %
        % \xrowht[()]{20pt}
        \cline{2-3}
        \xrowht[()]{20pt}
        & \begin{tabular}{@{}c@{}}
            $\lognormal(\boldmu, \Sigma)$ \\
            $\boldmu \in \R^D$, $\Sigma  \in \R^{D\times D}$
        \end{tabular}
        & \(\displaystyle \frac{1}{2} \log_2\left[ (2\pi e)^D \;  \vert \Sigma \vert \; e^{2 \sum_{d=1}^D \mu_d} \right] \)\\[2pt]
        \hline
        \xrowht[()]{20pt}
        \multirow{6}{*}{\begin{tabular}{@{}c@{}}Univariate \\ $\paramfull \in \R$ \\ $D=1$
        \end{tabular}}
        & \begin{tabular}{@{}c@{}}
            $\mathcal{N}(\mu, \sigma^2)$ \\
            $\mu \in \R, \sigma > 0$
        \end{tabular}
        &  \(\displaystyle \frac{1}{2} \log_2 \left[ 2 \pi e \; \sigma^2 \right] \) \\
        \cline{2-3}
        \xrowht[()]{20pt}
        & \begin{tabular}{@{}c@{}}
            $\Unif{a, b}$ \\ $a < b$
        \end{tabular}
        & $ \log_2 \left[ b-a \right]$ \\
        \cline{2-3}
        \xrowht[()]{20pt}
        & \begin{tabular}{@{}c@{}}
            $\lognormal(\mu, \sigma^2)$ \\
            $\mu \in \R, \sigma > 0$
        \end{tabular}
        &  \(\displaystyle \frac{1}{2} \log_2 \left[ 2 \pi e \; \sigma^2 \; e^{2 \mu} \right] \)\\
        \hline
    \end{tabular}
    \tablefoot{
      As introduced in Sect.~\ref{subsec:obs_model_and_proba}, $\mathcal{N}$ denotes a Gaussian distribution, Unif a uniform distribution and $\lognormal$ a lognormal distribution.
      Vol$(\mathcal{C})$ is the volume of a set $\mathcal{C}$,
      $\vert \Sigma \vert$ is the determinant of a covariance matrix $\Sigma$.
    }
\end{table}

\subsection{Conditional differential entropy: effects of observations}%
\label{subsec:conditional_entropy}

Observations are performed in order to infer physical parameters~$\paramfull{}$.
In Sect.~\ref{subsec:obs_model_and_proba}, we described observations that include noise.
Observing a vector $\obsvectsubset{}$ thus does not permit one to determine the physical conditions $\paramfull$ with infinite precision.
However, it can reduce the uncertainty on the physical parameters $\paramfull$.

The conditional differential entropy $\condh{\paramfull}{\obsfullsubset}$ quantifies the expected uncertainty remaining on $\paramfull$ when $\obsfullsubset$ is known, \ie, after a future observation.
It is defined as
\begin{align}
    \label{eq:cond_diff_entropy}
    \condh{\paramfull}{\obsfullsubset}
    & = \mathbb{E}_{\paramfull{},\obsfullsubset} \left[ - \log_2 \pi \left(\paramfull \, \vert \,\obsfullsubset\right)\right]  \\
    & =
    - \int
    \left[ \log_2 \pi \left(\paramvect{} \, \vert \, \obsvectsubset{}\right) \right]
    \; \pi \left(\paramvect{}, \obsvectsubset{} \right)
    \, \diff \paramvect{} \, \diff \obsvectsubset{}.
    \label{eq:cond_diff_entropy_diff_entropies}
\end{align}
The conditional differential entropy $\condh{\paramfull}{\obsfullsubset}$ is a mean value characterizing all the possible joint realizations of the observations and the physical parameters.
It is therefore not a function of a specific realization $\obsvectsubset{}$ of the $\obsfullsubset$ random variable.
Instead, it quantifies how a future observation $\obsvectsubset{}$ of $\obsfullsubset$ would affect the uncertainty on the physical conditions $\paramfull$ on average.
This average is computed with respect to the joint distribution of physical parameters $\paramfull$ and observations $\obsfullsubset$.
The conditional differential entropy can thus be evaluated prior to any observation and estimation.
It can be shown that
\begin{equation}
  \label{eq:cond_joint_entropy}
  \condh{\paramfull}{\obsfullsubset} = \h{\paramfull{},\obsfullsubset}
    - \h{\obsfullsubset}.
\end{equation}
This means that the remaining uncertainty on $\paramfull$, once $\obsfullsubset$ is known, is the information jointly carried by both $\Theta$ and $\obsfullsubset{}$ minus the information brought by $\obsfullsubset{}$ alone.
In other words, knowing $\obsfullsubset$ provides additional information to estimate $\paramfull$.
This implies that the conditional differential entropy is always lower or equal to the differential entropy:
\begin{align}
    \label{eq:conditional_entropy_lowerthan_entropy}
    \condh{\paramfull}{\obsfullsubset}
    \leq
    \h{\paramfull}
    .
\end{align}
This inequality becomes an equality if and only if $\paramfull$ and
$\obsfullsubset$ are independent.
This can occur for instance in the low S/N regime, when additive noise
completely dominates the line intensity.
Conversely, if there exist a bijection between $\paramfull{}$ and $\obsfullsubset{}$, \eg in the absence of noise and with a bijective $\truef$ in~\eqref{eq:obs_model_abstract}, then $\condh{\paramfull}{\obsfullsubset}$ is equal to $-\infty$.

The example of Sect.~\ref{subsec:example} shows how different values of $\obsfull$ yield different uncertainties on $\paramfull$.
The lower panel in Fig.~\ref{fig:gaussian-example} shows that, among the three observed $y$, lower values of $y$ lead to a tighter distribution and thus to lower uncertainties on $\paramfull$.
The remaining uncertainty on $\paramfull$ is $-2.01$, $-1.11$, or $-0.58$\,bits after observing $y=0.5$, $1$, or $1.5$, respectively.
The conditional differential entropy $\condh{\paramfull}{\obsfull}$ averages over all possible observations $y$.
Using~\eqref{eq:cond_diff_entropy_diff_entropies} and the lognormal formulae from Table~\ref{tab:entropy_formulae}, in this case, $\condh{\paramfull}{\obsfull} = -1.08 - 0.07 = -1.15$\,bits.
The latter value is the mean uncertainty on $\paramfull$ when observing $\obsfull$, averaged on all possible values of $\obsfullsubset$.

The differential entropy $\condh{\paramfull}{\obsfull}$ is related to the error in estimating $\paramfull$ from the $\obsfull$ data, and in particular to the root mean squared error.
For instance, in an estimation procedure, decreasing the entropy by 1 bit improves the estimation precision%
\footnote{
    In this paper, the precision is considered to be homogeneous with the inverse of a standard deviation.
    This differs from the traditional definition in statistics, where it corresponds to the inverse of a variance.
}
by a factor 2 in the Gaussian case.
Appendix~\ref{sec:what_is_a_bit} illustrates the notion of a difference of one bit between two probability distributions.
An interpretation valid in the general case will be presented in the second paper of this series.

\subsection{Mutual information}%
\label{sec:mi}

\begin{figure}
    \centering
    \includegraphics[width=0.45\textwidth]{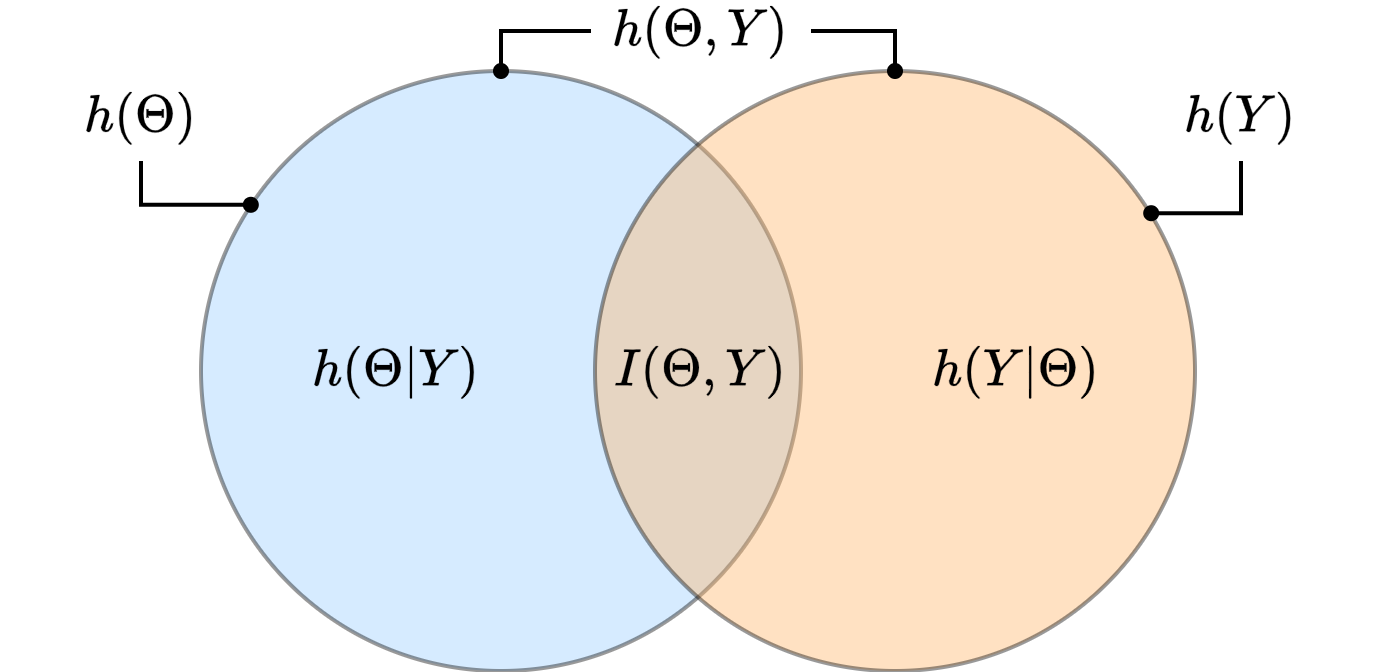}
    \caption{
        Venn diagram representation of the differential entropy $\h{\paramfull}$ (and $\h{\obsfull}$),
        of the conditional differential entropy $\condh{\paramfull}{\obsfull}$ (and $\condh{\obsfull}{\paramfull}$),
        and of the mutual information $\mi{\paramfull}{\obsfull}$.
    }%
    \label{fig:entropy_venn}
\end{figure}

The mutual information $\mi{\paramfull}{\obsfullsubset}$~\citep[sect. 8.6]{coverElementsInformationTheory2006} is often preferred for a simpler interpretation.
It quantifies the information on $\paramfull$ that is gained by knowing $\obsfullsubset$:
\begin{align}
    \label{eq:mi}
    \mi{\paramfull}{\obsfullsubset}
    =
    \h{\paramfull} - \condh{\paramfull}{\obsfullsubset}
    .
\end{align}
Figure~\ref{fig:entropy_venn} shows a Venn diagram that illustrates the relationships between differential entropy, conditional differential entropy and mutual information.
It illustrates~\eqref{eq:cond_joint_entropy} and~\eqref{eq:mi}.

Mutual information is always positive, as implied by~\eqref{eq:conditional_entropy_lowerthan_entropy}.
A high mutual information indicates that knowing $\obsfullsubset$ considerably lowers the uncertainty on $\paramfull$.
If we consider different distributions of a given physical parameter (\eg, corresponding to different physical regimes), represented by different random variables $\paramfull$, the mutual information is delicate to compare as it depends on the initial uncertainty.
Indeed, it is easier to provide information on the physical parameter if the latter is highly uncertain than if it is already precisely constrained.

The mutual information is invariant to invertible transformations of $\paramfull$ or $\obsfullsubset$ separately.
Its value is thus identical whether integrated intensities are considered in linear scale, logarithm scale or with a $\asinh$ transformation as in~\citet{gratierDissectingMolecularStructure2017}.
Conversely, non-bijective transformations result in a loss of information, and thus decrease the mutual information.
For instance, an integrated intensity is obtained with a non-invertible integration of the associated line profile, and thus contains less information.

In the example from Sect.~\ref{subsec:example}, the value of mutual information is $\mi{\paramfull}{\obsfull} = 1.22$\,bits, \ie, the difference between $\h{\paramfull} = 0.07$\,bits and $\condh{\paramfull}{\obsfull} = -1.15$\,bits.
This means that observing $\obsfull$ increases the information on $\paramfull$ by $1.22$\,bits on average.
Equivalently, observing $\obsfull$ improves the precision on $\paramfull$ by a factor $2^{1.22} \simeq 2.3$, on average.

%%%%%%%%%%%%%%%%%%%%%%%%%%%%%%%%%%%%%%%%%%%%%%%%%%%%%%%%%%%%%%%%%%%%%%%%%%%

\subsection{Finding the lines that best constrain physical parameters}%
\label{sec:optimization_pb}

Constraining a physical parameter is commonly defined as reducing the uncertainty associated with it.
In information theory, this uncertainty is quantified by the conditional entropy $\condh{\paramfull}{\obsfull}$.
The best subset $s_K$ of $K$ lines for a given physical regime is then the solution of the discrete optimization problem
\begin{align}\label{eq:optim_problem_formulation_condh}
    s_K
    =
    \argmin_{s \in \mathcal{S}_K} \;
    \condh{\paramfull}{\obsfullsubset}
    ,
\end{align}
with $\mathcal{S}_K$ the set of all possible subsets of $K$ lines.
Using the relationship \mbox{$\condh{\paramfull}{\obsfullsubset} = \h{\paramfull} - \mi{\paramfull}{\obsfullsubset}$}, the problem can be restated as maximizing mutual information such that an equivalent formulation is
\begin{align}\label{eq:optim_problem_formulation}
    s_K
    =
    \argmax_{s \in \mathcal{S}_K} \;
    \mi{\paramfull}{\obsfullsubset}
    .
\end{align}
This optimization problem is solved by comparing mutual information values for all subsets $s \in \mathcal{S}_K$.
The entropy and mutual information values are heavily dependent on the choice of prior on the $\paramfull$ distribution.

Solving~\eqref{eq:optim_problem_formulation} requires the ability to evaluate the mutual information for each pair $\left(\paramfull, \, \obsfullsubset \right)$.
In real-life applications, the shape of the distribution on $\left(\paramfull, \obsfullsubset\right)$ can be complex or unknown.
In such cases, the mutual information does not have a simple closed-form expression, unlike the simple cases listed in Table~\ref{tab:entropy_formulae}.
It then needs to be evaluated numerically with a Monte Carlo estimator $\miEst{\paramfull}{\obsfullsubset{}}$ from a set of $N$ pairs $\left(\paramvect{n}, \obsvectsubset{n}\right)$. %_{n=1}^N$.

The Monte Carlo estimator $\miEst{\paramfull}{\obsfullsubset{}}$ considered in the remainder of this work is the ``Kraskov estimator''~\citep{kraskovEstimatingMutualInformation2004}.
This estimator does not make assumptions on the shape of the joint distribution on $\left(\paramfull, \obsfullsubset\right)$.
It can thus capture both linear and nonlinear relationships between lines $\obsfullsubset$ and physical parameters $\paramfull$.
See Appendix~\ref{subsec:estimation} for more details on this estimator and the derivation of the associated error bars.

The set of $N$ pairs $\left(\paramvect{n}, \obsvectsubset{n}\right)$ can be made up of real observations or simulated observations.
This paper considers simulated observation.
The considered approach involves $3$ steps:
i) drawing $N$ physical parameters vectors $\paramvect{n}$ from a distribution $\pi(\paramvect{})$,
ii) evaluating the ISM model $\truef$ on each physical parameter $\paramvect{n}$ for all lines,
iii) applying the noise model $\noiseGeneral$ to obtain simulated noisy observations $\obsvect{n}$.
In the second paper of this series, the method is applied to a set of real observations.

%%%%%%%%%%%%%%%%%%%%%%%%%%%%%%%%%%%%%%%%%%%%%%%%%%%%%%%%%%%%%%%%%%%%%%%%%%%

\section{Application to simulated PDRs observed with IRAM~30m EMIR}%
\label{sec:dataset}

Mutual information, introduced in Sect.~\ref{sec:entropy}, allows one to evaluate the constraining power of ionic, atomic and molecular lines.
The general method presented in Sect.~\ref{sec:optimization_pb} allows one to determine which lines are the most informative to constrain the physical properties of an emitting object.
This method can be applied to any astrophysical model that computes line intensities from a few input parameters, \eg, radiative transfer codes simulating interstellar clouds, emission lines from protoplanetary disks, or stellar spectra synthesis models.
It can also be applied to any other spectroscopic observations.

In this section, we introduce two synthetic cases of PDRs.
In both cases, we resort to a fast and accurate emulator of the Meudon PDR code, and simulate noise using the characteristics of the EMIR receiver at the IRAM~30m.
With these two cases, we will show how mutual information can provide insights for ISM physics understanding, and apply the proposed line selection method.
As the results of the proposed approach heavily depend on various aspects -- \eg, the instrument properties, the integration time or the observed environment -- we depict these two cases in detail.

The Meudon PDR code is first presented along with a fast and accurate emulator.
Then, the details of the generation of the sets of models are introduced, namely, the physical parameter distribution and the observation simulator.
Overall, we consider two situations with distinct physical parameter distributions.

\subsection{The Meudon PDR code}%
\label{subsec:meudon_pdr}

The Meudon PDR code\footnote{\url{https://ism.obspm.fr}}~\citep{lepetitModelAtomicMolecular2006} is a one-dimensional stationary code that simulates a PDR, \ie, neutral interstellar gas illuminated with a stellar radiation field.
It permits the investigation, \eg, of the radiative feedback  of a newborn star on its parent molecular cloud, but it can also be used to simulate a variety of other environments.

% Here is an overview of how the code works.
%
The user specifies physical conditions such as the thermal pressure in the cloud~\Pth, the intensity of the incoming UV radiation field~\Gnaught~(scaling factor applied to the~\citealt{mathisInterstellarRadiationField1983} standard field), and the depth of the slab of gas expressed in visual extinctions~\AV.
The code then solves multiphysics coupled balance equations of radiative transfer, thermal balance, and chemistry for each point of an adaptive spatial grid of a one-dimensional slab of gas.
First, the code solves the radiative transfer equation, considering absorption in the continuum by dust and in the lines of key atoms and molecules such as \ch{H} and \HH~\citep{goicoechea2007penetration}.
Then, from the specific intensity of the radiation field, it computes the gas and grain temperatures by solving the thermal balance.
The code accounts for a large number of heating and cooling processes, in particular photoelectric and cosmic ray heating, and line cooling.
Finally, the chemistry is solved, providing the densities of about 200 species at each position.
About $3\,000$ reactions are considered, both in the gas phase and on the grains.
The chemical reaction network was built combining different sources including data from the KIDA database\footnote{\url{https://kida.astrochem-tools.org/}} \citep{wakelamKIneticDatabaseAstrochemistry2012} and the UMIST database\footnote{\url{http://udfa.ajmarkwick.net/}}~\citep{mcelroyUMISTDatabaseAstrochemistry2013} as well as data from articles.
For key photoreactions, cross sections are taken from~\citet{heaysPhotodissociationPhotoionisationAtoms2017} and from Ewine van Dishoeck's photodissociation and photoionization database\footnote{\url{https://home.strw.leidenuniv.nl/~ewine/photo/index.html}}.
The successive resolution of these three coupled aspects is iterated until a global stationary state is reached.

The code yields 1D-spatial profiles of density of many chemical species and of temperature of both grains and gas as a function of depth in the PDR.
From these spatial profiles, it also computes the line integrated intensities emerging from the cloud that can be compared to observations.
As of version 7 (released in 2024), thousands line intensities are predicted from species such as H$_2$, HD, H$_2$O, C$^+$, C, CO, \latexmol{13co}, \latexmol{c18o}, \latexmol{13c18o}, SO, \latexmol{hcop}, OH,  \latexmol{hcn}, \latexmol{hnc}, CH$^+$, CN or CS.
Although the Meudon PDR code was primarily designed for PDRs, it can also simulate the physics and chemistry of a wide variety of other environments such as diffuse clouds, nearby galaxies, damped Lyman alpha systems and circumstellar disks.

\subsection{Neural network-based emulation of the model}%
\label{subsec:model_emulation}

The numerical estimation of the mutual information requires drawing thousands of physical parameters $\paramvect{n}$ and evaluating the associated integrated intensities $\truefell(\paramvect{n})$ in order to achieve satisfying precisions for line ranking -- see, for instance, the experiment from App.~\ref{subsec:estimation}.
A single full run of the Meudon PDR code is computationally intensive and typically lasts a few hours for one input vector $\paramvect{}$.
Generating such a large set of models with the original code would therefore be very slow.
This is a recurrent limitation of comprehensive ISM models that received a lot of attention recently.
The most common solution is to derive a fast approximation of a heavy ISM code using
an interpolation method~\citep{gallianoDustSpectralEnergy2018,wuConstrainingPhysicalConditions2018,ramambasonInferringHIIRegion2022},
a machine learning algorithm~\citep{bronTracersIonizationFraction2021,smirnov-pinchukovMachineLearningacceleratedChemistry2022}
or a neural network~\citep{demijollaIncorporatingAstrochemistryMolecular2019,holdshipChemulatorFastAccurate2021,grassiReducingComplexityChemical2022,paludNeuralNetworkbasedEmulation2023}.

In this work, we use the fast, light (memory-wise) and accurate neural network approximation of the Meudon PDR code proposed in~\citet{paludNeuralNetworkbasedEmulation2023}.
This approximation is valid for \mbox{$\logd \Pth \in [5, 9]$}, \mbox{$\logd \Gnaught \in [0, 5]$}, \mbox{$\logd \AV \in [0, \logd(40)]$}.
As neural networks can process multiple inputs at once in batches, the evaluation of $10^3$ input vectors $\paramvect{}$ with this approximation lasts about 10\,ms on a personal laptop.
With the original code, performing that many evaluations would require about a week using high performance computing, \ie, about 60 million times longer even with much more computing power.
For the lines studied in this paper, the emulator results in an average error of about 3.5\% on the validity intervals, which is three time lower than the average calibration error at the IRAM~30m.
The error on mutual information values due to using the emulator instead of the original code is thus negligible.
For this reason and to simplify notation in the remainder of this paper, we denote $\truef$ this neural network approximation.

\subsection{Generating sets of models}%
\label{subsec:dataset_generation}

\begin{figure*}[h]
    \centering
    \includegraphics[width=\linewidth, trim={0 2.7cm 0 0}, clip]{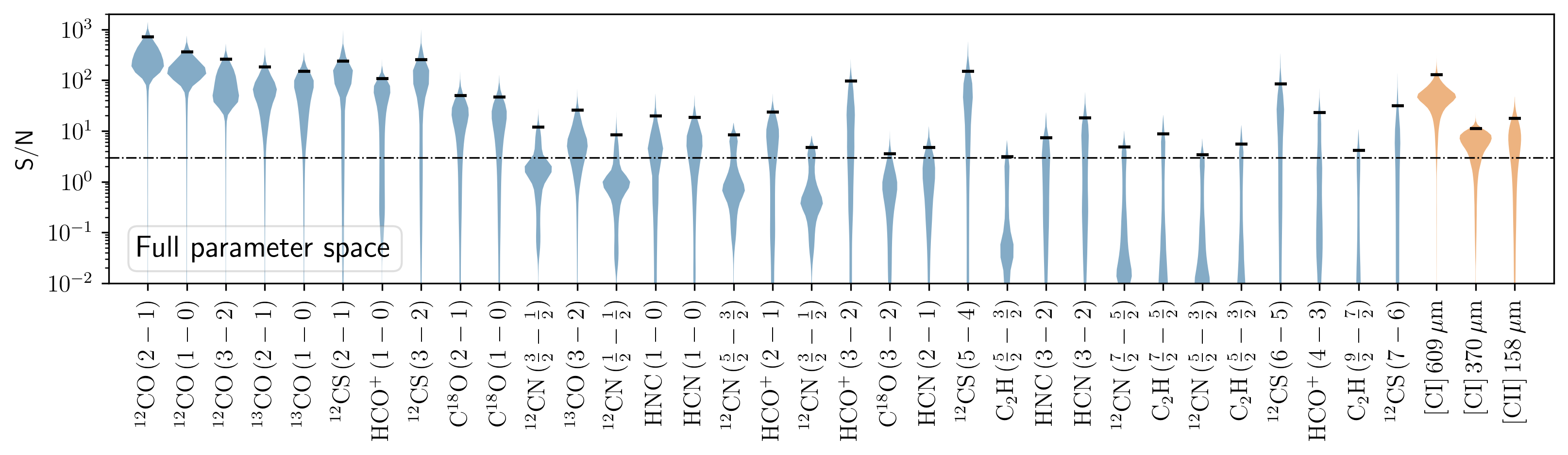}
    \includegraphics[width=\linewidth]{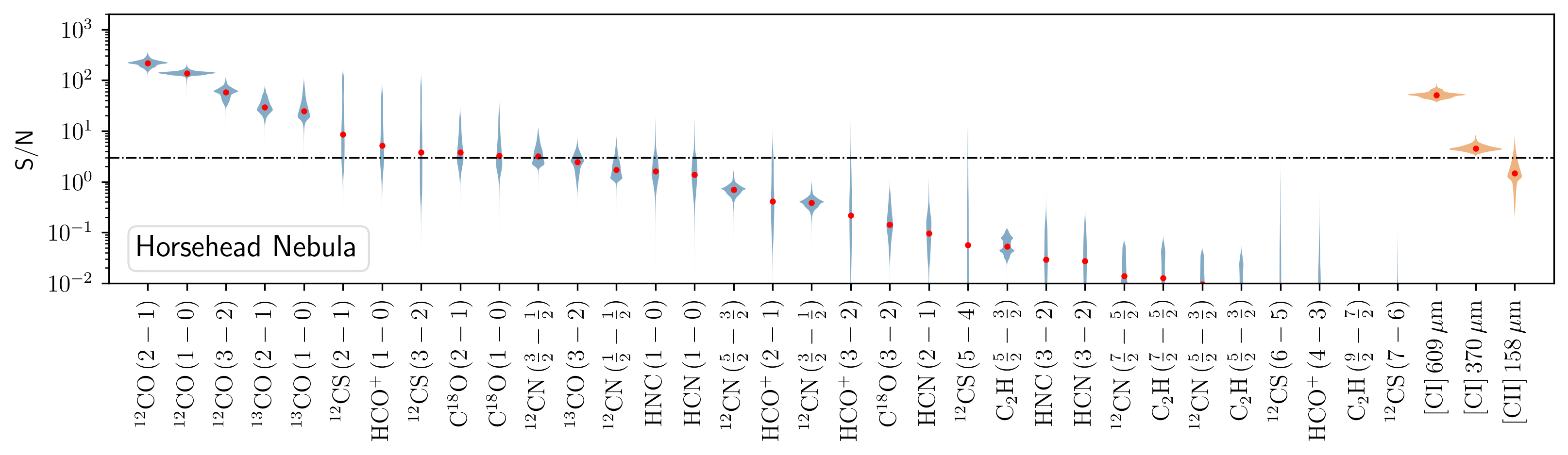}
    \caption{
        Violin plots of the S/N of the spectral lines considered in this study, with the S/N defined as $\truefell(\paramvect{}) / \sigma_{a,\ell}$.
        The EMIR lines are displayed in blue on the left, while the \ci{} and \cp{} lines are shown in orange on the right.
        Top: S/N distributions for a loguniform distribution on the full validity intervals on the physical parameters.
        The considered line filter only keeps lines that have a 99\% percentile S/N greater than 3.
        This threshold is indicated with the horizontal dashed black line, and the actual 99\% percentile S/N is shown with a short black line for each line.
        Bottom: S/N distributions in an environment similar to the Horsehead pillar, for the same lines.
        The lines are ranked by decreasing median S/N, indicated in red.
        }%
    \label{fig:snr_emir_lines}
\end{figure*}

To demonstrate the power of the approach presented in Sect.~\ref{sec:optimization_pb}, we apply it to a simulation of lines observed by the EMIR (Eight MIxer Receiver) heterodyne receiver.
This receiver operates in the 3\,mm, 2\,mm, 1.3\,mm and 0.9\,mm bands at the IRAM~30m telescope~\citep{carter2012emir}.
This application also includes the far infrared (FIR) \ci{} 370\,\microm{}, \ci{} 609\,\microm{} and \cp{} 157\,\microm{} lines.
These three lines are relevant for this application as their behavior is well understood within PDRs~\citep{kaufmanFarInfraredSubmillimeter1999}, especially their dependency on \Guv{}.

However, choosing which lines to include in the study is not the only critical choice.
Indeed, the values of mutual information and therefore the result of the optimization problem heavily depend on the prior distribution $\pi\left(\paramvect{}\right)$ on the physical parameters -- which, in particular, specifies the expected physical regime -- and the simulator of observation.

\subsubsection{Physical regimes and distribution of parameters}%
\label{subsec:physical_params_distri}

\begin{table} % [h]
    \centering
    \renewcommand{\arraystretch}{1.4}
    \addtolength{\tabcolsep}{-0.5mm}
    \caption{Summary of the parameters distribution for the two studied situations.}
    \begin{tabular}{ccc}
        \hline
        \hline
        \multirow{2}{*}{Situation} & Parameters & Parameters\\[-1mm]
        & bounds & distribution\\
        \hline
        \begin{tabular}{@{}c@{}}Full parameter\\[-1mm]space\end{tabular}
        & \begin{tabular}{@{}c@{}}$1 \leq \AV{} \leq 40$\\[-1mm]$1 \leq \Guv{} \leq 10^5$\\[-1mm]$10^5 \leq \Pth{} \leq 10^9$\end{tabular}
        & \begin{tabular}{@{}c@{}}Loguniform\\[-1mm]Loguniform\\[-1mm]Loguniform\end{tabular}\\
        \hline
        \begin{tabular}{@{}c@{}}Horsehead\\[-1mm]Nebula\end{tabular}
        & \begin{tabular}{@{}c@{}}$3 \leq \AV{} \leq 24$\\[-1mm]$10^1 \leq \Guv{} \leq 10^3$\\[-1mm]$10^5 \leq \Pth{} \leq 5\cdot10^6$\end{tabular}
        & \begin{tabular}{@{}c@{}}Power law ($\alpha=-2.24$)\\[-1mm]Power law ($\alpha=-1.05$)\\[-1mm]Loguniform\end{tabular}\\
        \hline
    \end{tabular}
    \label{tab:parameters-distribution}
\end{table}

The distribution $\pi(\paramvect{})$ on physical parameters represents the expected proportions of pixels in each physical regime within an observation.
This distribution has a crucial influence on ISM model predictions and thus on the mutual information values and line ranking.
It should therefore be carefully chosen.
In this paper, we study two situations, summarized in Table~\ref{tab:parameters-distribution}.

First, we consider a loguniform distribution over the whole validity space of the emulated ISM model.
As this option does not favor any physical regime, it is a common choice in ISM studies -- see, for example, \citet{behrensTracingInterstellarHeating2022,blancIZIInferringGas2015,thomasInterrogatingSeyfertsNebulaBayes2018,holdshipBayesianInferenceRates2018,joblinStructurePhotodissociationFronts2018}.
In other words, it assumes that all kinds of environments are equally likely, which is not the case in general in observed environments.
However, choosing the distribution of maximal entropy on $\log \AV$ and $\log \Gnaught$ permits us to average the lines informativity over different physical conditions without introducing any bias.
%
% it is a good choice to obtain informativity results averaged over different physical conditions.

Second, we consider a physical environment similar to the Horsehead pillar.
Real life observations of molecular clouds such as Orion~B~\citep{petyAnatomyOrionGiant2017} or OMC-1~\citep{goicoecheaMolecularTracersRadiative2019} typically contain more pixels corresponding to translucent gas than dense cores.
This is due to the fact that translucent gas fills a larger volume than dense cores in a galaxy.
To incorporate this physical knowledge in our study, we fit a power law distribution on \AV{} and \Gnaught{}~\citep{hennebelle2012turbulent}.
The associated exponents are adjusted on \mbox{ORION-B} data, following the method described in~\citet{clauset2009power}.

For a given situation, one can choose to simulate observations only within a particular environment (\eg, translucent clouds with \mbox{$3 \leq \AV{} \leq 6$}).
This physical a priori can then be used to refine the results.
In practice, any available physical knowledge is useful to integrate into the parameters prior distribution or the simulator of observation.

\subsubsection{Observation simulator}%
\label{subsec:obs_model}

\eqref{eq:obs_model_abstract} involves an abstract noise model~$\noiseGeneral$.
In this experiment, the considered noise model combines two sources of noise for each of the considered lines: one additive Gaussian and one multiplicative lognormal.
The additive noise corresponds to thermal noise, whereas the multiplicative noise corresponds to the calibration uncertainty.
For all lines, we compute the integrated line intensity over a velocity range of 10\,\kms.
Overall, for the $n^\text{th}$ element of the dataset ($1 \leq n \leq N$) and the $\ell^\text{th}$ line, the observation simulator reads
\begin{align}\label{eq:obs_model_instance}
    \obselt
    =
    \multnoise \truefell \left(\paramvect{n} \right) + \addnoise,
\end{align}
with
\begin{align}
    \label{eq:noise_models}
  \begin{dcases}
    \addnoise\sim \mathcal{N} (0,\,\sigma_{a,\ell}^2), \\
    \multnoise \sim \lognormal \left(-\frac{\sigma_m^2}{2},\,\sigma_m^2 \right).
  \end{dcases}
\end{align}
The standard deviation of the multiplicative noise $\sigma_m$ is set so that a $1\sigma$ uncertainty interval corresponds to a given percentage for the calibration error.
For instance, a 5\% calibration error leads to $\sigma_m = \log (1.05)$.
For EMIR lines, this percentage is assumed to be identical for the lines within the same band: 5\,\% at 3\,mm, 7.5\,\% at 2\,mm and 10\,\% at both 1.3\,mm and 0.9\,mm.
For the time being, the additive noise RMS levels $\sigma_{a,\ell}^2$ are set according to the \mbox{ORION-B} Large Program observations~\citep{einigDeepLearningDenoising2023}.
To do this, we resort to the IRAM~30m software that delivers the telescope sensitivity as a function of frequency.
We consider standard weather conditions at Pico Veleta and set the integration time per pixel to 24~seconds.
An increase of the integration time would amount to dividing the additive noise RMS $\sigma_{a,\ell}$ by the square root of the increase factor.

For FIR lines, we assume that the \cp{} line is observed with SOFIA and has an additive noise RMS of 2.25\,K per channel in addition of a 5\,\% calibration error~\citep{risacher2016upgreat, pabst2017c}.
We also assume that both \ci{} lines are observed at Mount Fuji observatory with an RMS of 0.5\,K and a 20\,\% calibration error~\citep{ikeda2002distribution}.
For all lines, the integration range is assumed to be 10\,\kms.

Important observational effects such as the beam dilution or the cloud geometry are disregarded in~\eqref{eq:obs_model_instance}.
As a consequence, we propose an alternative simulator of observations that accounts for such observational effects through a scaling factor~$\kappa$.
This scaling factor is assumed common to all lines such that
\begin{align}\label{eq:obs_model_instance_with_scaling}
    \forall 1 \leq \ell \leq L, \quad
    \obselt
    =
    \multnoise \kappa_n \truefell \left(\paramvect{n} \right) + \addnoise
    .
\end{align}
Beam dilution decreases line intensities, while an edge-on geometry increases line intensities compared to a face-on orientation.
Therefore, we consider that $\logd \kappa$ follows a uniform distribution
on $[-0.5, 0.5]$, which seems realistic when looking at extended sources like Orion~B.
See~\citet{shefferPDRMODELMAPPING2013} for a more thorough description of this scaling parameter.
This approach of including these effects in the observation simulator is a first order approximation.
In particular, the hypothesis of a shared $\kappa$ among all lines is only valid for optically thin lines. % and does not really hold for optically thick lines for which the geometrical increase is not the same for all lines.

In the remainder of this work, unless explicitly specified, the considered observation simulator is~\eqref{eq:obs_model_instance} -- without the $\kappa$ term.

\subsubsection{Considered lines}%
\label{subsubsec:lines}

In the simulated observations, the intensity of some lines is completely dominated by the additive noise.
The intensity of these lines is thus nearly independent of physical parameters $\paramfull{}$ and has a near-zero mutual information with them.
To avoid useless mutual information evaluations, we filter out uninformative lines based on their S/N.
We thus only study lines that have an S/N greater than 3 for at least 1\% of the full parameter space.
In total, $L = 36$ lines are considered: $33$ millimeter lines -- with multiple lines in each of the four frequency bands -- and the $3$ lines from atomic and ionized carbon.
For lines with hyperfine structure, the Meudon PDR code considers the transitions independently.
To simplify our systematic comparison, only the brightest transition is retained.
Summing the integrated intensities of all the transitions might lead to a more realistic approximation of the overall line.

Figure~\ref{fig:snr_emir_lines} shows the distribution of S/N level across the considered parameter space for each of the $L=36$ considered lines.
These lines include the first three low-$J$ transitions of $^{12}$CO, \latexmol{13co}, \latexmol{c18o}, the first four of \latexmol{hcop}, five of the first seven of $^{12}$CS, six lines of $^{12}$CN, two lines of \latexmol{hnc}, three lines of \latexmol{hcn}, and four lines of C$_2$H.
The first row contains S/N violin plots for a loguniform distribution on the validity intervals for the physical parameters $\paramvect{}$.
It shows that all the considered lines can have very low S/Ns for some regimes of the explored physical parameter space.
Below an S/N of 1--2, signal becomes difficult to distinguish from noise.
The second row contains S/N violin plots for a parameter space restricted to the range found in the Horsehead pillar.
In this use case, the line S/Ns cover fewer orders of magnitude.
For instance, in this case, the lines corresponding to the last 18 blue violin histogram have a very low S/N, and are thus unlikely to be informative.
This shows that the subset of informative lines could be further reduced in this case.
While dedicated filters could be performed for each use case, we maintain the same subset of $L = 36$ lines in all the studied use cases to simplify interpretations.

The considered noise properties of the EMIR receiver, of SOFIA, and of the Mt. Fuji observatory are not identical for all lines.
For instance, Fig.~\ref{fig:snr_emir_lines} shows similar range of S/N values for the ground state transition of \latexmol{12co} and \latexmol{13co}.
This might be surprising, since the ground state transition of \latexmol{12co} is known to be brighter than that of \latexmol{13co}~\citep{petyAnatomyOrionGiant2017}.
In this case, the additive noise standard deviation $\sigma_{a,\ell}$ of \latexmol{12co}\,\transp10 is much larger than that of \latexmol{13co}\,\transp10 because \latexmol{12co}\,\transp10 is located on the upper limit of the band at 3\,mm.
This results in their comparable S/Ns.
The same observation can be done for the [CII] line: although this line is usually much brighter than all the other considered lines, its S/N is close to 1 due to the considered noise properties of SOFIA.
Appendix~\ref{app:line-selection} provides the full list of considered lines and the associated noise characteristics.

%%%%%%%%%%%%%%%%%%%%%%%%%%%%%%%%%%%%%%%%%%%%%%%%%%%%%%%%%%%%%%%%%%%%%%%%%%%

\section{Simulation results and general applications}%
\label{sec:global-results}

In this section, we show general results and insights of our approach in the considered setting.
To do so, we evaluate the mutual information between the integrated intensity of a few ISM tracers with either the visual extinction $\AV{}$ or the far UV (FUV) illumination field $\Guv{}$.
First, we consider the impact of integration time, and thus of S/N, on the mutual information value.
Second, we show how the mutual information between line intensities and
\AV{} or \Guv{} changes with the values of \AV{} and \Guv{}, in order to better understand the physical processes that control the informativity of these lines.
Third, we illustrate how combining different lines can impact their mutual information with \AV{}.

The goal of this section is to demonstrate the approach potential and consistency with already known results.
Therefore, we restrict the analysis to two variables -- for visualization purposes -- and choose the two variables for which astrophysicists have the best intuition, namely the visual extinction $\AV{}$ or the UV field intensity~\Guv{}.
In particular, we do not present mutual information values for the thermal pressure~$\Pth$ although the proposed approach and code can perform these computations.
In addition, we restrict the experiment to univariate physical parameters as this greatly simplifies physical interpretations.
In other words, we compute mutual information for only one physical parameter ($\AV$ or $\Gnaught$) at a time, although the proposed approach and code can evaluate the mutual information for both $\AV{}$ and $\Guv{}$ simultaneously.
Analyzing less understood physical parameters such as the thermal pressure~$\Pth$ or evaluating the mutual information for multiple physical parameters at once is left for future work.

\subsection{Which S/N for a line to deliver its full physical potential?}
\label{subsec:mi_obs_time}

The mutual information $\mi{\paramfull}{\obsfull_\ell}$ between a line intensity $\obsfull_\ell$ and a given physical parameter $\paramfull$ not only depends on the intrinsic physical sensitivity of the lines with the considered physical parameter, but also on the mean S/N of the studied observation.
For a given line, the mean S/N is influenced by
1)~the corresponding species and its quantum transition,
2)~the physical conditions (\eg, kinetic temperature and volume density),
and 3)~the integration time with an observatory to reach a given noise level%
\footnote{The noise level for a given integration time depends on additional parameters such as the weather conditions for a ground observatory.}%
.

\begin{figure} %[H]
    \centering
    \includegraphics[width=\linewidth, trim={0 2mm 0 1mm}, clip]{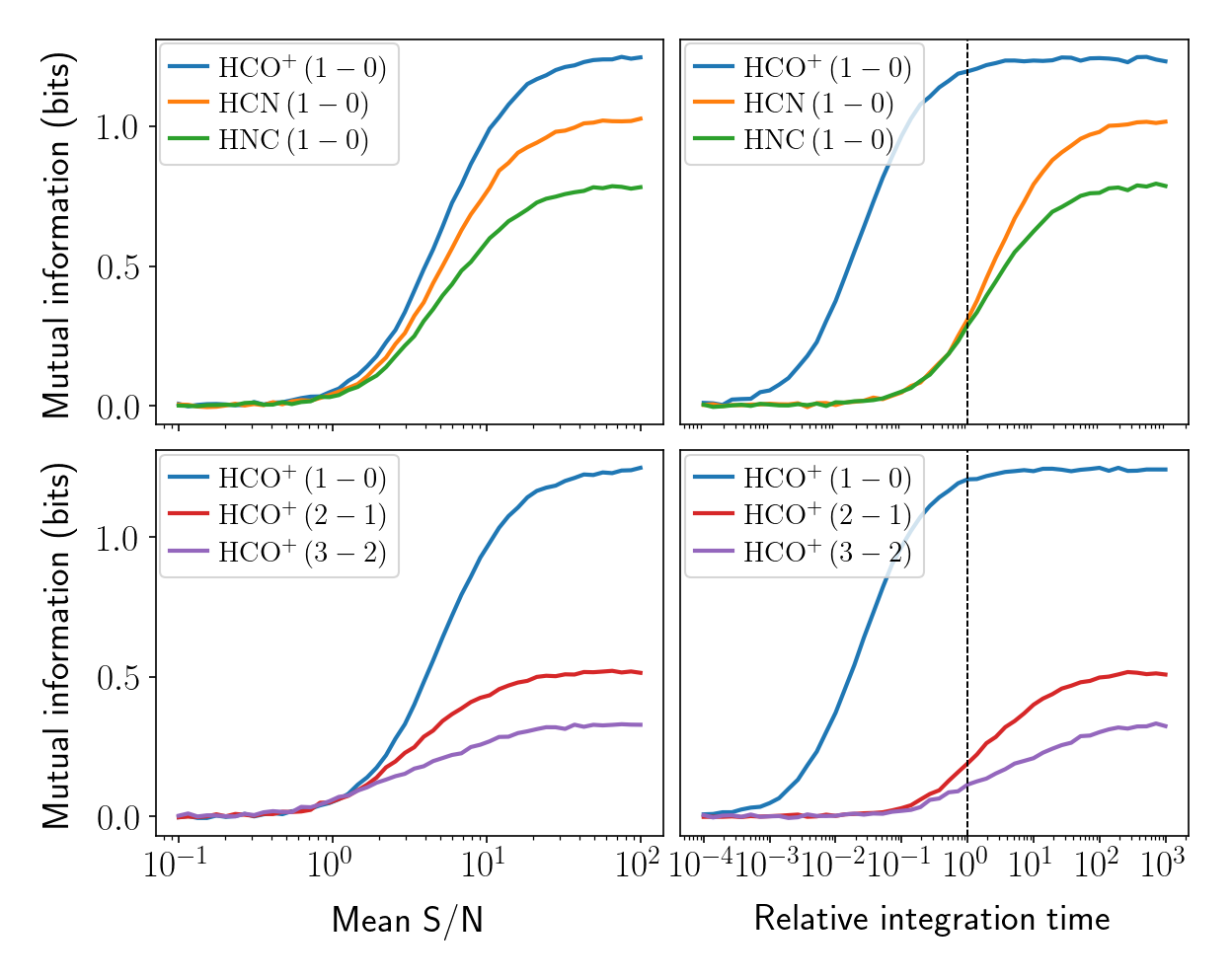}
    \caption{
        Evolution of mutual information between the visual extinction $\AV{}$ and integrated line intensities as a function of S/N (left column) and integration time (right column), for \mbox{$6 \leq \AV{} \leq 12$} (filamentary clouds).
        The top row shows the comparison between some chemical species, while the bottom row shows the comparison between the three lower energy transitions of \latexmol{hcop}.
    }
    \label{fig:profiles-a}
\end{figure}

Figure~\ref{fig:profiles-a} shows the influence of the mean S/N (left column) and the integration time (right column) on $\mi{\AV{}}{\obsfull_\ell}$ for several transitions of \latexmol{hcop}, \latexmol{hcn}, and \latexmol{hnc}.
The considered distribution $\pi(\paramvect{})$ on physical parameters is the one similar to the Horsehead Nebula (see Table~\ref{tab:use-cases}), restricted to filamentary gas ($6 \leq \AV{} \leq 12$).
The dotted vertical line in the right column shows the typical integration time per pixel in the \mbox{ORION-B} dataset.
For each line, the mutual information varies with mean S/N and time following an S-shape.
Low S/N values lead to zero mutual information because the line intensity is dominated by additive noise.
The inflection point of the S-curve is located at S/N about 3.
A given line reaches its full informativity potential when the curve starts to saturate, \eg, S/N $\sim 10$ for all lines in this case.
For large S/N, the mutual information converges to a finite value that depends on the line micro-physical characteristics.
This value is finite because a given \AV{} value is combined with many values of the thermal pressure and UV illumination in this example.

Using the proposed method, the integration time can be set to achieve a target mean S/N and mutual information.
For instance, according to Fig.~\ref{fig:profiles-a}, $\mi{\AV{}}{\obsfull_\ell}$ has already reached its maximum value for~\latexline{hcop}{1}{0} in the filamentary gas part of \mbox{ORION-B} dataset.
An increase of the integration time would thus not increase the informativity of this line, \ie{}, would not improve the precision in an estimation of \AV{} from~\latexline{hcop}{1}{0}.
Conversely, a 100-fold increase of the integration time would improve the mutual information for the \latexline{hcn}{1}{0} and \latexline{hnc}{1}{0} lines by 0.7 and 0.5\,bits, respectively, and would lead to maximum precision in an estimation of \AV{} with these lines.
Higher energy transitions of \latexmol{hcop} could also be fully exploited with such an increase of the integration time.
As a reference, the next generation of multibeam receivers currently foreseen in millimeter radio astronomy are expected to bring a 25-fold sensitivity improvement without increasing the integration time.
Appendix~\ref{app:additional-results} provides the same figures of evolution of mutual information with the integration time for the $36$ considered lines, in translucent gas, filamentary gas and dense cores.
It also displays results with respect to the intensity of the UV radiative field $\Guv{}$.

\begin{figure} %[H]
    \centering
    \includegraphics[width=\linewidth, trim={0 2mm 0 1mm}, clip]{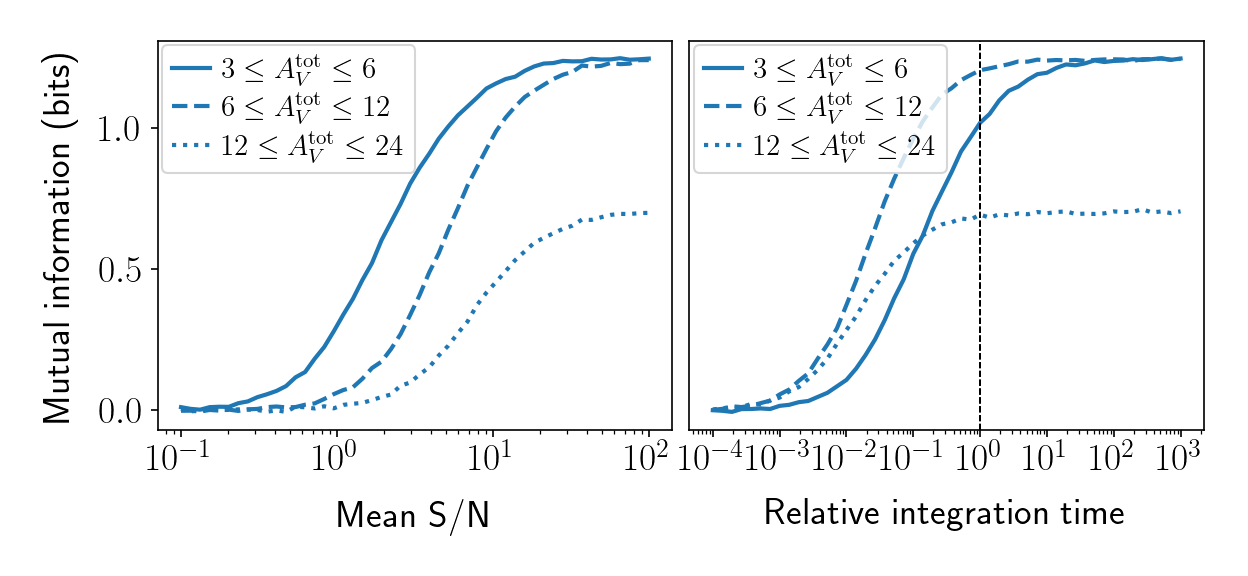}
    \caption{
        Evolution of mutual information between the visual extinction $\AV{}$ and integrated line intensities as a function of S/N (left column) and integration time (right column) for different $\AV{}$ regimes, using the example of the \latexline{hcop}{1}{0} line.
    }
    \label{fig:profiles-b}
\end{figure}

Figure~\ref{fig:profiles-b} shows how $\mi{\AV{}}{\obsfull_\ell}$
evolves with mean S/N for \latexline{hcop}{1}{0} in the Horsehead Nebula (see Table~\ref{tab:use-cases}) in three physical subregimes: translucent, filamentary, and dense core gas.
The inflection point of the S-shape curve happens at an S/N of about 2, 5, and 10, respectively.
Comparing the maximum value of mutual information for different regimes is hazardous here because the distribution of the \AV{} values (and thus the associated entropy) intrinsically depends on the studied physical regime.
If a considered physical regime is broad, the mutual information between a given line and \AV{} is likely to be higher than for another more localized regime even if the line is a better tracer of \AV{} in the latter.

%%%%%%%%%%%%%%%%%%%%%%%%%%%%%%%%%%%%%%%%%%%%%%%%%%%%%%%%%%%%%%%%%%%%%%%%%%%%%%

\subsection{In which physical regimes is a given line informative?}%
\label{subsec:maps_mi}

\begin{figure*}[t]
    \centering
    \includegraphics[width=\textwidth]{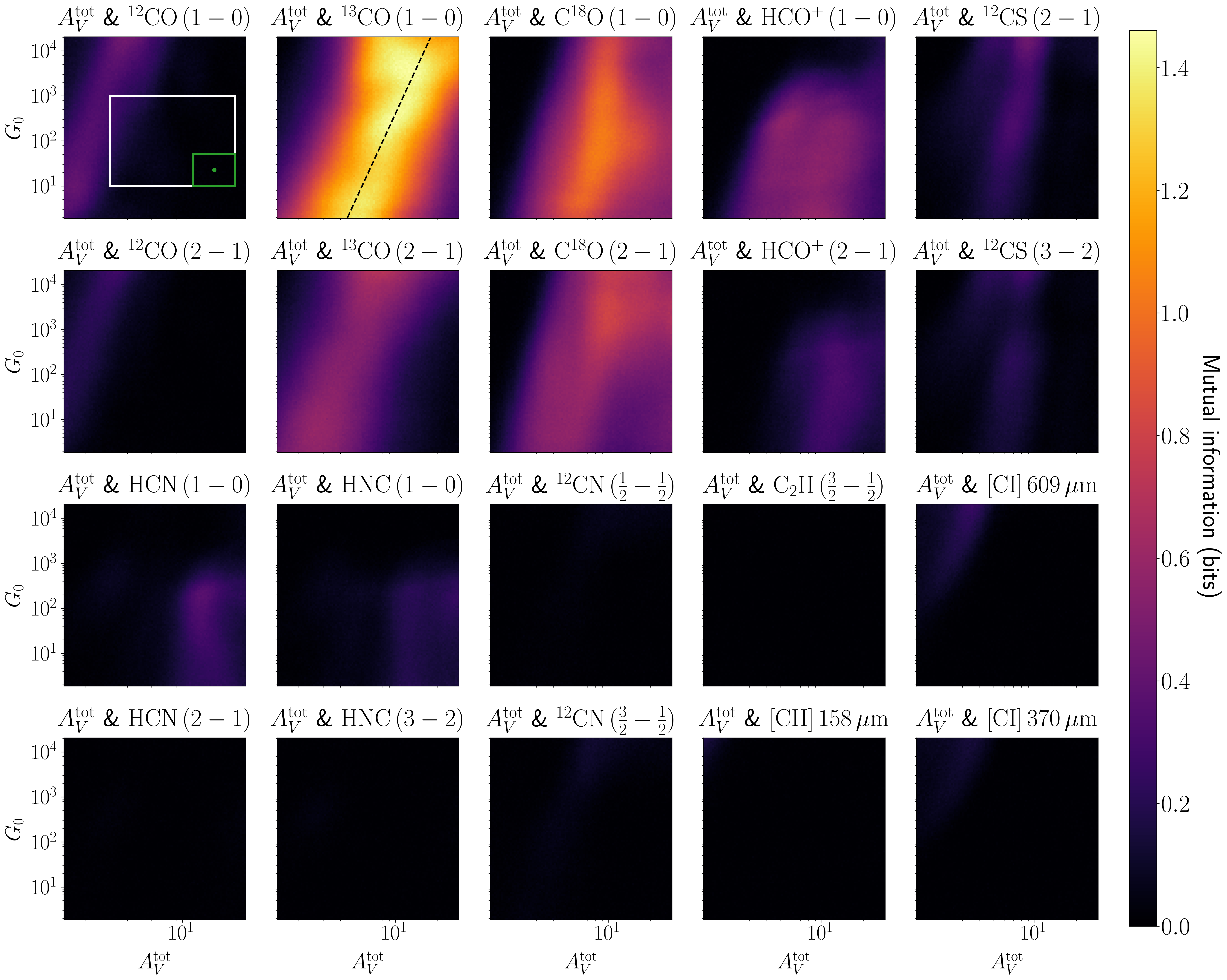}
    \caption{
        Maps of mutual information of individual lines with the visual extinction in function of the actual visual extinction \AV{} and intensity of the UV radiative field \Gnaught{}.
        The results are computed for the pressure following a loguniform distribution between $10^5$ and $5 \times 10^6$ K cm$^{-3}$.
        The red rectangle on the first panel shows the dimensions of the sliding window, while the white rectangle delimits the parameter space characterizing the Horsehead Nebula.
        The dashed black line on the \latexline{13co}{1}{0} panel corresponds to a constant $\Guv{}^{0.15} / \AV{}$ ratio.
    }%
    \label{fig:maps-grid-av}
\end{figure*}

In this section, we show how mutual information can provide insights for ISM physics understanding.
We showed in Fig.~\ref{fig:profiles-b} that the mutual information between a physical parameter and a line intensity may significantly vary with the physical regime.
The three large physical regimes used in the previous section were defined based on a priori astronomical knowledge.
This may result in the omission of processes that occur in smaller and intermediate regimes.
To overcome this issue, we introduce the notion of maps of the mutual information between a physical parameter (either \AV{} or \Guv{}) and line intensities as a function of both \AV{} and \Guv{}.
To do this, we filter the (\logd\,\AV{}, \logd\,\Guv{}) space with a sliding window of constant width, and consider loguniform distribution for each parameter.
This width corresponds to a factor 2 for \AV{} and a factor about 5.2 for \Guv{}, \ie, seven independent windows (without overlap) for each parameter.
Then, we compute the mutual information between the line intensities, simulated with parameters in the sliding window, and either \AV{} or \Guv{}.
The additive noise in the simulated spectra corresponds to the integration time corresponding to the \mbox{ORION-B} observations, \ie, 24 seconds per pixel.
After describing the obtained maps of mutual information with \AV{} and \Guv{}, we explain them with maps of predicted line intensities~$\truefell(\paramvect{})$.

Here, the values of mutual information can be compared from one value of the (\AV{}, \Guv{}) space to another because the sampling of this space is regular and the size of the sliding window is kept fixed.
For the same reasons, the values of mutual information can also be compared from one line to another at a constant value of (\AV{}, \Guv{}).
Similarly, for a given line and value of (\AV{}, \Guv{}), $\mi{\AV{}}{\obsfull_\ell}$ and $\mi{\Guv{}}{\obsfull_\ell}$ can be compared.

The considered prior $\pi(\paramvect{})$ for each parameter is always loguniform in this section.
In this very special case, a mutual information value of 1 bit for one physical parameter may be interpreted as a division of the standard deviation on the estimation of $\log \AV{}$ or $\log \Guv{}$ by a factor 2.
For instance, if the considered physical parameter is $\Guv{}$ and its mutual information $\mi{\log \Guv{}}{y}$ with some line $y$ is 1 bit, then the  standard deviation of the conditional distribution $\pi(\log \Guv{} \vert y)$ is a factor 2 lower than the one of the prior $\pi(\log \Guv{})$.
For more general prior distributions, this interpretation does not hold.
The second paper of this series will provide an interpretation for the general case.

\subsubsection{How relevant are individual ISM lines to constrain \AV{}?}

We here wish to identify
1)~which lines are the most relevant to estimate the
visual extinction~\AV{},
and 2)~in which part of the (\logd\,\AV{}, \logd\,\Guv{}) space.
Figure~\ref{fig:maps-grid-av} shows maps of mutual information between the intensity of 20 individual lines and~\AV{}.
The size of the sliding window is shown in the \latexmol{12co} map as a red rectangle while the range of the parameters within the Horsehead Nebula is represented with a white rectangle as a reference.

Among the presented lines, the most informative ones for estimating \AV{} on average are the lines of \latexmol{13co} and \latexmol{c18o} followed by \latexmol{hcop}.
The lines of \latexmol{12co}, \latexmol{hcn}, \latexmol{12cs}, and \ci{} are also informative but on more restricted regions of the (\AV{}, \Guv{}) space.
The $J=2-1$ transitions have systematically lower mutual information with \AV{} than the $J=1-0$ transitions, which is due to lower mean S/N -- as shown on Fig.~\ref{fig:snr_emir_lines}.

The three CO isotopologues give high values of the mutual information for most of the (\AV{}, \Guv{}) space.
For translucent clouds, the first two \latexmol{13co} lines are the most informative.
For dense clouds (large \AV{}), the first two \latexmol{13co} and \latexmol{c18o} lines are the most informative.
Finally, the fine structure \ci{} lines and the ground state transition of \latexmol{12co} have the highest mutual information values (even though these values are low) for the upper left corner, which corresponds to highly illuminated diffuse clouds.

Although the ground state transitions of \latexmol{hcn} and \latexmol{hnc} are among the most informative lines in the high-\AV{}, low-\Guv{} regime, we might have expected them to be even more informative in this physical regime since they are used as tracers of the dense cores.
Their relatively low informativity is explained by low mean S/N values.
As shown in Fig.~\ref{fig:profiles-a}, the integration time is too short to exploit the full potential of these lines.

We also observe that the mutual information with \AV{} is roughly constant with respect to the ratio $\Gnaught^{0.15} / \AV$ for multiple lines.
This ratio corresponds to a straight line in the $(\log \AV, \log \Guv{})$ space, and is displayed in Fig.~\ref{fig:maps-grid-av}.
That is particularly clear for the \latexmol{12co}, \latexmol{13co} and \latexmol{c18o} lines.
In the upper left corner, where the $\Gnaught^{0.15} / \AV$ ratio is maximum, the mutual information is low.
It increases as this ratio decreases, reaches a maximum and then decreases.

\begin{figure*}[t]
    \centering
    \includegraphics[width=\textwidth]{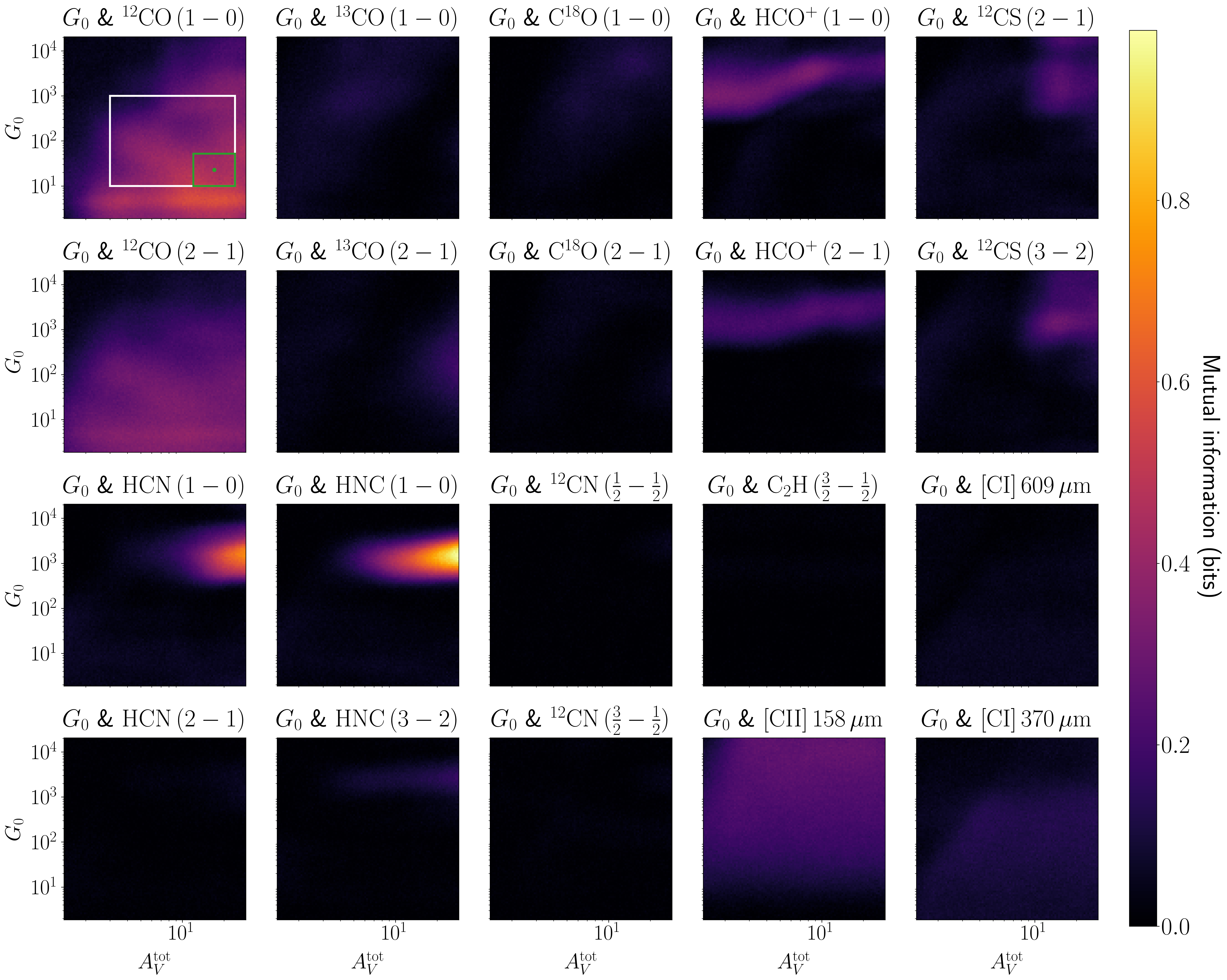}
    \caption{
        Maps of mutual information of individual lines with the UV radiative field in function of the actual visual extinction \AV{} and intensity of the UV radiative field \Gnaught{}.
        The results are computed for the pressure following a loguniform distribution between $10^5$ and $5 \times 10^6$ K cm$^{-3}$.
        The red rectangle on the first panel shows the dimensions of the sliding window, while the white rectangle delimits the parameter space characterizing the Horsehead Nebula.
    }
    \label{fig:maps-grid-g0}
\end{figure*}

\subsubsection{How relevant are individual ISM lines to constrain \Guv{}?}

We now apply the same approach on the FUV illumination \Guv{}.
Figure~\ref{fig:maps-grid-g0} shows maps of mutual information between the intensity of the same 20 individual lines and \Guv{}.
For most molecular lines except those of \latexmol{12co}, the mutual information values are lower for \Guv{} than for \AV{}.
This indicates that the considered lines are more informative for \AV{} than for \Guv{}, \ie, that achieving a good precision on \Guv{} is harder than on \AV{}.
This result is consistent with~\citet{gratierQuantitativeInferenceColumn2021}.

For most of the (\AV{}, \Guv{}) space, the most informative lines are \cp{}, \latexmol{12co} lines and, to a lesser extent, [CI] lines.
This is due to the fact that these five lines have a high mean S/N -- with the considered noise properties -- and are mostly emitted at the surface of the cloud, thus being sensitive to \Guv{}.
%
% The informativity of each of these five lines decreases for large values of \Guv{}.
%
% Unlike [CI] and [CII] lines, the informativity of the \latexmol{12co} lines decreases at low $\AV{}$ regions.
%
% In contrast with the mutual information between \latexmol{12co} lines and \AV{}, the mutual information between \latexmol{12co} lines and \Guv{} is larger for the $J=2-1$ transition than for $J=1-0$.
%
%
For highly illuminated clouds ($\Guv \in [10^3, 10^4]$), especially at low \AV{}, the most informative transitions are the ones of \latexmol{hcop}.
This is probably related to the fact that \latexmol{hcop} is easily excited by electrons at the surface of the clouds.
The mutual information of the \latexmol{hcn} and \latexline{hnc}{1}{0} intensities with \Guv{} reach high values compared to other species (more than 0.8\,bits) for \Guv{} around $2\times10^3$ and $Av > 20$.
Finally, the \latexmol{12cs} transitions are the most informative in the upper right corner, \ie, at both high \AV{} and \Guv{}.

\subsubsection{What are the underlying reasons?}

\begin{figure*}[t]
    \centering
    \includegraphics[width=\textwidth]{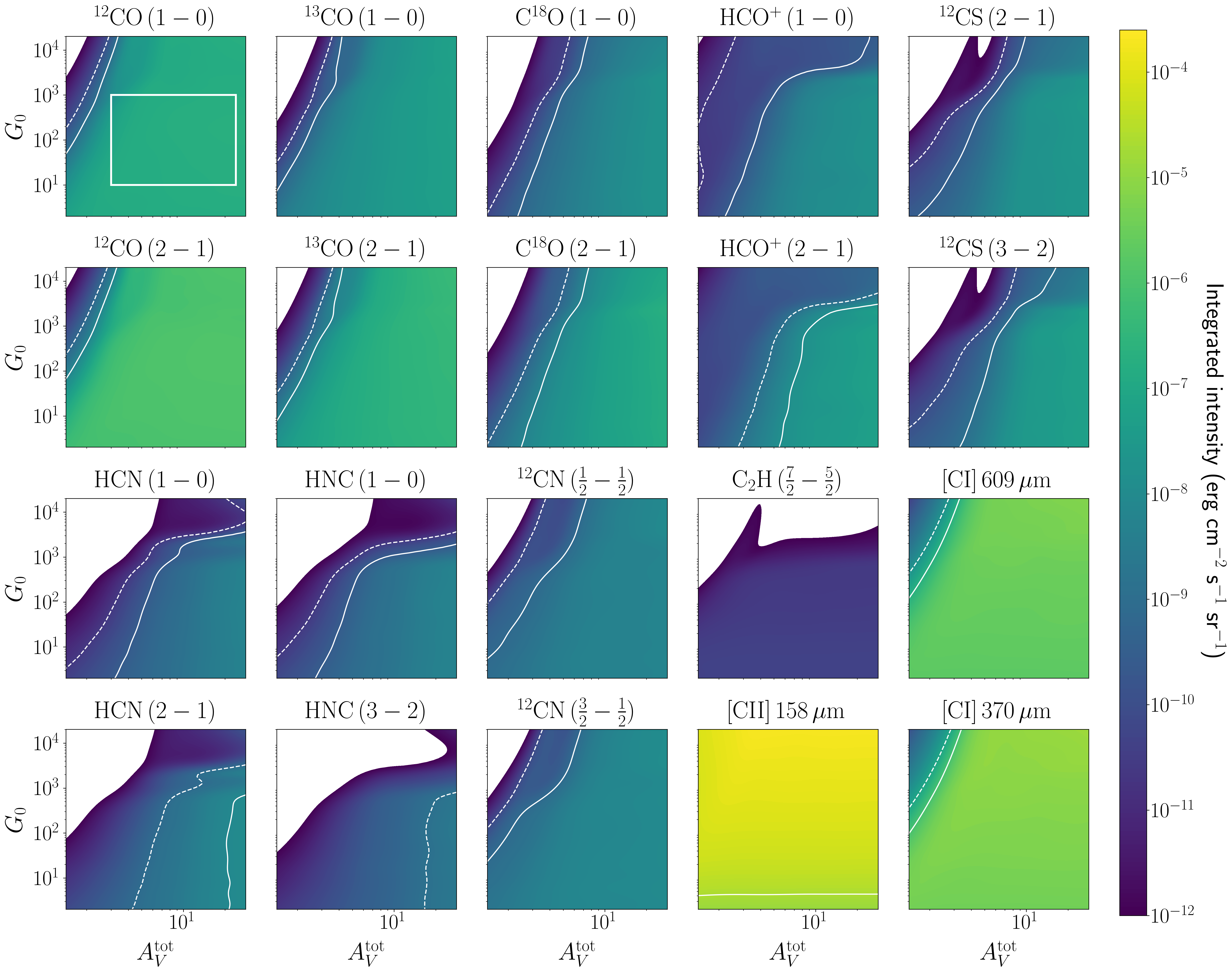}
    \caption{
        Predicted integrated intensities $\truefell(\paramvect{})$ as a function of \AV{} or \Gnaught{}, with $\Pth = 10^5$ K cm$^{-3}$.
        The white full line represents the standard deviation $\sigma_{a,\ell}$ of the additive noise from~\eqref{eq:obs_model_instance} for the \mbox{ORION-B} observations~\citep{petyAnatomyOrionGiant2017}.
        The white dashed line indicates the standard deviation with a ten times longer integration time (deeper integration use case).
        The regions with integrated intensities lower than $10^{-12}$ erg cm$^{-2}$ s$^{-1}$ sr$^{-1}$ are shown in white for better visibility of the higher intensities.
        The white rectangle on the first panel delimits the parameter space characterizing the Horsehead Nebula.
    }%
    \label{fig:predicted_maps}
\end{figure*}

In order to better understand these mutual information maps, Fig.~\ref{fig:predicted_maps} shows the integrated intensities $\truefell(\paramvect{})$ as a function of \AV{} and \Gnaught{}.
These predicted intensities are computed for \mbox{$\Pth = 10^5$}~K\,cm$^{-3}$, while the mutual information maps are computed for a pressure following a loguniform distribution on the \mbox{$[10^5, 5 \times 10^6]$ K cm$^{-3}$} interval.
However, they capture the main physical phenomena that drive mutual information.
In a nutshell, this figure shows that to be informative for a physical parameter, a line needs both a good S/N and a large gradient with respect to the physical variable of interest.
Since the gradient information might not be visible on Fig.~\ref{fig:predicted_maps}, Appendix~\ref{app:gradients} provides maps of the gradients of the log integrated intensities.

While the \cp{} line (last row) is the brightest of all, it has near-zero mutual information with \AV{} in all regimes.
As \cp{} mostly exists at the surface of the cloud, the predicted integrated intensity almost does not depend on visual extinction.
It only has a slight dependency at $\AV \sim 1$\,mag, which is the typical visual extinction where carbon becomes mostly neutral in a PDR~\citep{rollig2007photon} (it is then included in molecules such as CO).

After the \cp{} line, the two \ci{} lines are the brightest.
Their intensity first increases as $\Gnaught^{0.15} / \AV$ decreases in the top left corner (shallow and highly illuminated clouds) as the cloud progressively forms more atomic carbon, and then saturates as carbon mostly exists in molecules in darker clouds.
This explains why the \ci{} lines have a \mbox{$0.2 - 0.3$} bit mutual information with \AV{} in this region, and lower mutual information values (\mbox{$0.1 - 0.2$}) for \Guv{}.
Out of this top left corner, like \cp{}, atomic C mostly exists at the surface of the cloud, which is why the predicted integrated intensities of the two \ci{} lines almost do not depend on visual extinction and have a near-zero mutual information value with \AV{}.
However, the intensity of \ci{} lines increases slightly with \Guv{}, and the intensity of \cp{} increases quickly with \Guv{}, because \latexmol{12co} is photodissociated and C is ionized as \Guv{} increases.
This explains why these three lines have a high mutual information value with \Guv{}.

In the upper left corner (shallow and highly illuminated clouds), most of the molecular lines are very faint and have a large gradient orthogonal to the $\Gnaught^{0.15} / \AV$ direction.
In this high $\Gnaught^{0.15} / \AV$ regime, a small positive change in \AV{} or negative change in \Gnaught{} results in a large increase of the integrated intensities.
Increasing \AV{} favors the formation of molecules in the deeper parts of the cloud, and decreasing \Guv{} decreases photodissociation.
In this regime, the mutual information with \AV{} or \Guv{} is near-zero for most lines as they are drowned in noise.
There are two exceptions.
First, the \latexmol{12co} lines have the highest mean S/N as \latexmol{12co} is the first molecule to form in such clouds.
Second, the \latexmol{hcop} lines are just below the noise standard deviation for $\Pth{} = 10^5$ K cm$^{-3}$ but are brighter for higher pressures.

The first two $^{12}$CO lines show a similar pattern over the full (\AV{}, \Guv{}) space: their intensities first increase as $\Gnaught^{0.15} / \AV$ decreases, as the molecules form in the cloud, and then saturates as they become optically thick for large enough \AV{}.
The transition between the high intensity gradient due to the increase in the formation of the molecule, and the saturation due to optical thickness occurs at relatively low S/N.
These two lines thus have highest informativity on \AV{} in regions at low values of \AV{} along $\Gnaught^{0.15} / \AV$.
The precision in inferring \AV{} remains low because of the relatively low S/N.
The saturation value then slightly depends on \Guv{}, which is why the mutual information between these two lines and \Guv{} (out of the upper left corner) is non-zero.

As \latexmol{13co} is less abundant than \latexmol{12co}, the intensities of its first two lines become bright enough and then saturate for larger values of \Av.
There is a wide \AV{} interval for which these two lines have simultaneously a high S/N and a large gradient, which yields a high mutual information.
The first two \latexmol{c18o} lines show a similar pattern for darker clouds.
All this combined shows that combining the first lines of these three CO isotopologues can yield high mutual information with \AV{} over most of the (\AV{}, \Guv{}) space.
Finally, the sensitivity of the \latexmol{hcop}, \latexmol{hcn}, and \latexmol{hnc} lines to large \Guv{} values is related to their large gradient of intensities combined to a high enough S/N in these regions.

\subsection{What is the influence of combining lines?}
\label{subsec:maps_mi_combination}

\begin{figure*}
    \centering
    \includegraphics[width=\linewidth, trim={0 1.5cm 0 0}, clip]{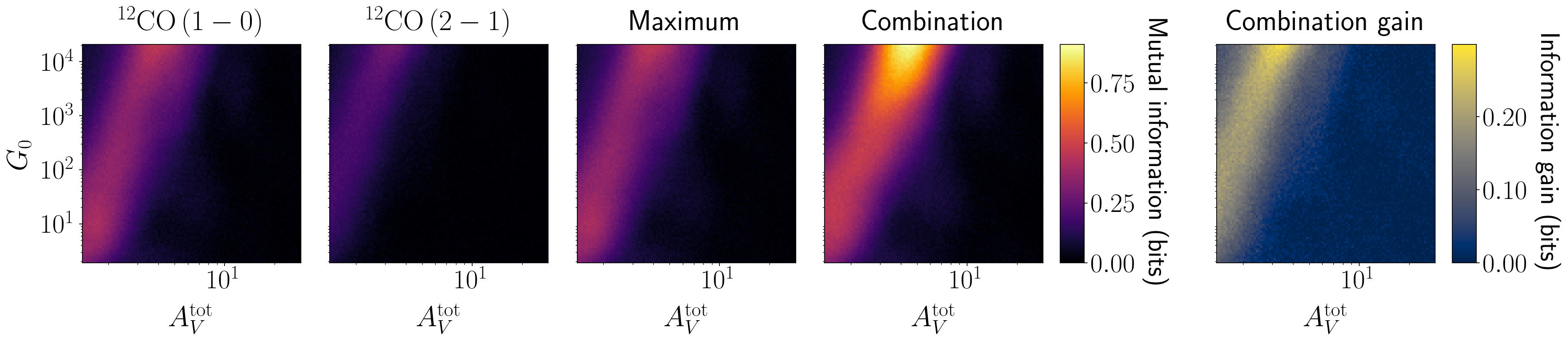}
    \includegraphics[width=\linewidth, trim={0 1.5cm 0 0}, clip]{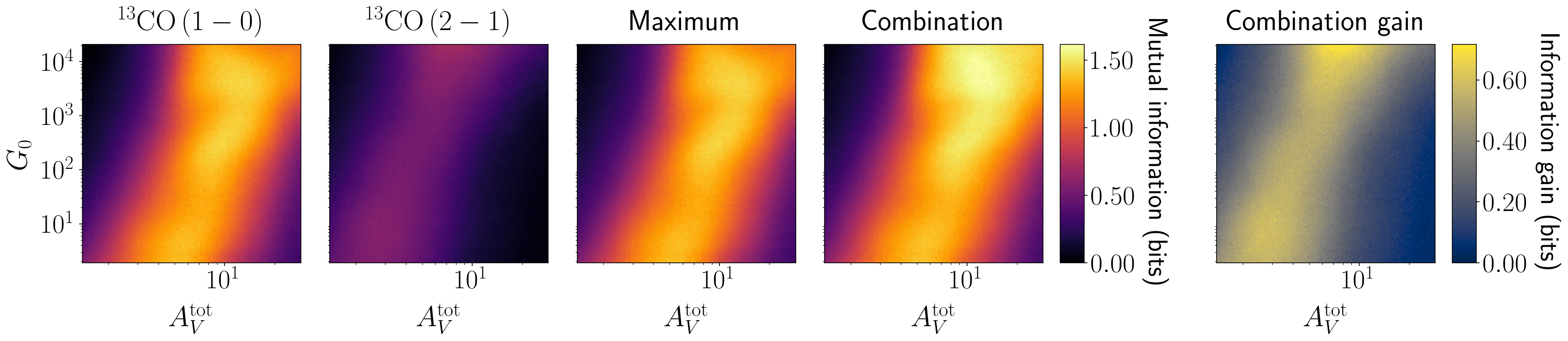}
    \includegraphics[width=\linewidth, trim={0 1.5cm 0 0}, clip]{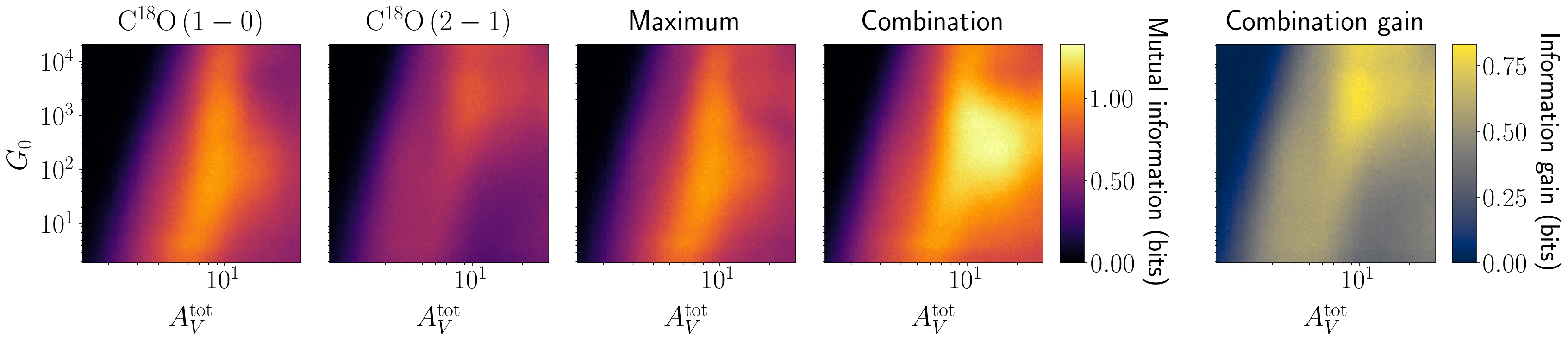}
    \includegraphics[width=\linewidth]{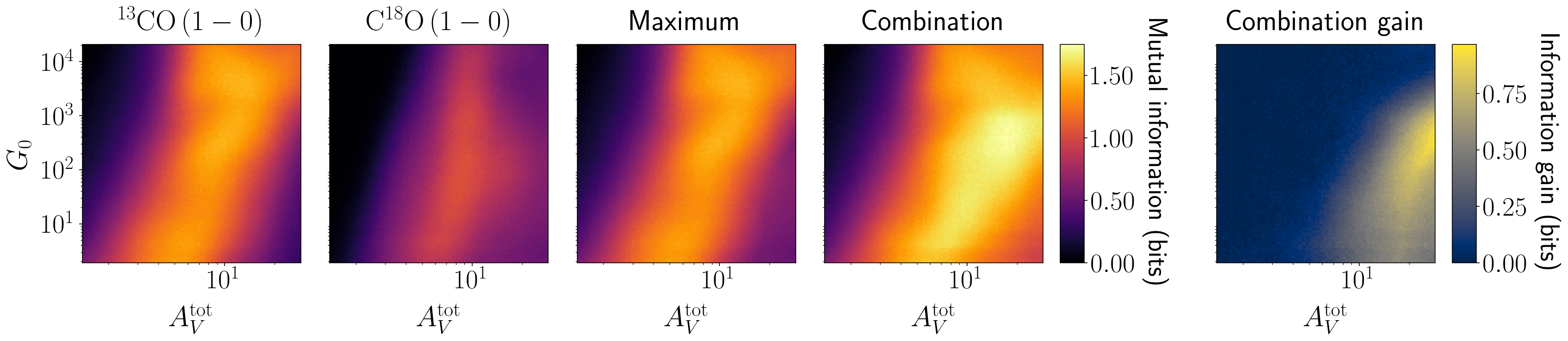}
    \caption{
        Mutual information maps between \AV{} and CO isotopologues lines.
        The first and second columns show the maps for individual lines while the third and fourth columns show the pixel-wise maximum and the combination maps, respectively.
        The last column shows the difference between the last two maps and corresponds to the amount of information gained by combining lines instead of considering only the most informative individual line.
    }%
    \label{fig:maps_mi}
\end{figure*}

The previous section shows how mutual information between individual line intensities and one physical parameter can be understood from a physical viewpoint.
However, using maps of predicted integrated intensities to determine informativity quickly becomes tedious for combinations of lines or combinations of physical parameters.
In particular, which lines to combine to improve informativity, or how informative a combination of lines can be, is unclear with such a simple scheme.
Mutual information allows one to effortlessly and quantitatively answer these questions.

Figure~\ref{fig:maps_mi} shows maps of mutual information for two lines of the three main CO isotopologues, first individually and then combined.
It also shows the highest mutual information for individual lines per physical regime.
As this value is always lower or equal than the mutual information provided by the line combination, it permits estimating an information gain, which is the amount of additional information that is obtained by combining lines.
The three first rows show the two first transitions of \latexmol{12co}, \latexmol{13co}, and \latexmol{c18o}, respectively.
% lines of the three main CO isotopologues
%
Here again, the values of mutual information can be compared at
constant (\AV{}, \Guv{}) values because the conditions of computation of the mutual information are identical.

For \latexmol{12co}, the first two transitions have similar patterns, and trace the same physical regimes in comparable ways.
They are therefore essentially redundant, as their combination does not provide any significant information gain.

For \latexmol{13co}, the second transition becomes informative at higher $\Gnaught{}^{0.15} / \AV{}$ values (towards lower $\AV$ values) than the first transition, and is thus complementary as it does not trace the same regimes.
Therefore, combining the two low-$J$ lines leads to a significant increase of the mutual information with \AV{}.
This confirms the physical insight that higher $J$ lines of $^{13}$CO allow us to better constrain the excitation conditions and thus the column density~\citep[see][]{roueff2024bias}.

Similarly, the first two lines of \latexmol{c18o} are informative in distinct regimes.
Although the \latexmol{c18o} low-$J$ lines considered individually provide little information on very dark cloud conditions, their combination doubles this information (from about 0.5 to more than 1\,bit for $\AV{}>10$\,mag).
This can be related to the fact that the \latexmol{c18o} lines ratio is sensitive to the molecule excitation temperature which is close to the kinetic temperature for such a low dipole moment molecule.

The last row of Fig.~\ref{fig:maps_mi} shows the combination of the \latexline{13co}{1}{0} and \latexline{c18o}{1}{0} lines.
It reveals that this combination brings much information on \Av{} in dense regions, up to almost 1\,bit.
This example shows that combining lines can extend the space of parameters where these lines are useful to constrain a given parameter.
%
% On the other hand, information theory assure that this gain is nonnegative, which means that it can't reduce this space extent.

\section{Line selection on the Horsehead Nebula}%
\label{sec:results_line_selection}

\begin{table} % [h]
    \centering
    \caption{
      Summary of the considered use cases.
    }
    \label{tab:use-cases}
    \renewcommand{\arraystretch}{1.3}
    \begin{tabular}{ccc}
        \hline
        \hline
        Use case
        & with $\kappa$
        & $\times$ integ. time\\
        \hline
        Reference
        & no
        & 1\\
        Deeper integration
        & no
        & 10\\
        Uncertain geometry
        & yes
        & 1\\
        \hline
    \end{tabular}
    \tablefoot{
        These use cases are introduced in Sect.~\ref{subsec:dataset_generation}.
        They are settings in which our line selection approach is applied to highlight specific aspects.
        The $\kappa$ parameter~\eqrefp{eq:obs_model_instance_with_scaling} includes observational uncertainties.
        The integration time factor is the ratio between the actual integration time and the one of the \mbox{ORION-B} dataset, used as a reference.
    }
\end{table}

\begin{figure*}[t]
    \centering
    \begin{subfigure}{0.46\linewidth}
        \includegraphics[width=\linewidth]{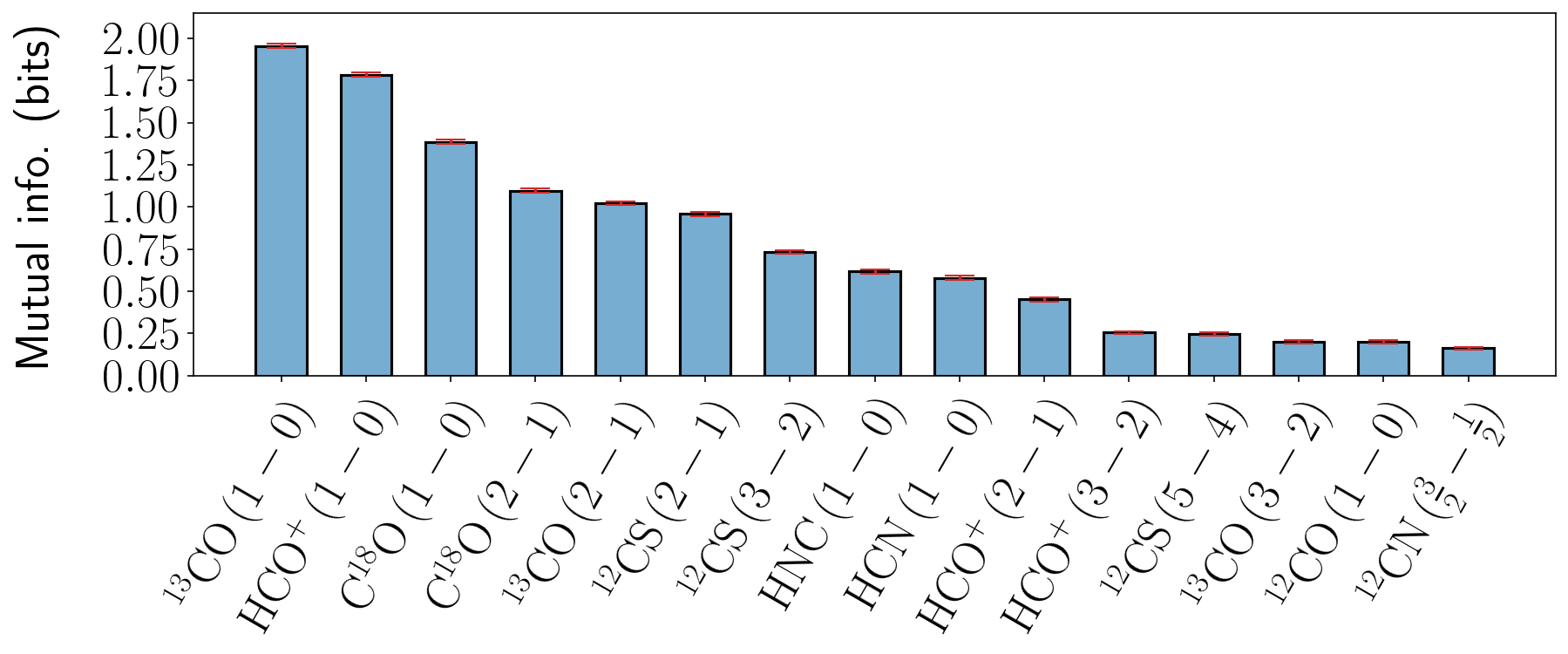}
        \includegraphics[width=\linewidth]{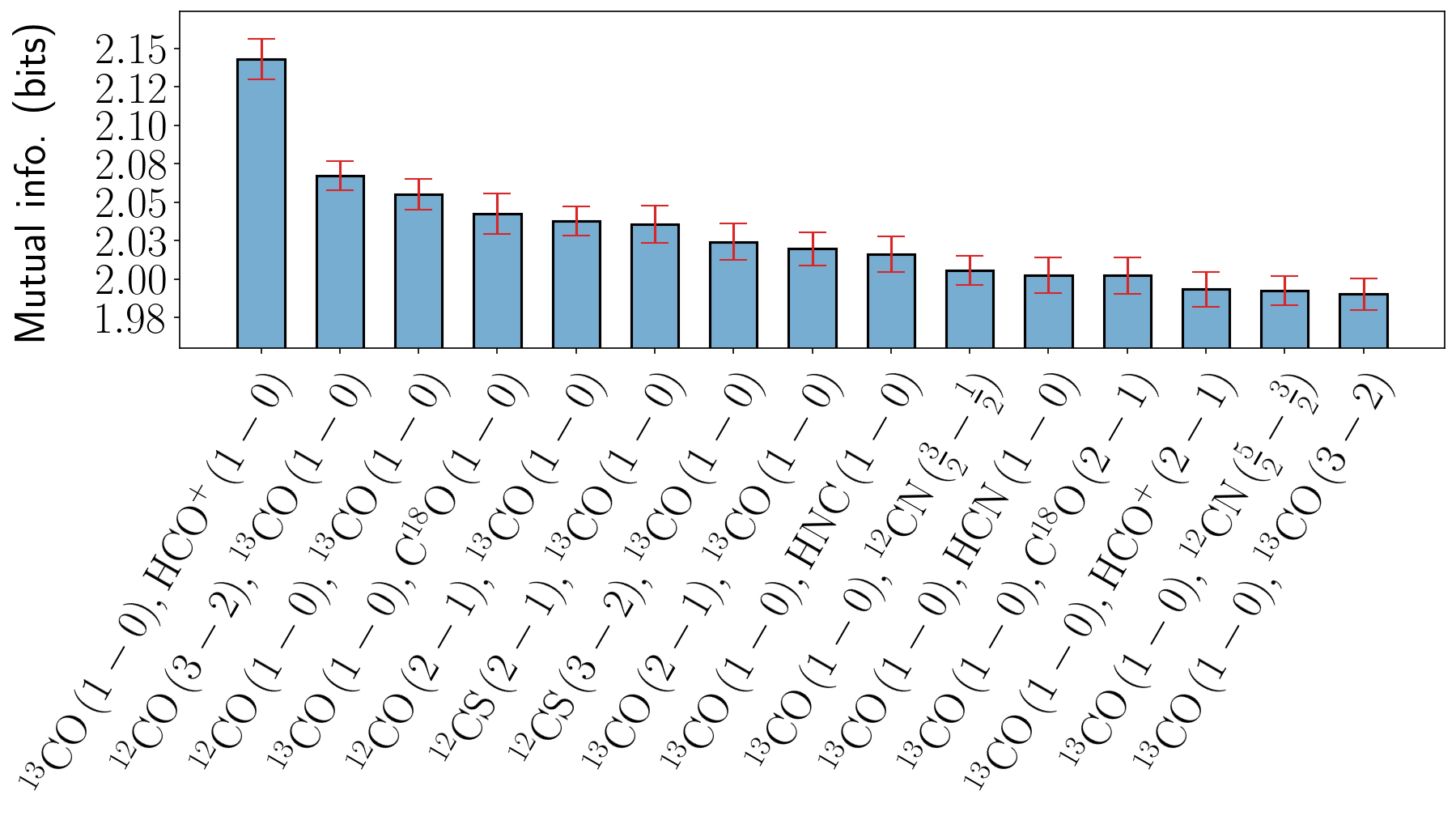}
        \caption{
            All \AV{} environments $(3 \leq \AV \leq 24)$.
        }
        \label{fig:av_selection:all}
    \end{subfigure}
    \hspace{5mm}
    \begin{subfigure}{0.46\linewidth}
        \includegraphics[width=\linewidth]{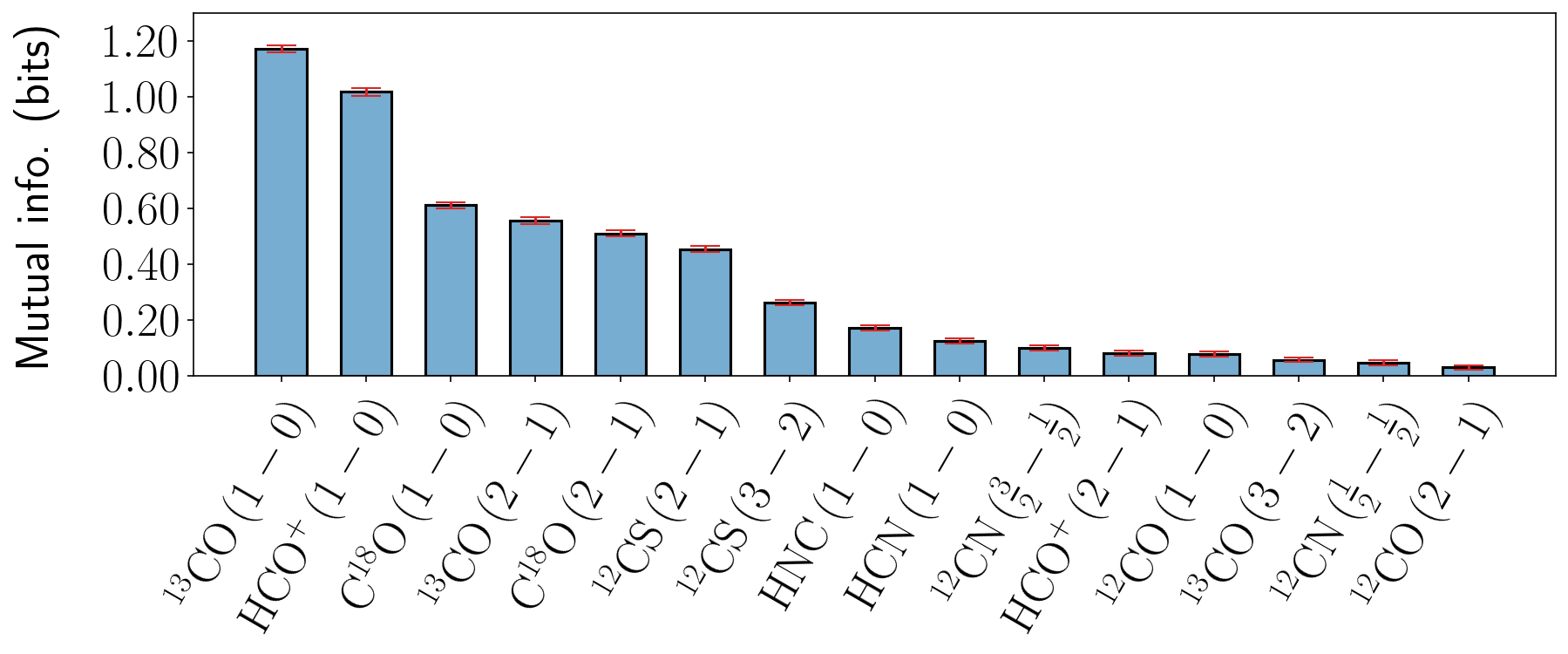}
        \includegraphics[width=\linewidth]{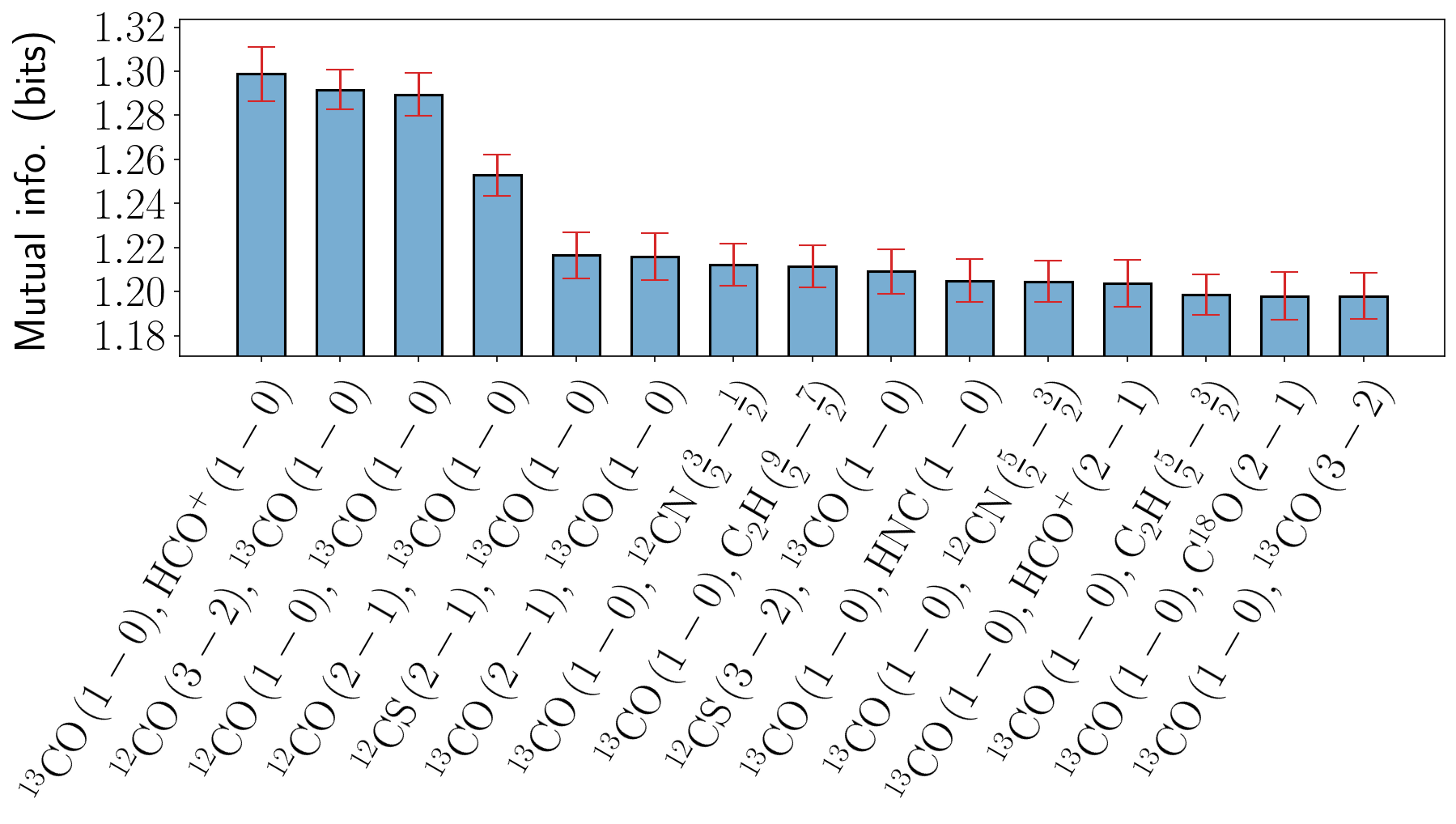}
        \caption{Results for translucent gas $(3 \leq \AV \leq 6)$.}
        \label{fig:av_selection:3_6}
    \end{subfigure}
    \par\medskip
    \begin{subfigure}{0.46\linewidth}
        \includegraphics[width=\linewidth]{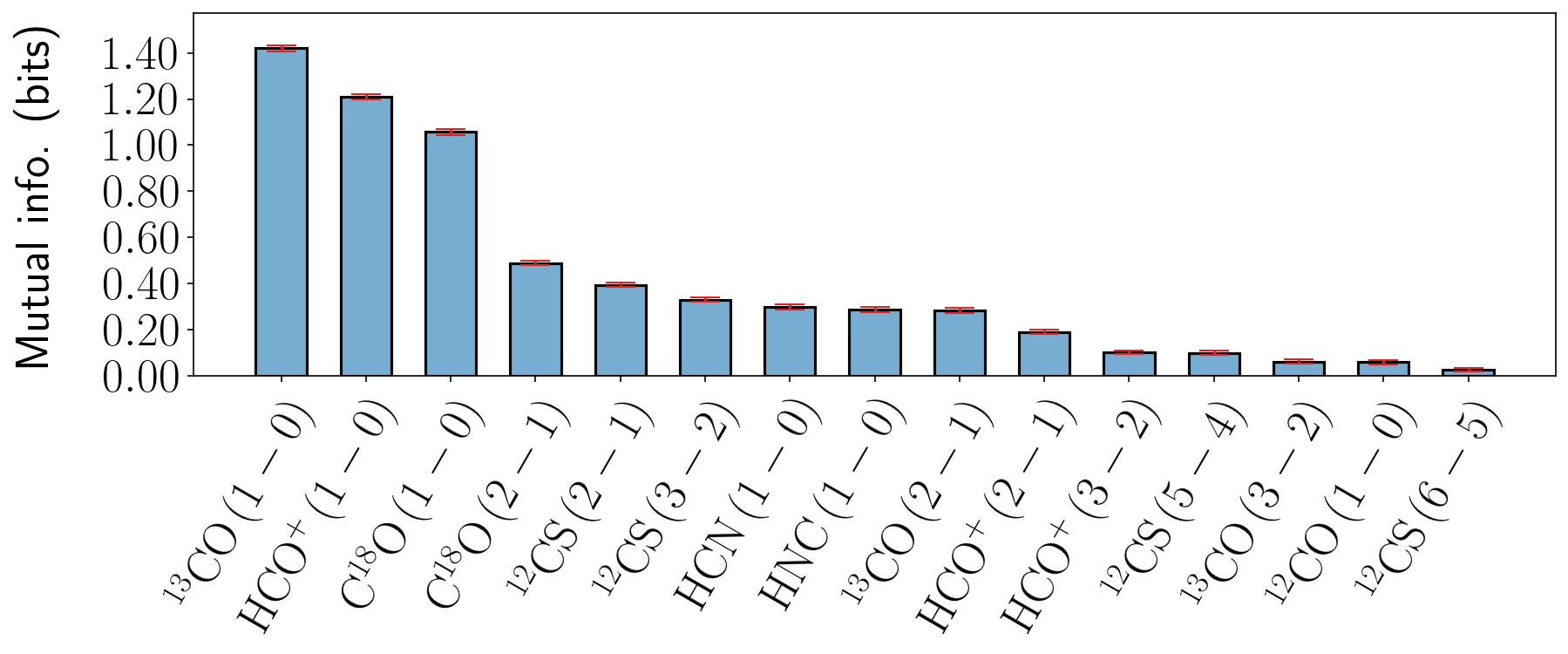}
        \includegraphics[width=\linewidth]{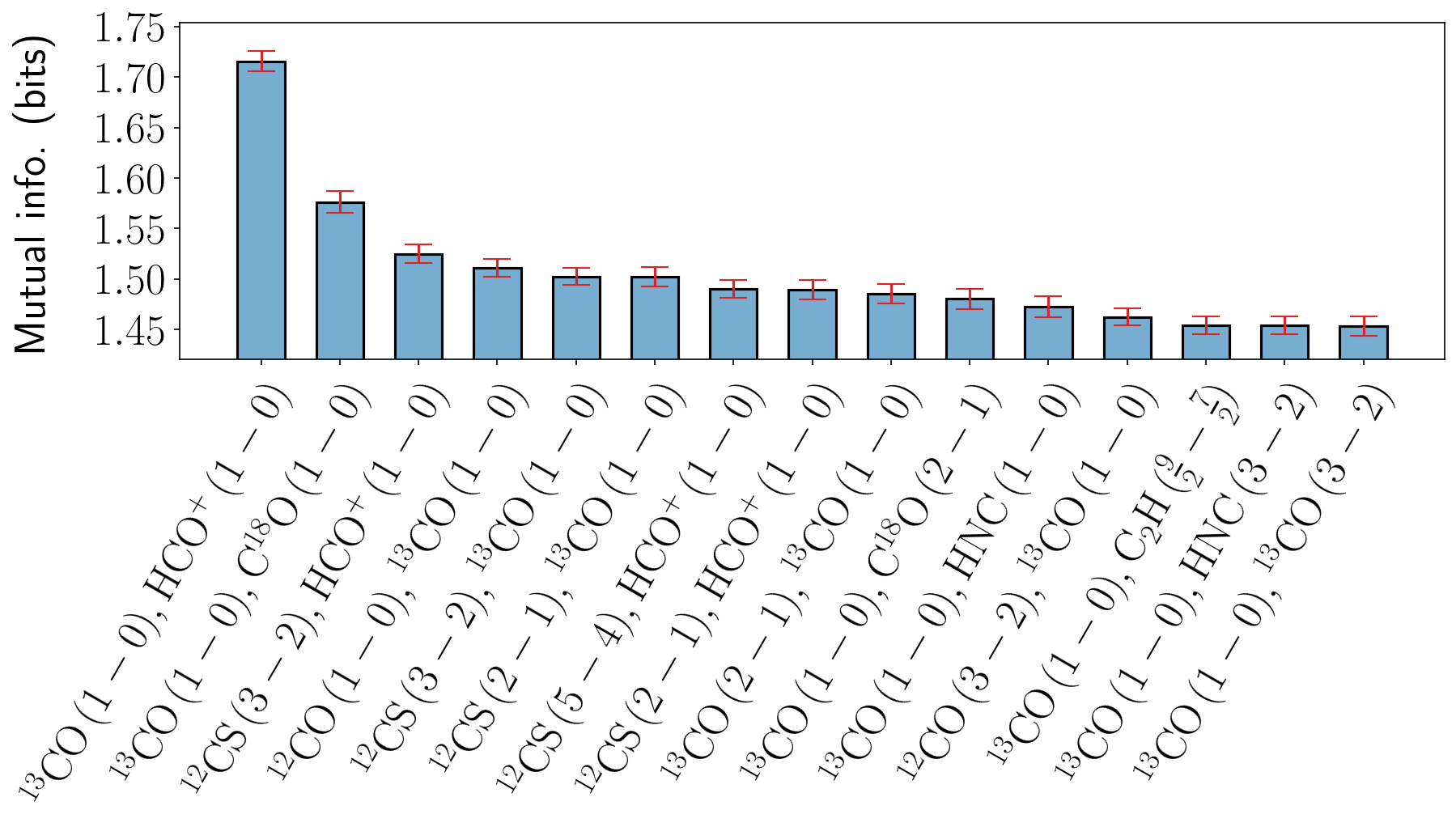}
        \caption{Results for filamentary gas $(6 \leq \AV \leq 12)$.}
        \label{fig:av_selection:6_12}
    \end{subfigure}
    \hspace{5mm}
    \begin{subfigure}{0.46\linewidth}
        \includegraphics[width=\linewidth]{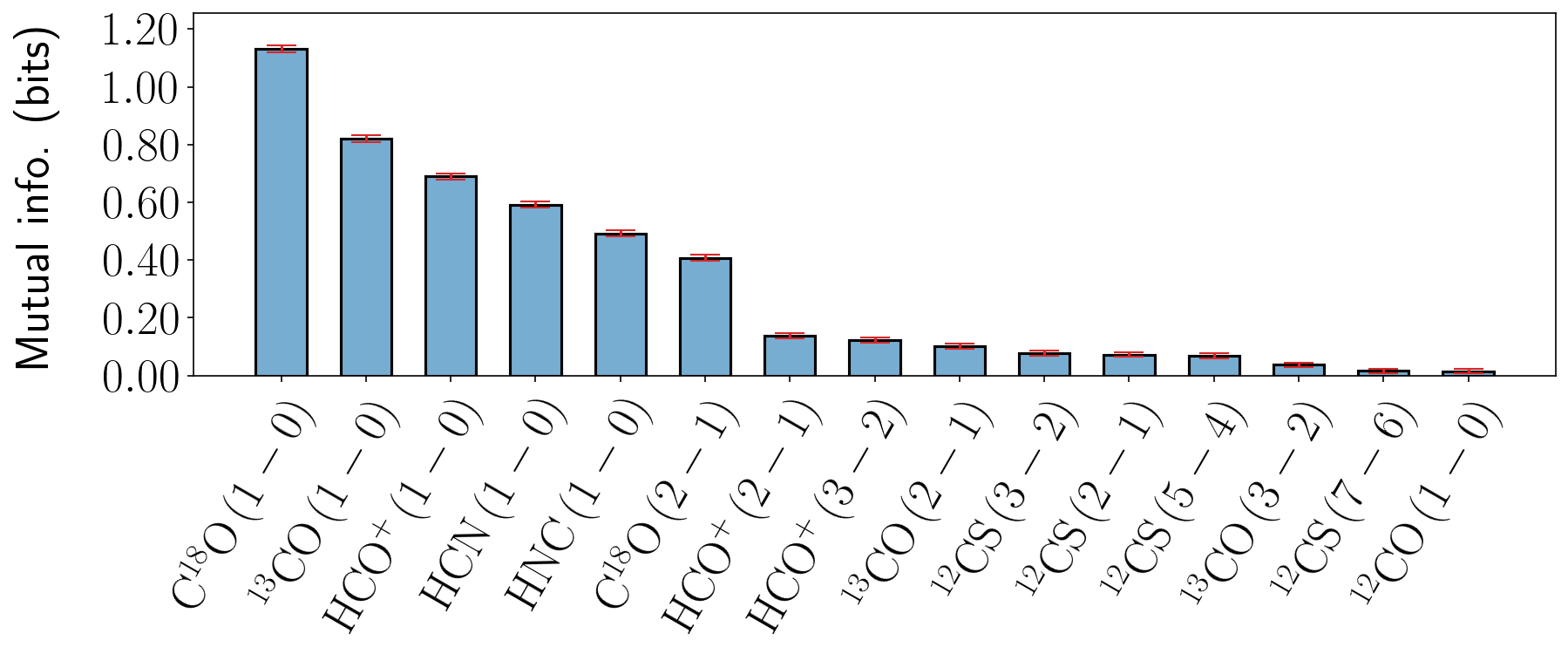}
        \includegraphics[width=\linewidth]{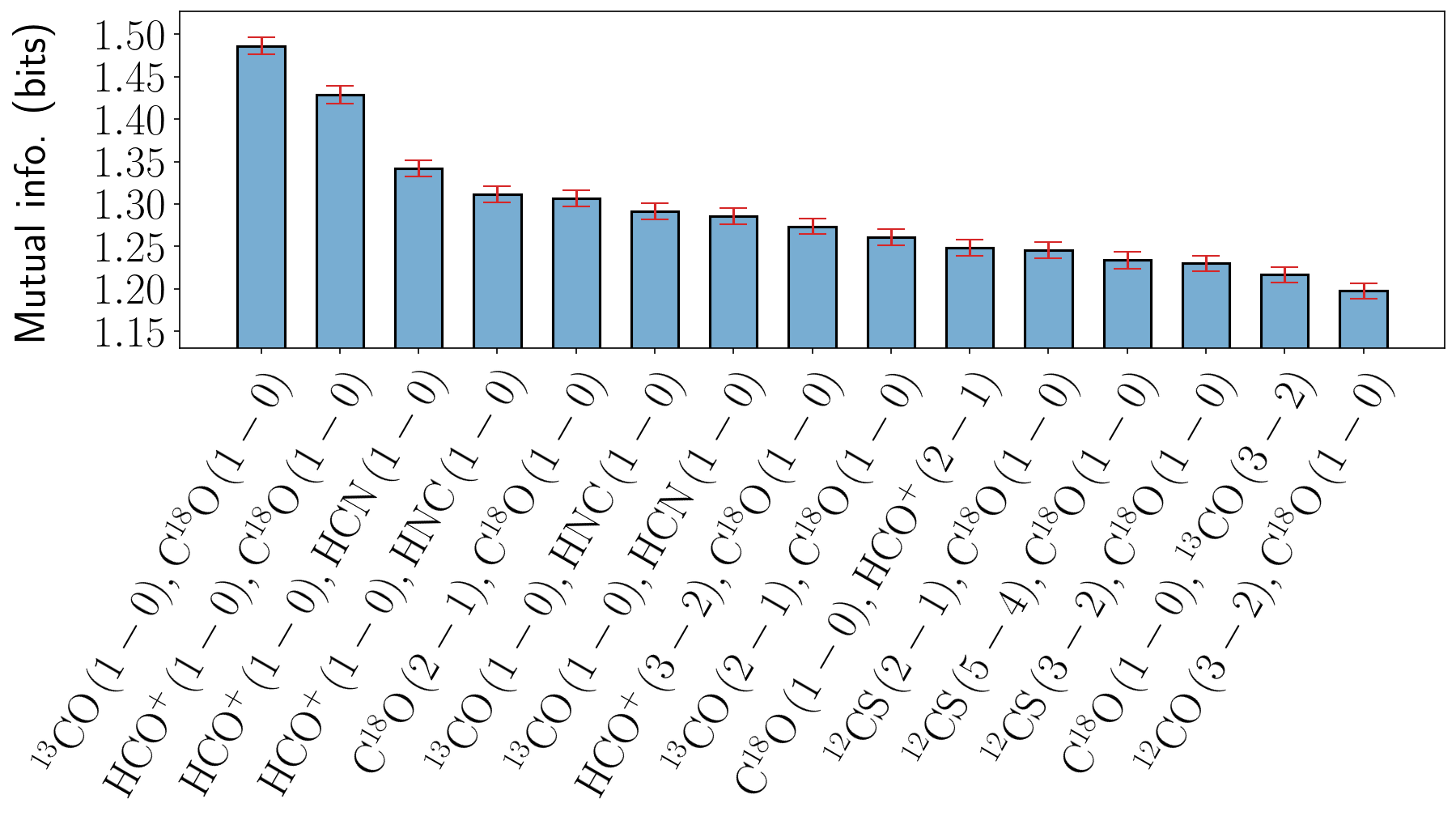}
        \caption{Results for dense cores $(12 \leq \AV \leq 24)$.}
        \label{fig:av_selection:12_24}
    \end{subfigure}
    \caption{
        Line selection for \AV{} for different regimes of \AV{}, in an environment similar to the Horsehead pillar, for the reference use case (reference integration time and no scaling factor $\kappa$ in the simulator of observation).
        %
        % The mutual information values of the best 15 line couples are less scattered than for individual lines.
        %
        For better visualization, for couples of lines, we set the plot lower limit to the highest mutual information of individual EMIR lines.
        % For better visualization, for couples of lines, we ``zoom in'' on the highest values by setting the plot lower limit to the highest mutual information of individual EMIR lines.
        %
        The figure thus provides direct information on the gain achieved by combining lines compared to individual lines.
    }%
    \label{fig:av_selection}
\end{figure*}

In this section, we apply the line selection method introduced in Sect.~\ref{sec:optimization_pb} to determine the best (combination of) lines to constrain \AV{} or \Guv{}.
For simplicity, we restrict ourselves to the space of parameters present in the Horsehead Nebula (see Table~\ref{tab:parameters-distribution}), mostly observed with EMIR at the IRAM~30m telescope.
We first analyze which lines are the most sensitive to \AV{} in the case where the S/N is set by the integration time per pixel achieved in the \mbox{ORION-B} Large Program.
Hereafter, we refer to this framework as the ``reference use case''.
Secondly, we consider how the line ranking changes when integrating ten times longer.
We then assess the importance of additional causes of uncertainty such as the inclination of the source on the line of sight or the beam dilution when trying to infer \Guv{}.
Finally, we quantify the gain of analyzing two lines with respect to just analyzing their ratio.
To make these studies, we generate three sets of simulated observation $(\paramvect{n}, \obsvect{n})_{n=1}^N$ with $N = 10^4$ as described in Sect.~\ref{sec:optimization_pb}.
Table~\ref{tab:use-cases} lists the detailed characteristics of the considered use cases.

The results are discussed for all the values of \AV{} present in the Horsehead
\mbox{$(3 \leq \AV \leq 24)$}, and for three physical subregime, namely translucent clouds with \mbox{$3 \leq \AV \leq 6$}, filamentary gas with \mbox{$6 \leq \AV \leq 12$}, and dense cores with \mbox{$12 \leq \AV \leq 24$}.
In contrast with the results presented in the previous section, the values of mutual information can not be easily compared from one physical regime to the other because the distribution of $\paramfull{}$ differs from one regime to the other.
However, the values of mutual information can be compared within one regime, for individual lines or combination of lines and for \AV{} and \Guv{}.

\subsection{Best lines to infer \AV{} for the reference use case}%
\label{subsec:situation_B}

Figure~\ref{fig:av_selection} shows the mutual information between the visual extinction \AV{} and the intensity of either one line or a line couple, ranked by decreasing order of the mutual information.
Only the first 15 most informative lines or couples are displayed for readability.
Red error bars on the mutual information allow one to assess the significance of the line ranking (see App.~\ref{subsec:estimation} and App.~\ref{app:variance} for details on their computation).

In the case of the Horsehead Nebula featuring large variations of \AV{} ($3 \leq \AV \leq 24$\,mag), the most informative individual lines are the ground state transitions of \latexmol{13co}, \latexmol{hcop} and \latexmol{c18o}, followed by the second transition of \latexmol{c18o} and \latexmol{13co}.
The $^{12}$CO lines are individually poorly informative.
These results are consistent with the mutual information maps from Fig.~\ref{fig:maps-grid-av}.
The most informative couples of lines here simply combine the single most informative individual line, \ie, the ground state transition of \latexmol{13co}, with another line.
In particular, the most informative couple of lines (ground state transitions of \latexmol{13co} and \latexmol{hcop}) combines the two most informative individual lines.
However, this combination only improves the mutual information by about 0.2\,bits.
In other words, using only \latexline{13co}{1}{0} to infer \AV{} instead of the combination of any line couples results in a limited loss of information.

Figures~\ref{fig:av_selection:3_6},~\ref{fig:av_selection:6_12} and~\ref{fig:av_selection:12_24} show the line rankings for the three sub-regimes of \AV{}.
In each of these sub-regimes, the ground state transition of \latexmol{13co} is among the top two most informative individual lines, but it falls behind \latexmol{c18o} for the highest \AV{} as it becomes optically thick.
Conversely, the ground state transition of \latexmol{c18o} improves its ranking as \AV{} grows, because its S/N increases and it remains optically thin.
In the translucent regime, one of the most informative couple of lines is $\left(\latexline{13co}{1}{0}, \latexline{12co}{1}{0}\right)$, even though \latexline{12co}{1}{0} is individually relatively uninformative in this regime.
This can be explained by the fact that, for a single line, the excitation of the line shows a degeneracy between column density and gas temperature.
A highly optically thick line, such as \latexline{12co}{1}{0}, provides information on the gas temperature, and thus helps lifting this degeneracy~\citep{roueffC18O13CO12CO2021,roueff2024bias}.

These results are consistent with~\citet{gratierQuantitativeInferenceColumn2021}.
We both obtain that for the Horsehead Nebula, the three most informative line to trace the extinction include \latexline{13co}{1}{0} and \latexline{hcop}{1}{0} for translucent gas.
We also both find that they include the \latexline{13co}{1}{0} and \latexline{c18o}{1}{0} for filamentary gas.

\subsection{Best lines to infer \Guv{} for the reference use case}%
\label{subsec:situation_B_G0}

\begin{figure*}[ht]
    \centering
    \begin{subfigure}{0.46\linewidth}
        \includegraphics[width=\linewidth]{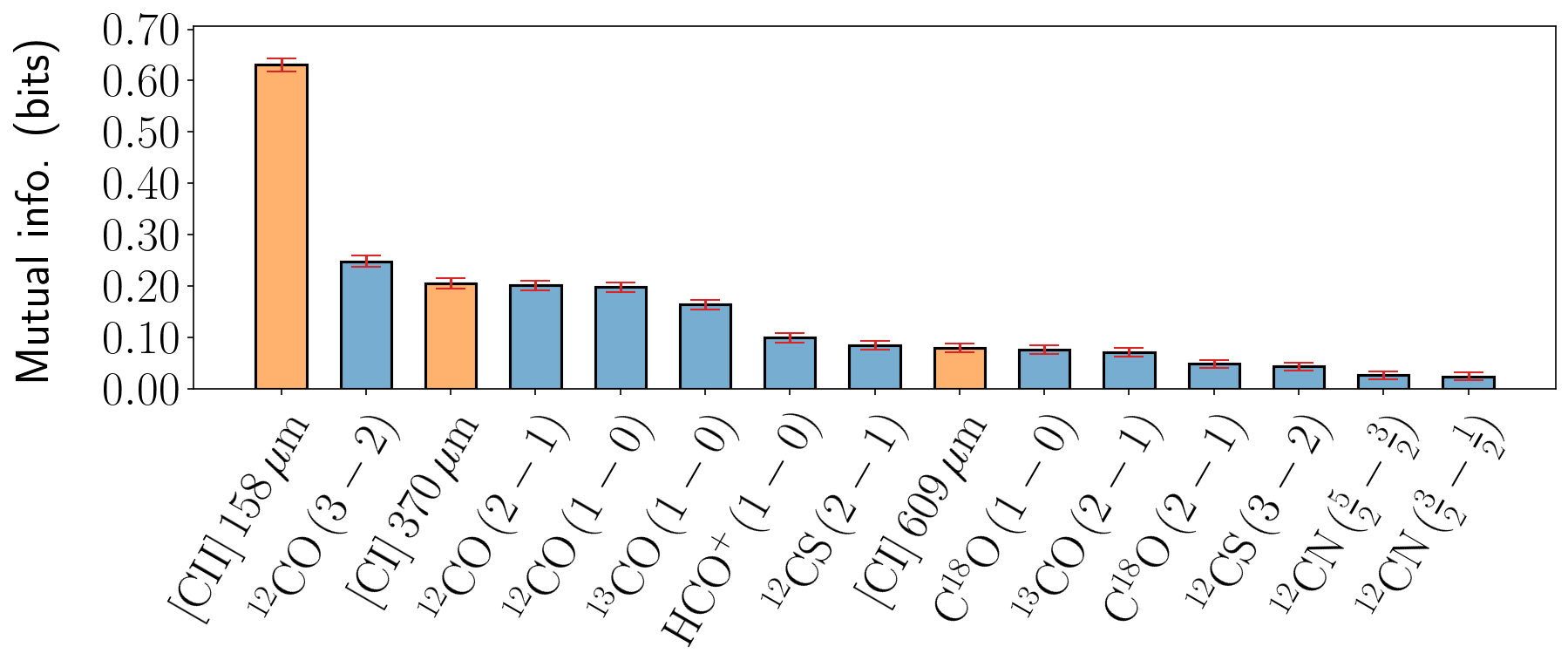}
        \includegraphics[width=\linewidth]{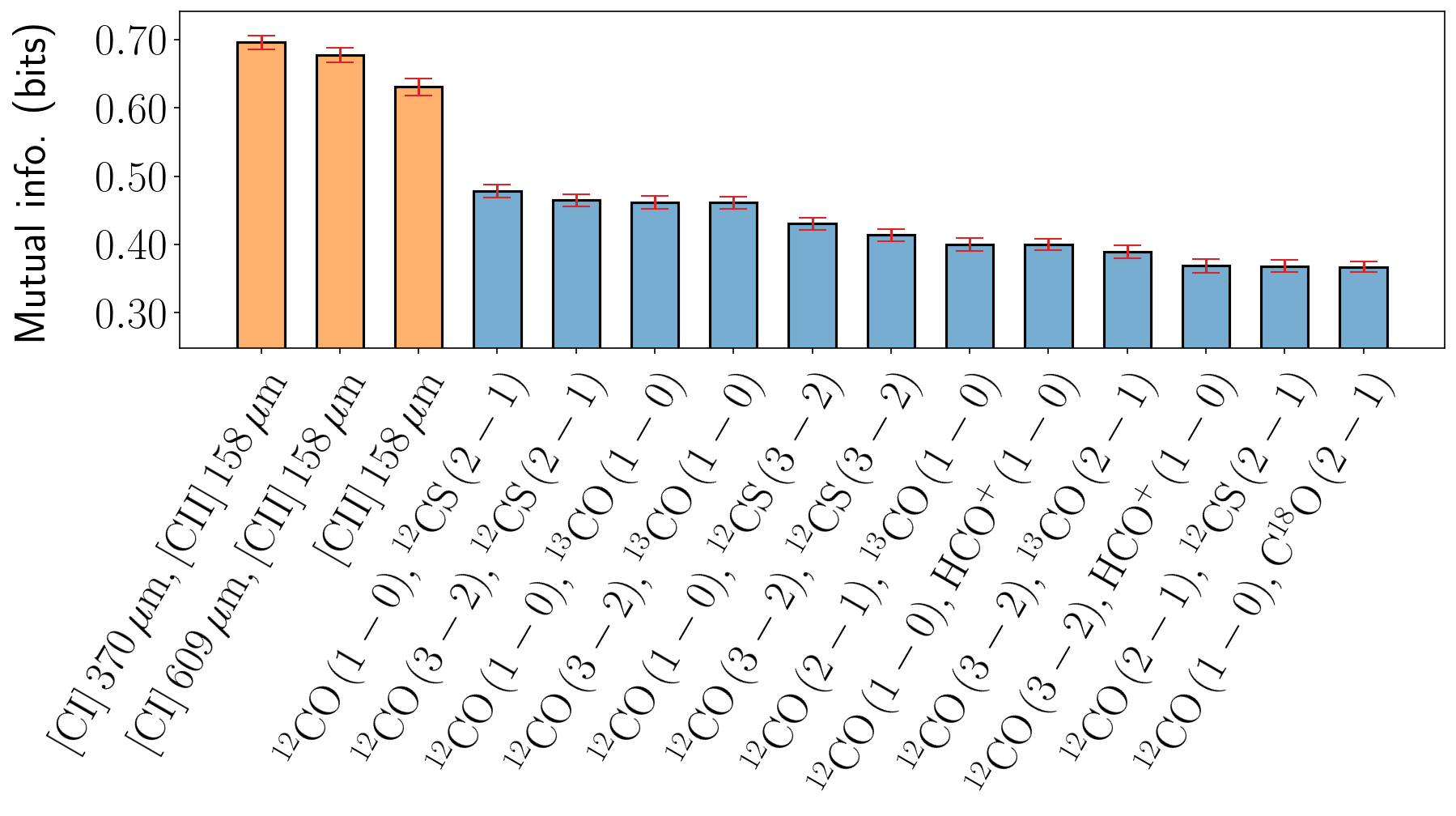}
        \caption{All \AV{} environments $(3 \leq \AV \leq 24)$.}
    \end{subfigure}
    \hspace{5mm}
    \begin{subfigure}{0.46\linewidth}
        \includegraphics[width=\linewidth]{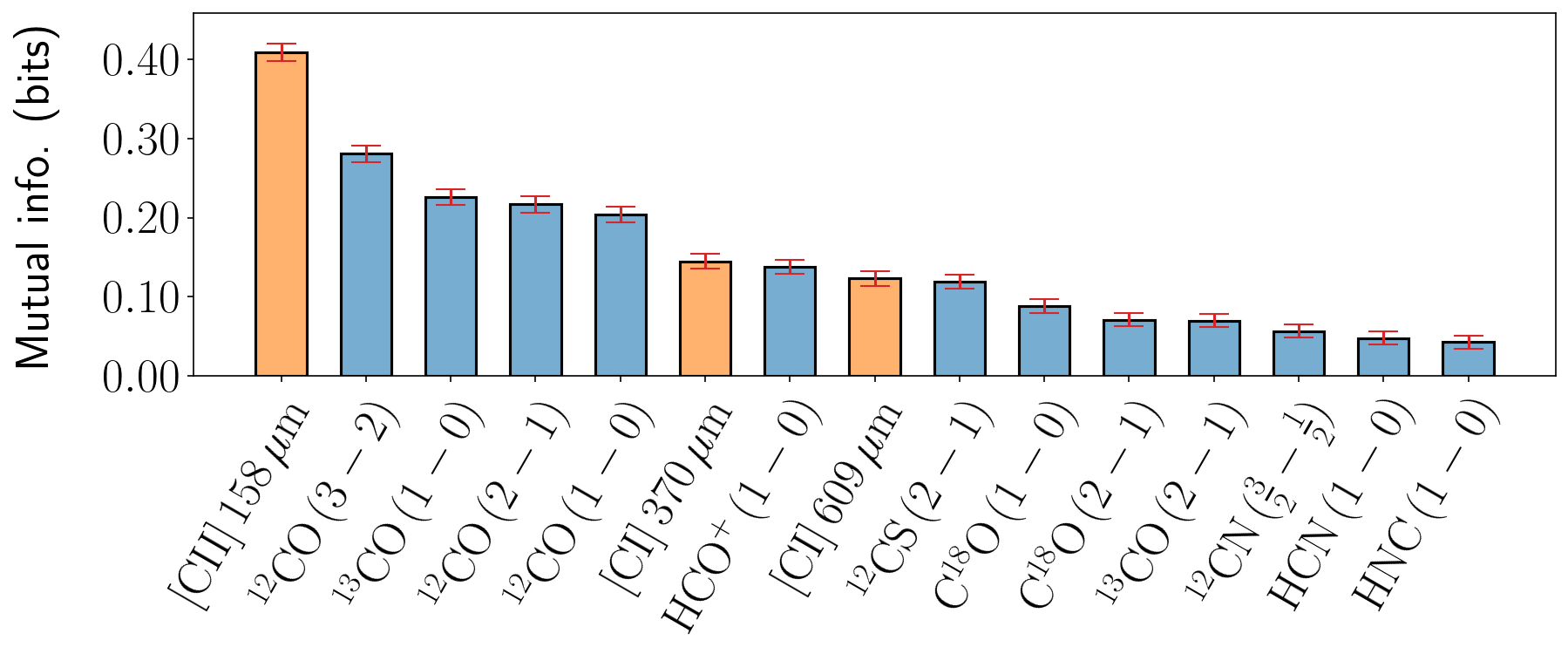}
        \includegraphics[width=\linewidth]{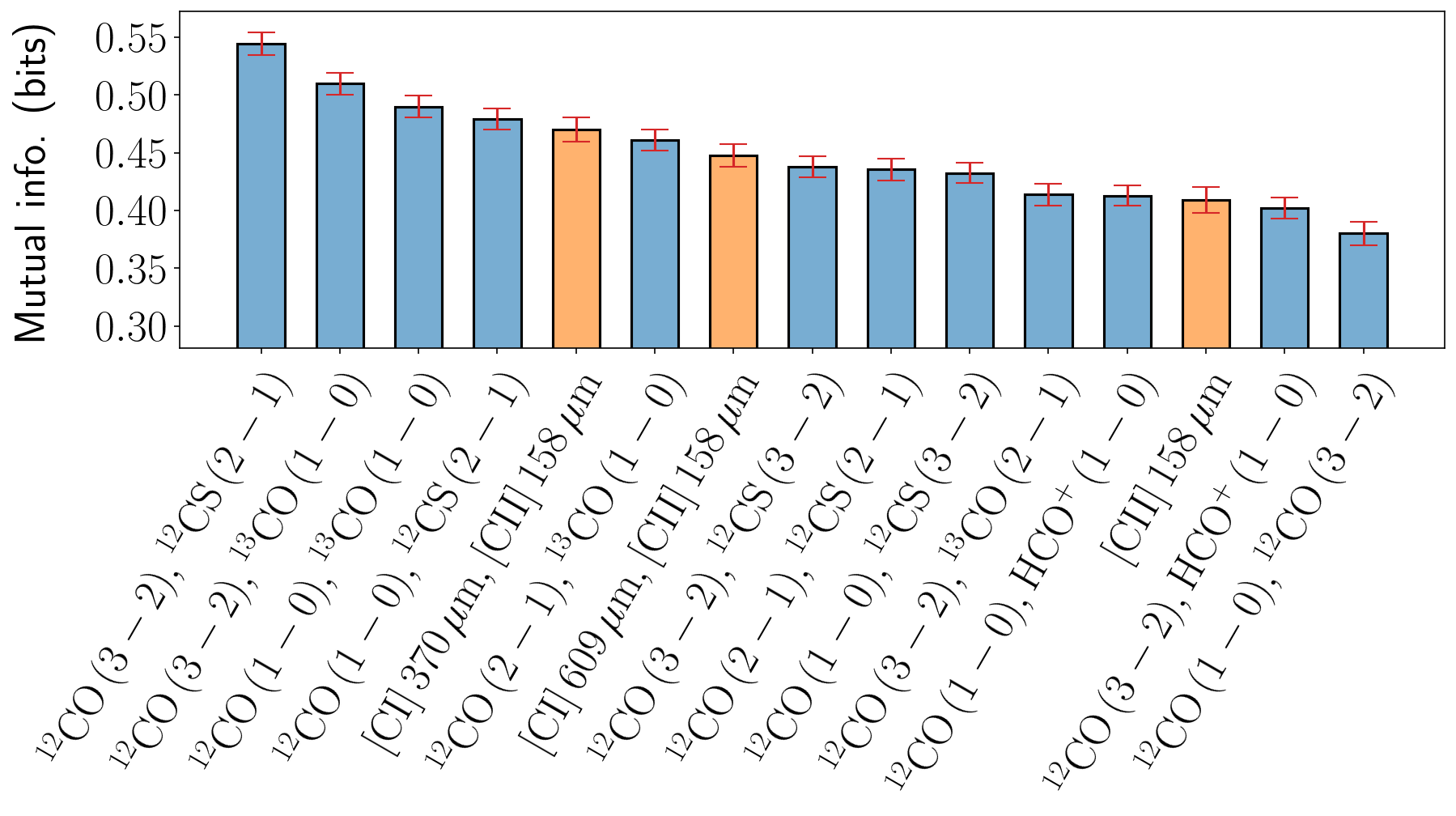}
        \caption{Results for translucent gas $(3 \leq \AV \leq 6)$.}
    \end{subfigure}
    \par\medskip
    \begin{subfigure}{0.46\linewidth}
        \includegraphics[width=\linewidth]{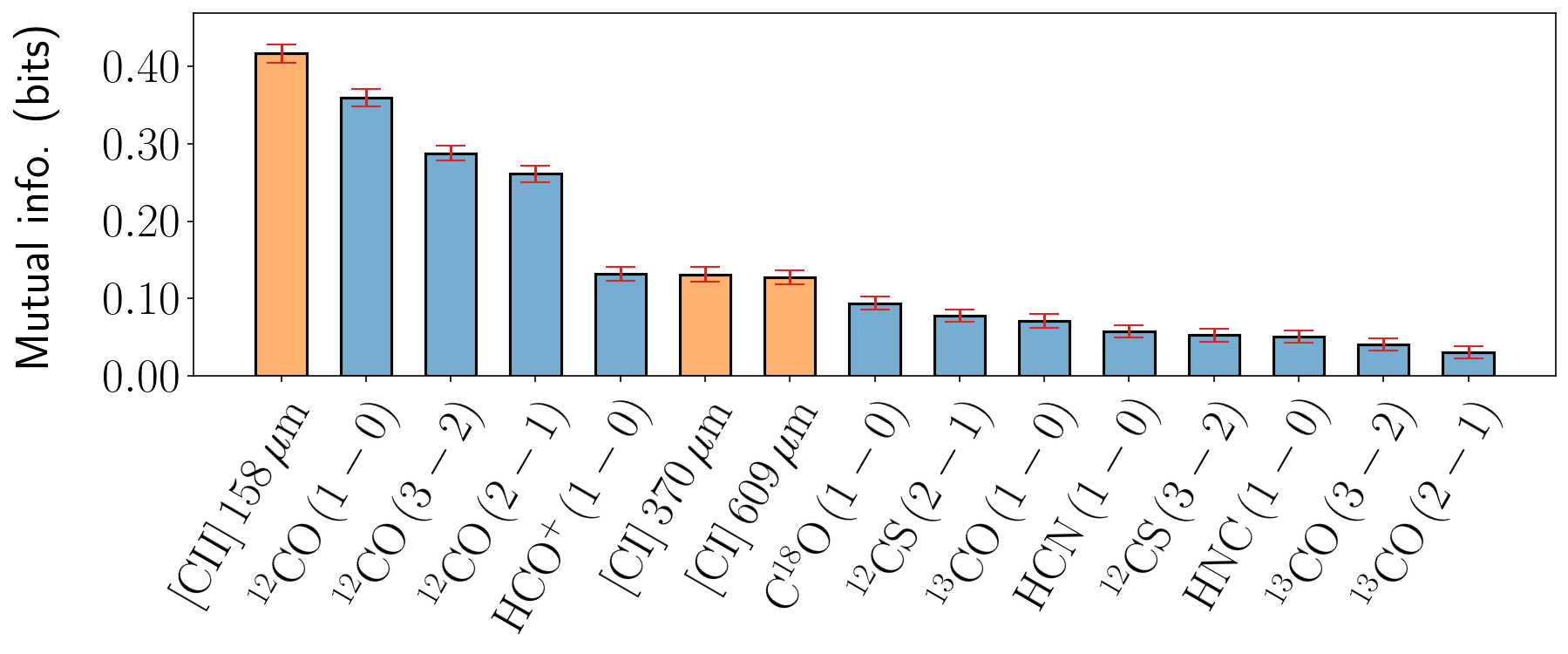}
        \includegraphics[width=\linewidth]{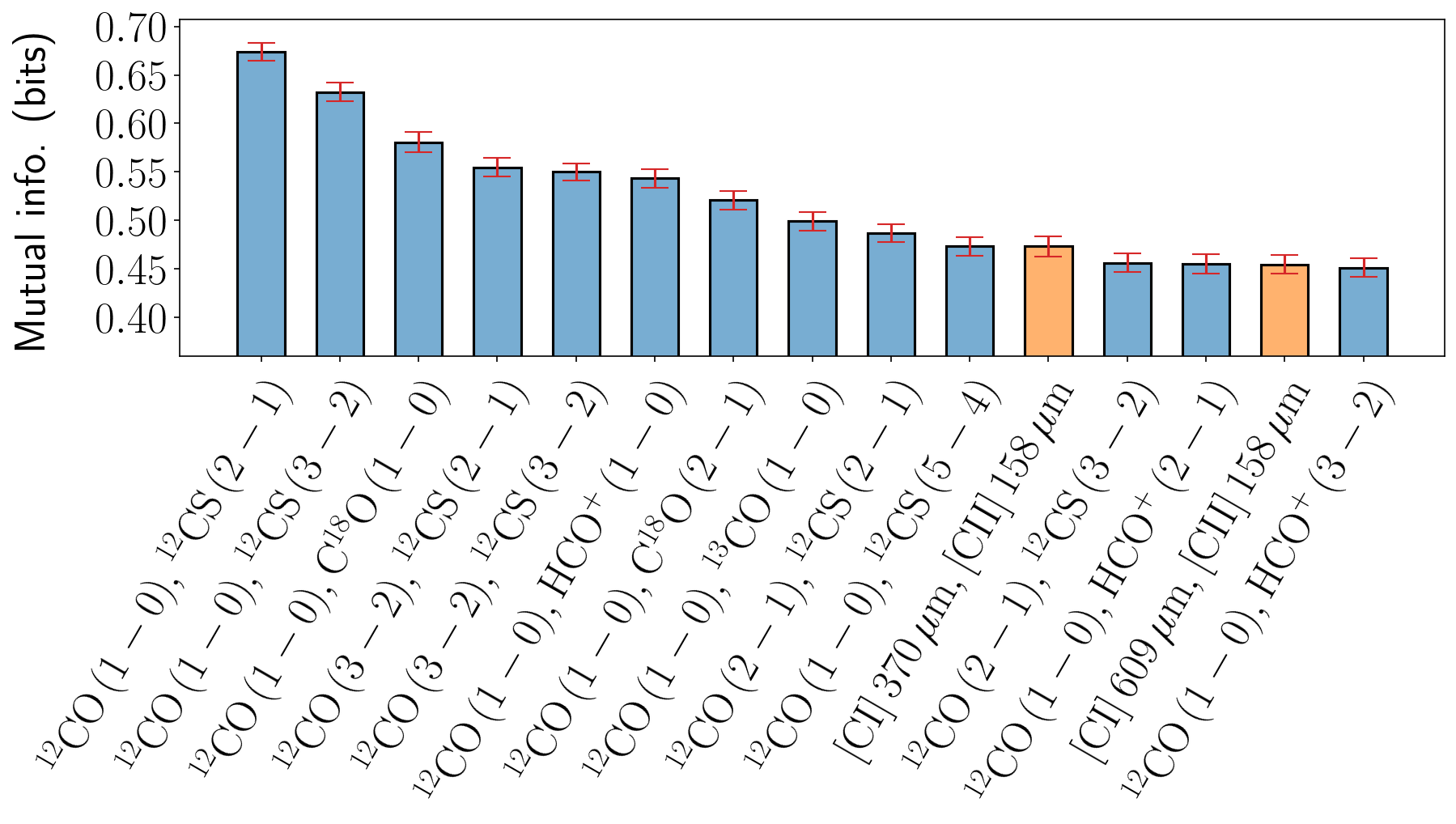}
        \caption{Results for filamentary gas $(6 \leq \AV \leq 12)$.}
    \end{subfigure}
    \hspace{5mm}
    \begin{subfigure}{0.46\linewidth}
        \includegraphics[width=\linewidth]{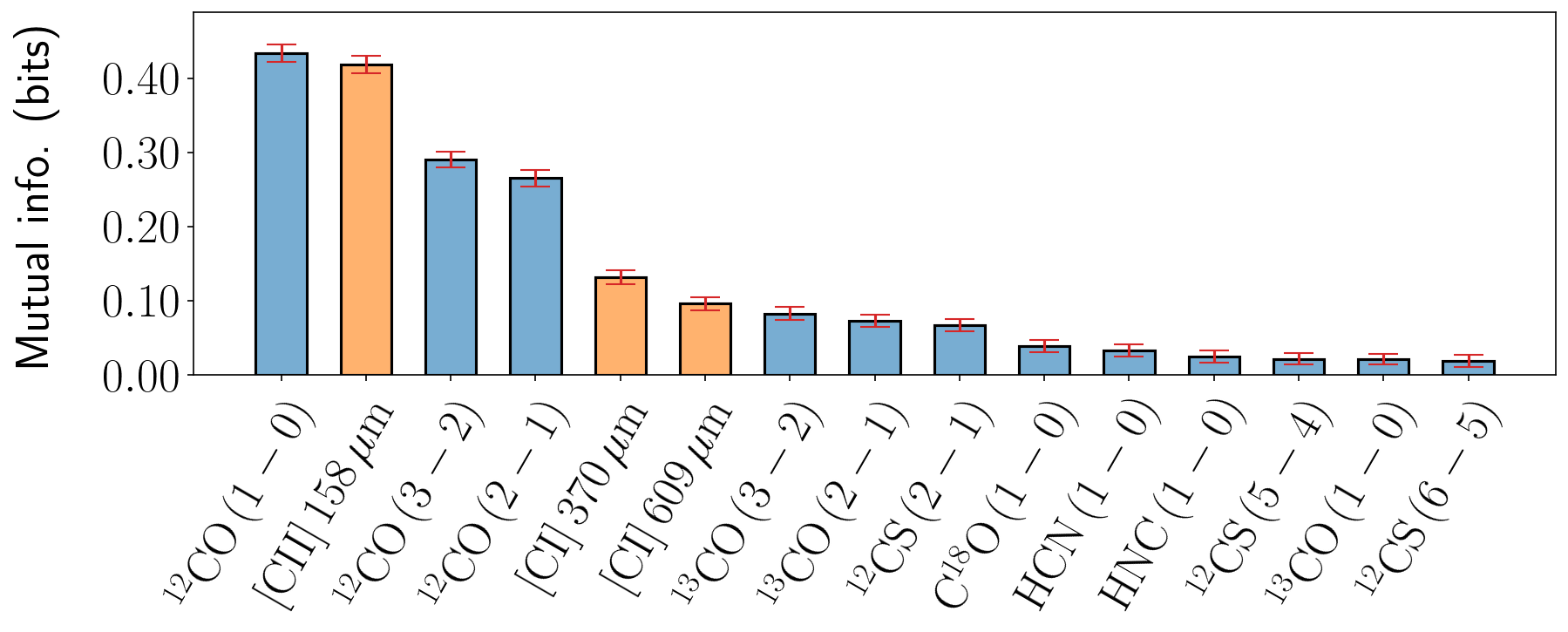}
        \includegraphics[width=\linewidth]{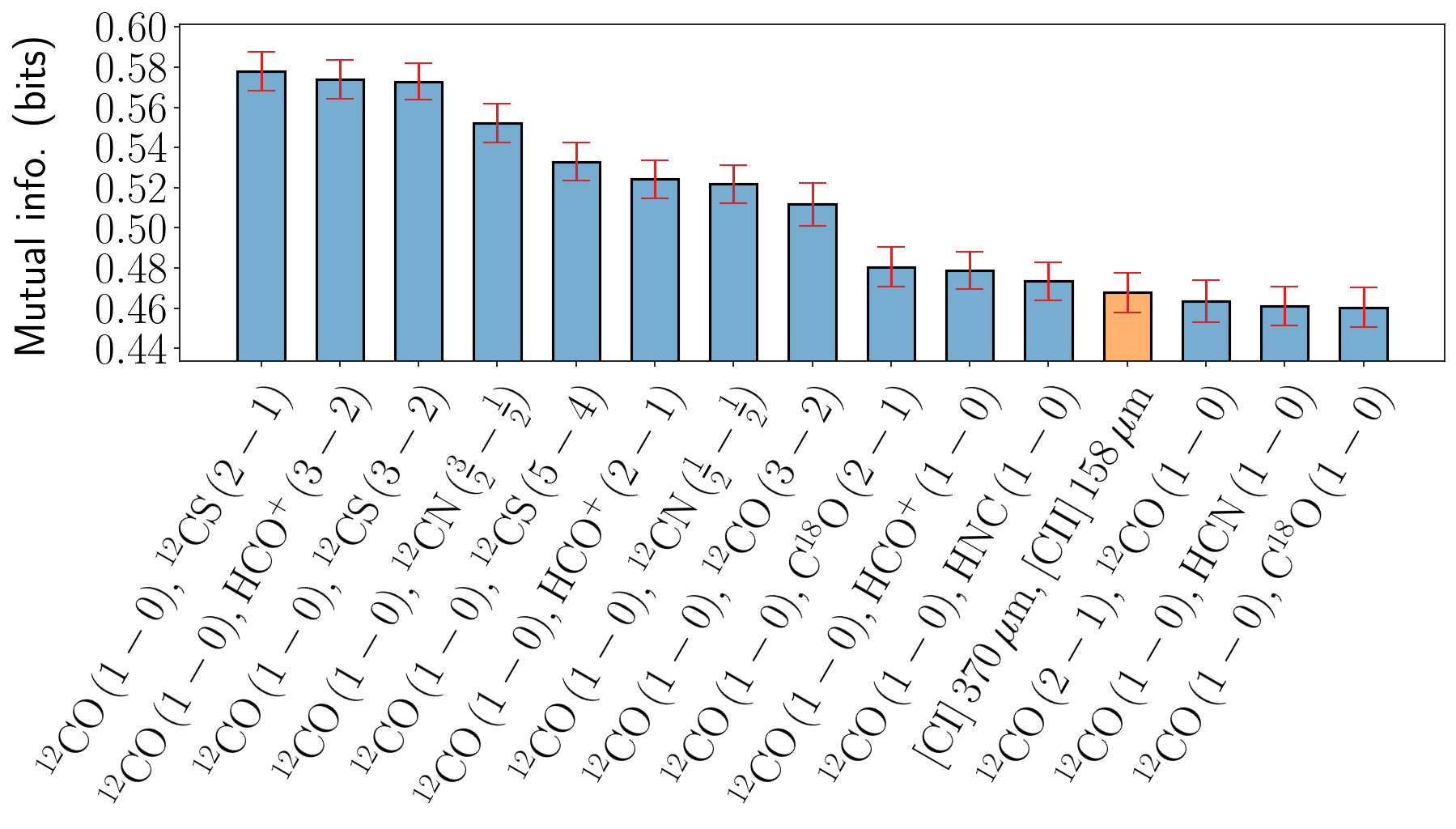}
        \caption{Results for dense cores $(12 \leq \AV \leq 24)$.}
    \end{subfigure}
    \caption{
        Line selection for \Guv{} for different regimes of \AV{}, in an environment similar to the Horsehead pillar, for the reference use case (reference integration time and no scaling factor $\kappa$ in the observation simulator).
        Orange bars correspond to \ci{} and \cp{} lines.
    }%
    \label{fig:g0_selection}
\end{figure*}

Figure~\ref{fig:g0_selection} shows the mutual information between the incident UV radiative field intensity \Guv{} and the intensity of individual or couples lines, sorted by decreasing mutual information.
The mutual information with \Guv{} is always lower than 0.65\,bits.

The seven most informative lines are the \cp{}, \ci{} and \latexmol{12co} lines.
While \AV{} is related to the cloud depth, \Guv{} is a physical quantity defined at the cloud surface.
It is therefore intuitive that the most informative lines for \Guv{} are those that exist in the outer layers of the cloud.
At the ionization front, the carbon is mostly in ionized state, and after the photodissociation front converts to C and then to mostly CO.

When mixing all kinds of gas, the \cp{} line is the most informative one.
The mutual information of \latexmol{12co} lines increases with the regime of \AV{}, and \latexline{12co}{1}{0} becomes the most informative line to infer \Guv{} towards dense cores.
In this regime, the \latexline{12co}{1}{0} line is optically thick.
The intensity at which it saturates mostly depends on the kinetic temperature~\citep{kaufmanFarInfraredSubmillimeter1999}, and thus on \Gnaught{}.
However, looking at pairs of lines, some combinations of molecular lines are more informative than any combination of the \cp{} and \ci{} lines.
This result is encouraging for ISM studies since \cp{} and \ci{} lines can no longer be observed with Herschel and SOFIA.
In particular, to the best of our knowledge, there is currently no instrument that can observe the \cp{} line, and this should not change in the coming years.

\subsection{Effect of integration time on the best lines to infer \Av}%
\label{subsec:add_noise_var_effect}

\begin{figure*}
    \centering
    \begin{subfigure}{0.46\linewidth}
        \includegraphics[width=\linewidth]{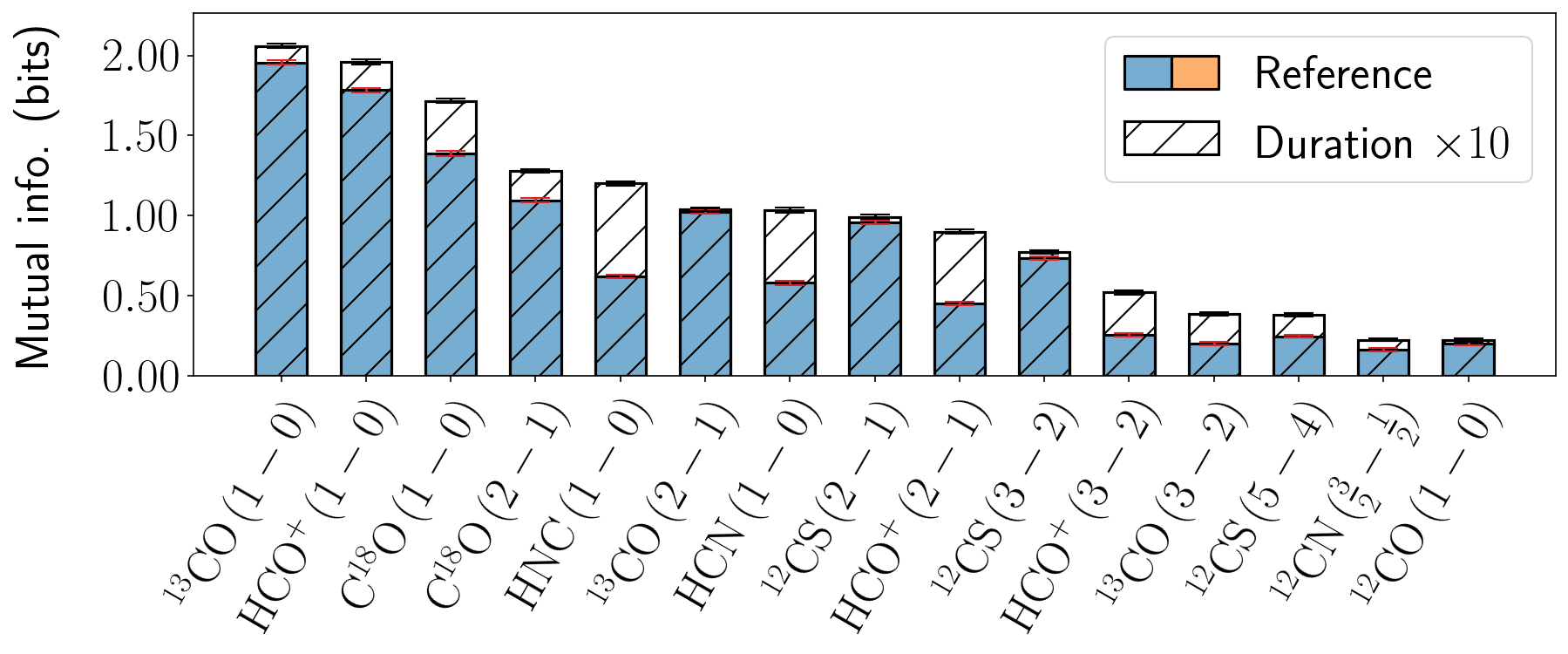}
        \includegraphics[width=\linewidth]{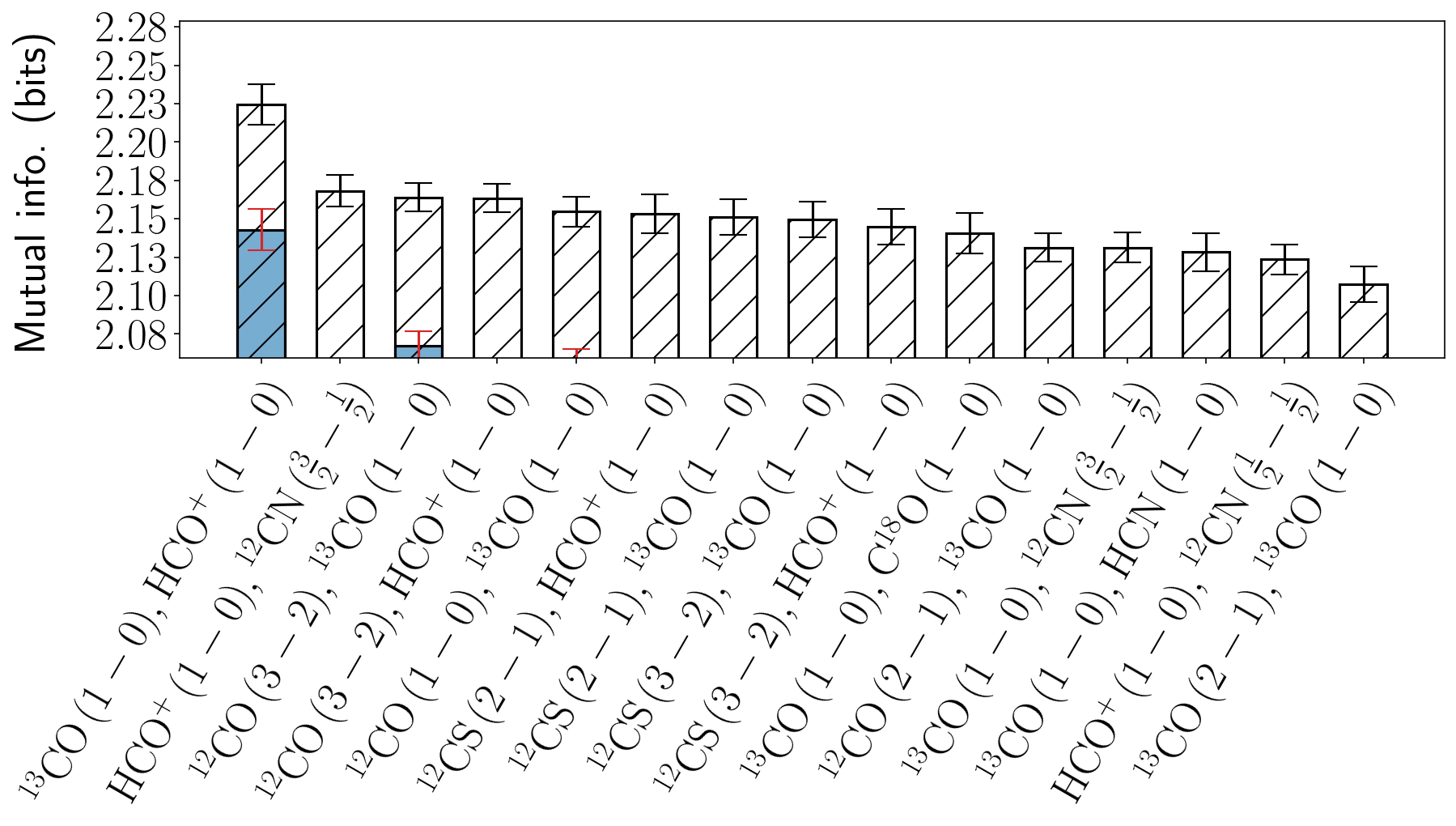}
        \caption{
            All \AV{} environments $(3 \leq \AV \leq 24)$.
        }
        \label{fig:obs_time_av_selection:all}
    \end{subfigure}
    \hspace{5mm}
    \begin{subfigure}{0.46\linewidth}
        \includegraphics[width=\linewidth]{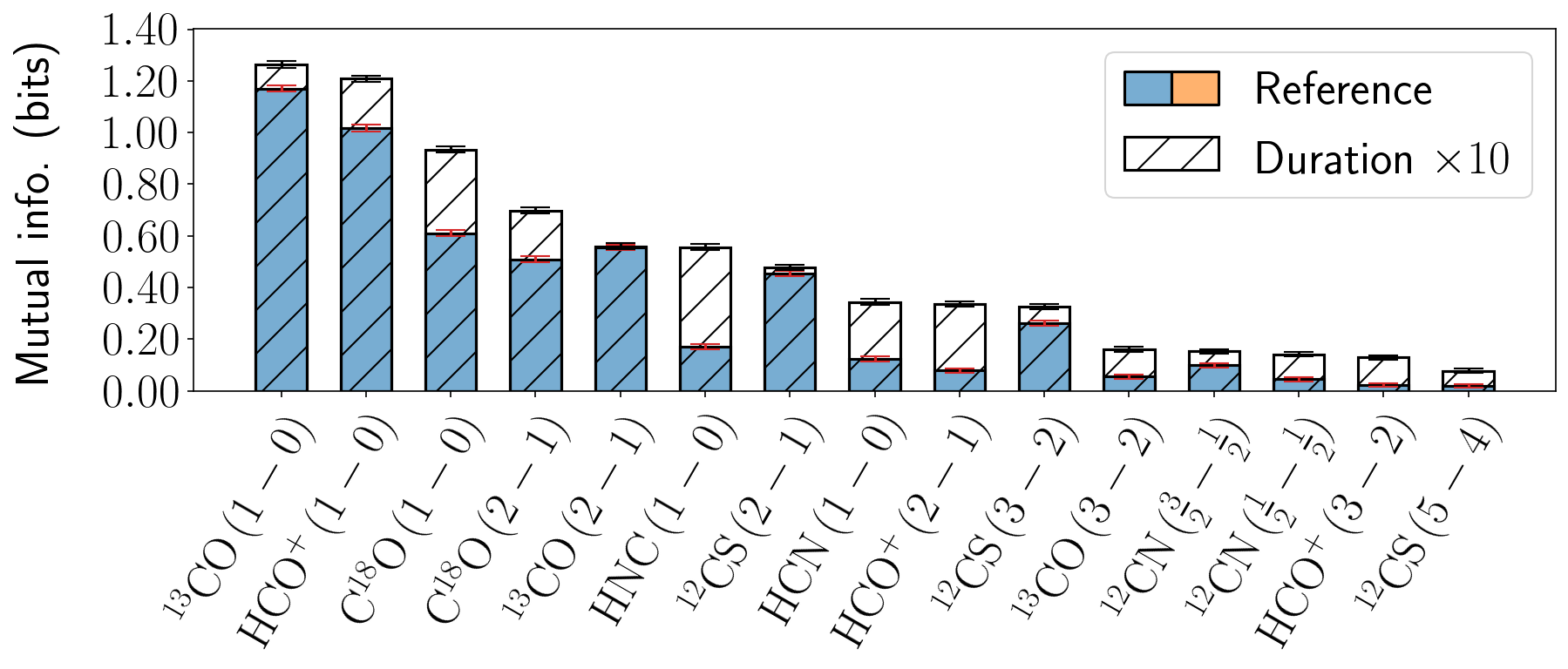}
        \includegraphics[width=\linewidth]{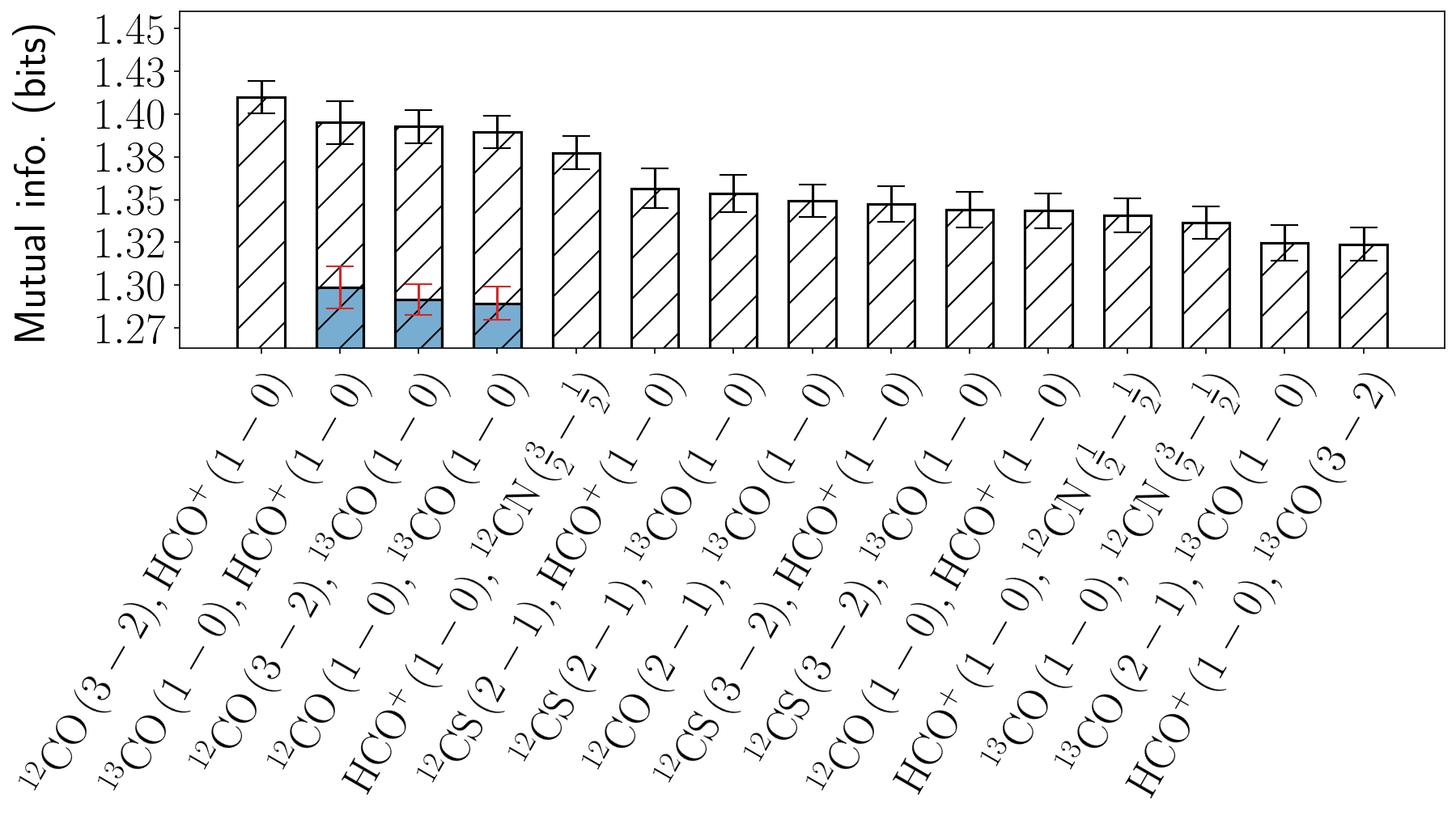}
        \caption{Results for translucent gas $(3 \leq \AV \leq 6)$.}
        \label{fig:obs_time_av_selection:3_6}
    \end{subfigure}
    \par\medskip
    \begin{subfigure}{0.46\linewidth}
        \includegraphics[width=\linewidth]{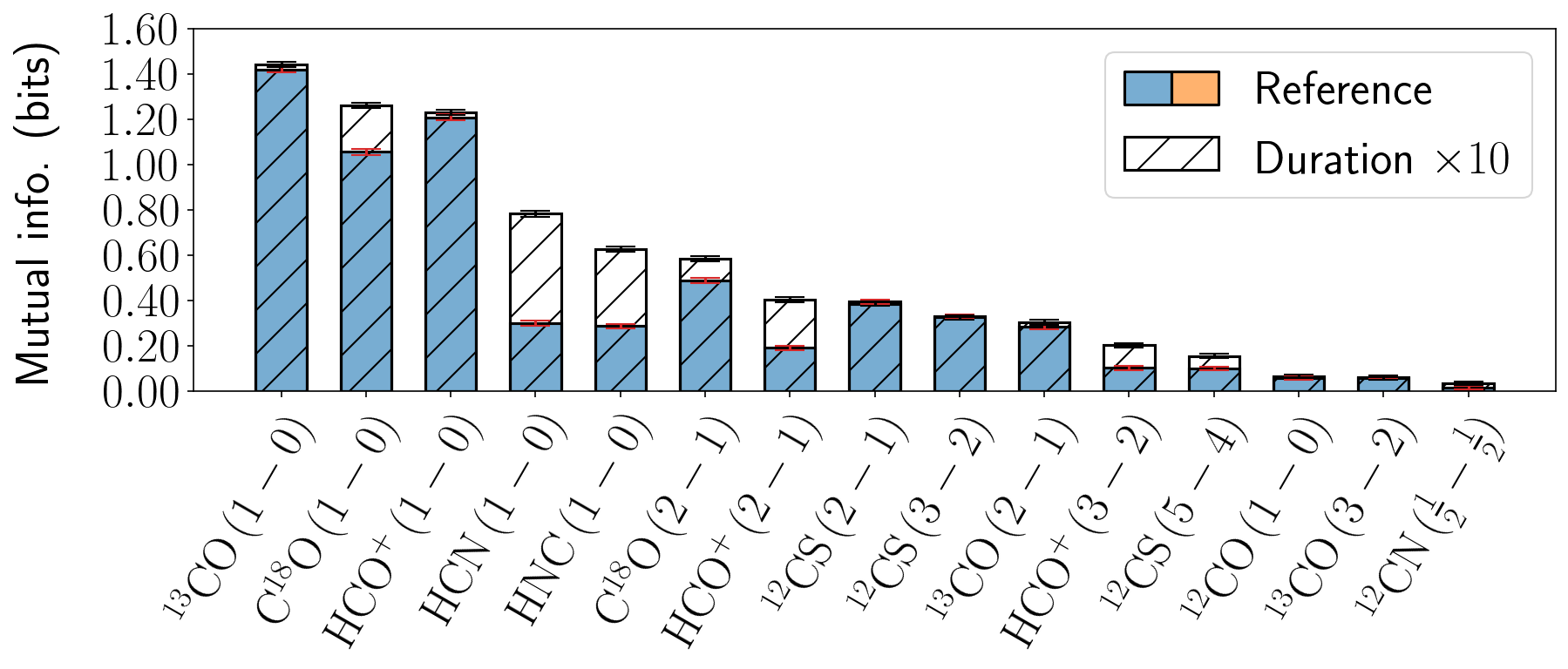}
        \includegraphics[width=\linewidth]{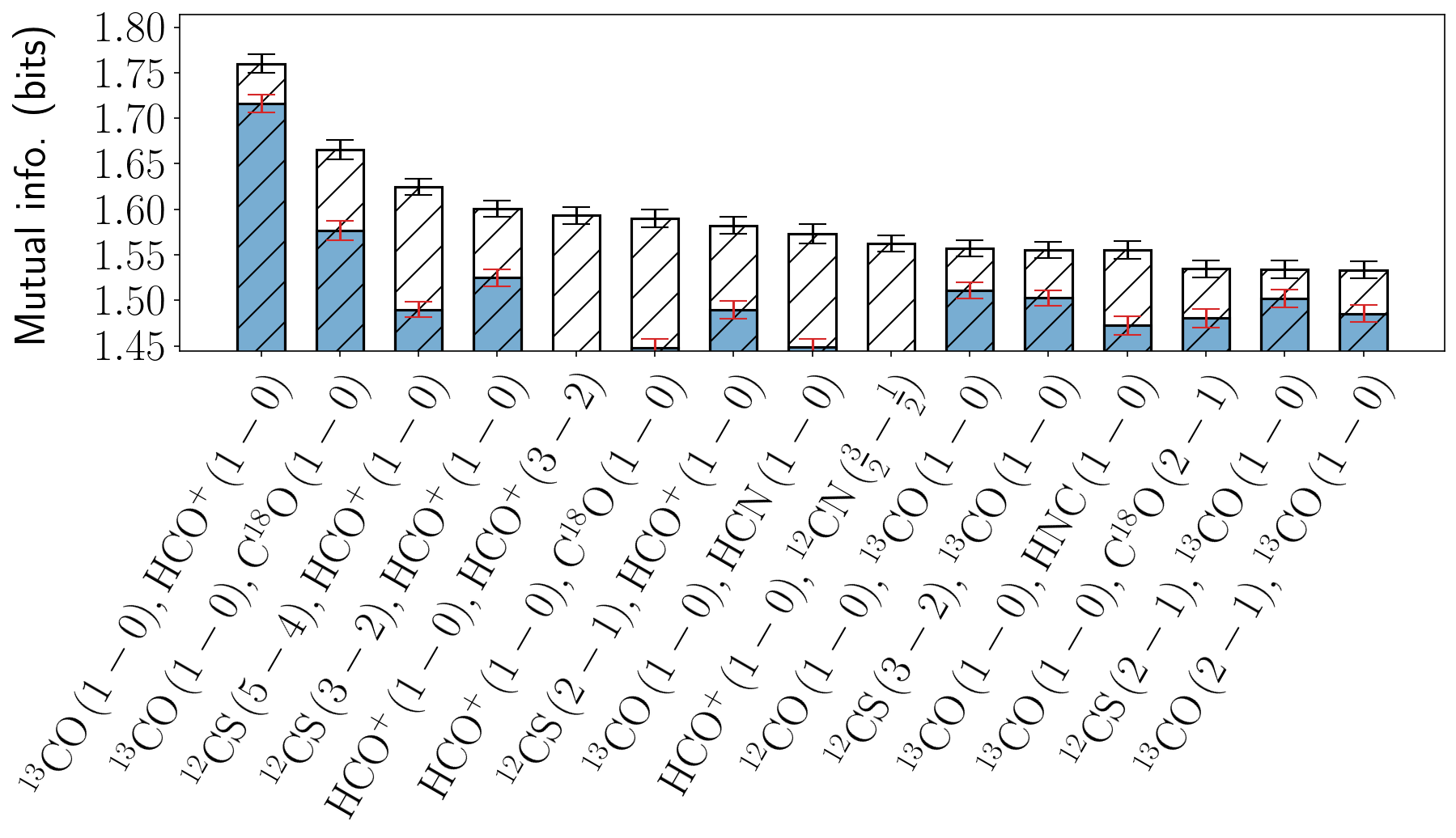}
        \caption{Results for filamentary gas $(6 \leq \AV \leq 12)$.}
        \label{fig:obs_time_av_selection:6_12}
    \end{subfigure}
    \hspace{5mm}
    \begin{subfigure}{0.46\linewidth}
        \includegraphics[width=\linewidth]{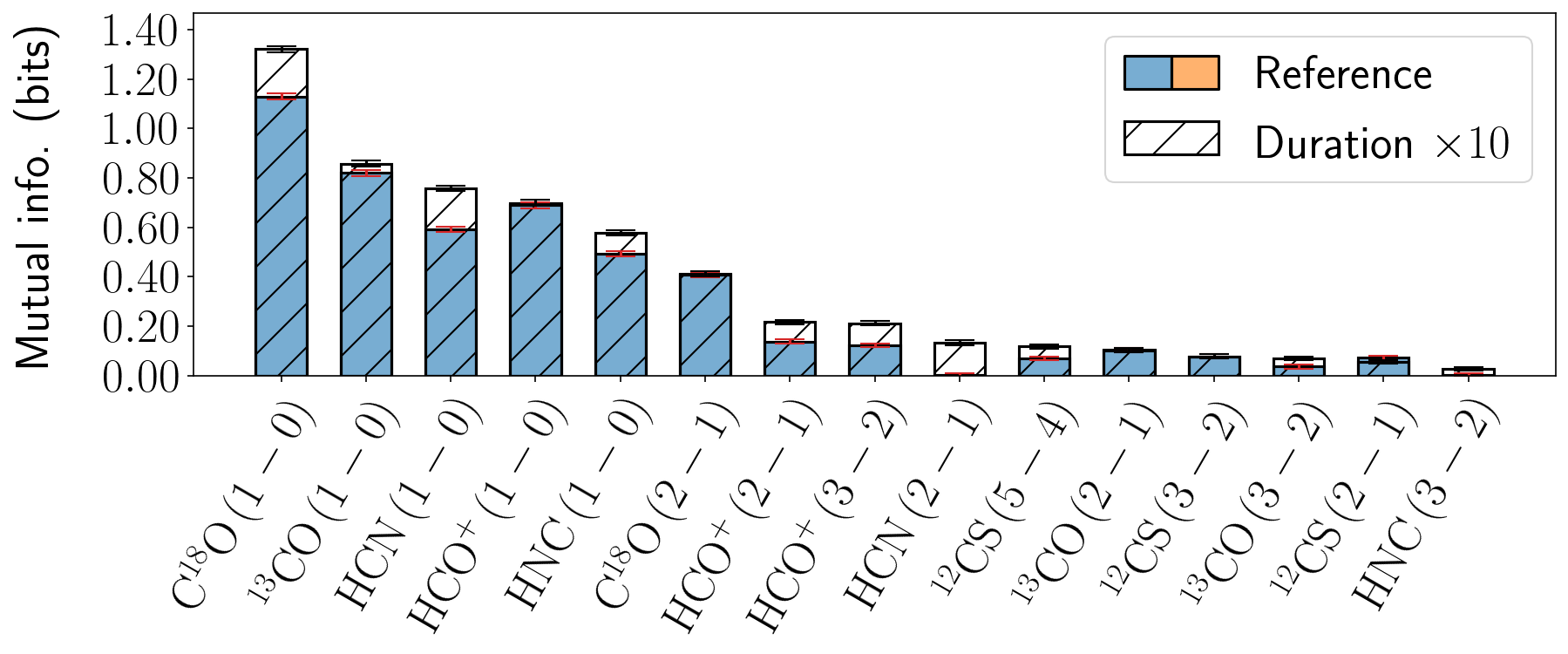}
        \includegraphics[width=\linewidth]{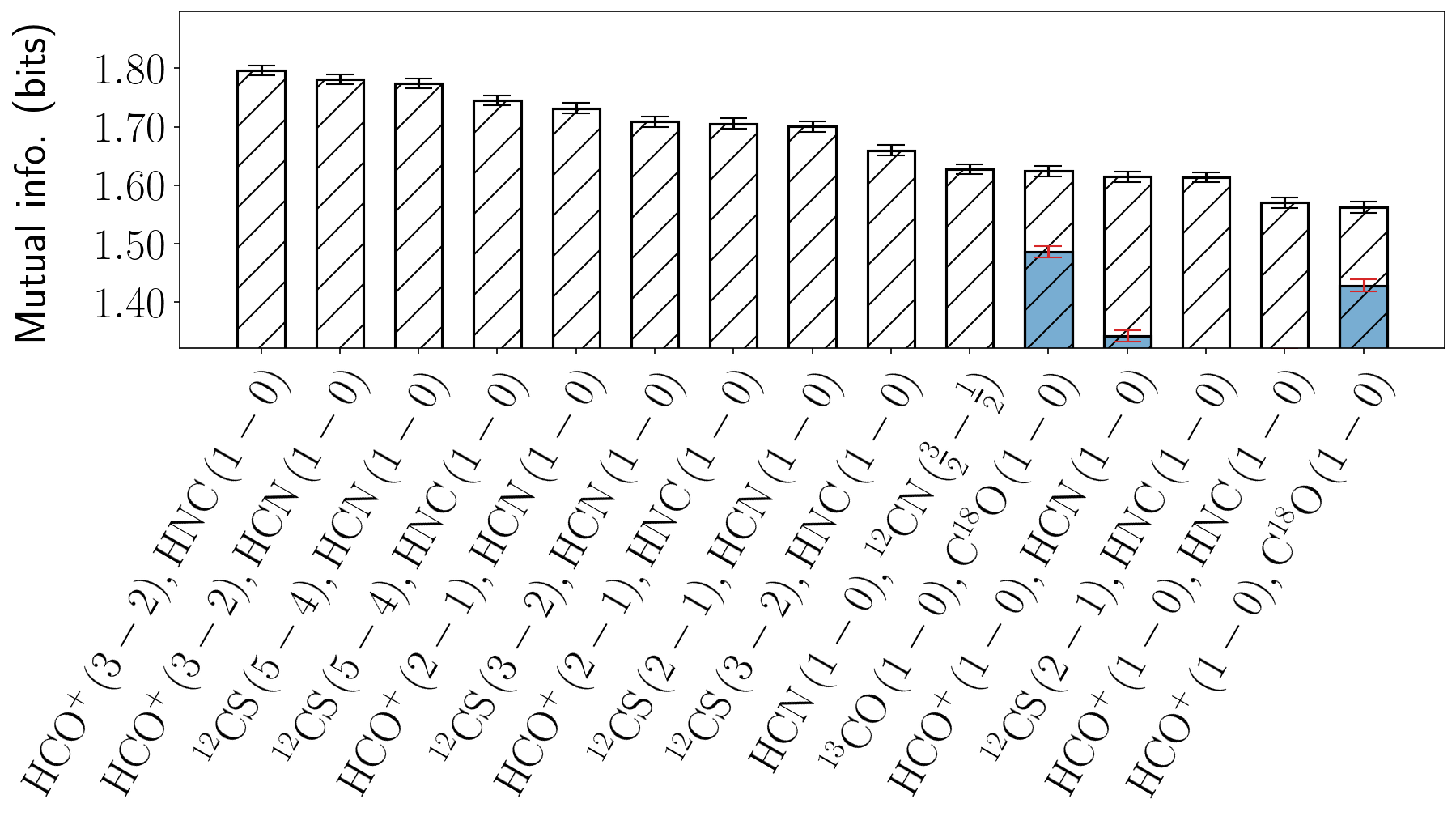}
        \caption{Results for dense cores $(12 \leq \AV \leq 24)$.}%
        \label{fig:obs_time_av_selection:12_24}
    \end{subfigure}
    \caption{
        Line selection for \AV{} for different regimes of \AV{}, in an environment similar to the Horsehead pillar, for the deeper integration use case (10 times longer observing duration and no scaling factor $\kappa$ in the observation simulator).
        The hatched bars correspond to the results obtained for the reference use case, \ie, with the reference integration time (see Fig.~\ref{fig:av_selection}).
    }%
    \label{fig:obs_time_av_selection}
\end{figure*}

We here check the impact of a 10-fold increase of the integration time (deeper integration use case) on the line ranking.
For concision, only the results for \AV{} are analyzed.

Figure~\ref{fig:obs_time_av_selection} compares the mutual information between the line intensities and \AV{} for the reference and the deeper integration use case.
As expected, the mutual information increases or saturates with the integration time.
Saturation is almost reached for the \latexmol{13co}, \latexmol{hcop}, and \latexmol{12cs} lines, when they are considered alone.
In contrast, this increase is larger for combinations of two lines than for individual
lines.
Moreover, the mutual information increase varies as a function of the line or couple of lines.

For individual lines, the S/N improvement mostly benefits the ground state transition of \latexmol{c18o}, \latexmol{hnc}, \latexmol{hcn}, as well as \latexline{hcop}{2}{1}, with an approximate 0.5\,bits increase in mutual information.
These lines all have a median S/N of about 1 in the reference case as shown in Fig.~\ref{fig:snr_emir_lines}.
Improving the S/N thus has a strong impact on their informativity.
Conversely, the ground state transition of \latexmol{13co}, \latexmol{hcop}, and \latexmol{12co}, along with \latexline{12cs}{2}{1}, only have an improvement of about 0.1\,bits.
These lines all have a median S/N of at least 10 in the reference case. Despite these differences, the three overall most informative individual lines remain the ground state transition of \latexmol{13co}, \latexmol{hcop} and \latexmol{c18o}.
At higher S/Ns, some higher energy transitions, such as those of \latexmol{hcop} and \latexmol{12cs}, provide more information than the lowest one.
This justifies the use of the 2\,mm and 1\,mm atmospheric bands.

For couples of lines, the top three most informative couples remain identical in all regimes, except in dense cores where the ranking completely changes.
Indeed, combinations involving $\latexline{hcn}{1}{0}$ or $\latexline{hnc}{1}{0}$ and $\latexline{hcop}{1}{0}$, or the $\left(\latexline{hcn}{1}{0}, \latexline{12cs}{5}{4}\right)$ couple, gain more than 0.7\,bits of mutual information and become some of the most informative couples.
This can be explained by the fact that
1)~\latexmol{hnc} and \latexmol{hcn} become more abundant in dense cores,
2)~these lines have large values of critical densities~\citep[higher than $10^6$\,\pccm, see][table~2.4]{tielens2005physics},
and 3)~the significant increase in integration time enables these lines to become informative.
Significantly increasing the integration time, and therefore the S/N, is thus useful to increase the informative potential of lines, even though they were already detected in the reference case.

\subsection{Effect of uncertain geometry on the best lines to infer \Guv{}}%
\label{subsec:scaling_factor}

\begin{figure*}[h]
    \centering
    \begin{subfigure}{0.46\linewidth}
        \includegraphics[width=\linewidth]{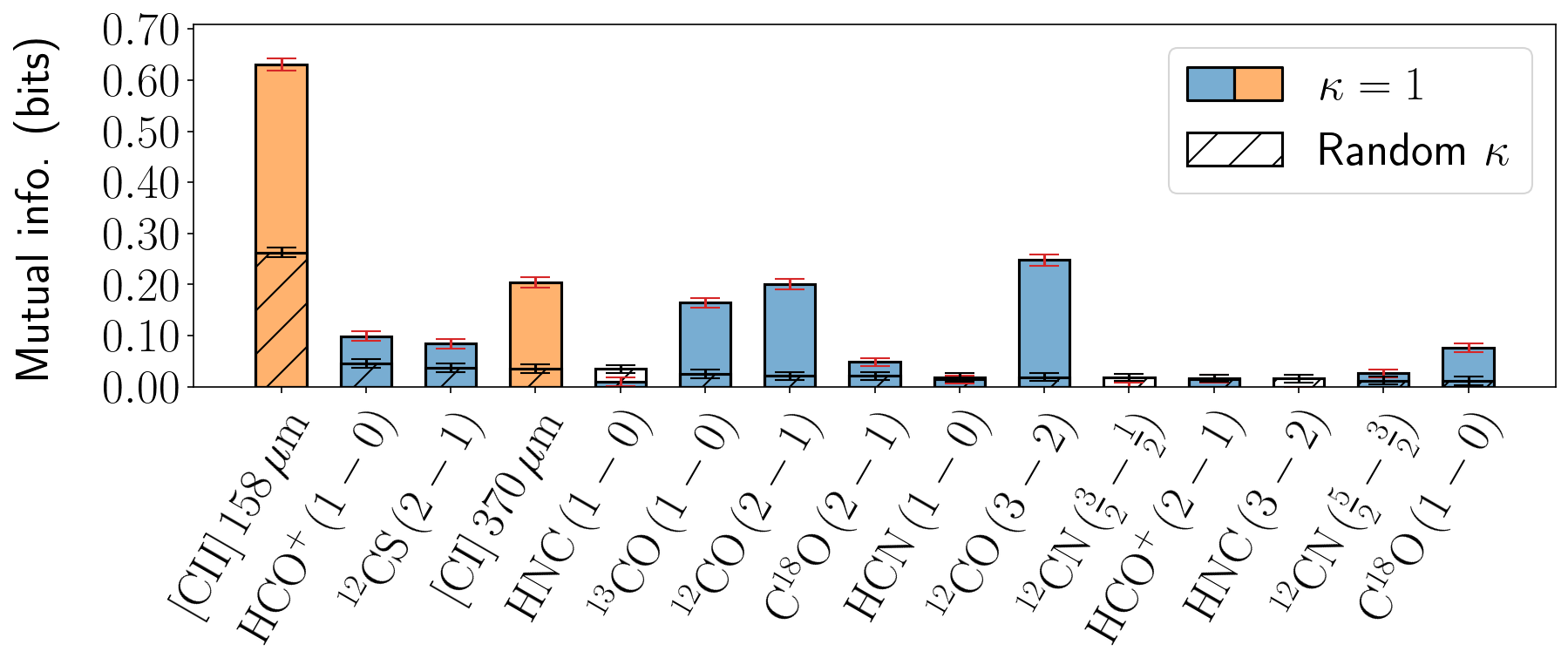}
        \includegraphics[width=\linewidth]{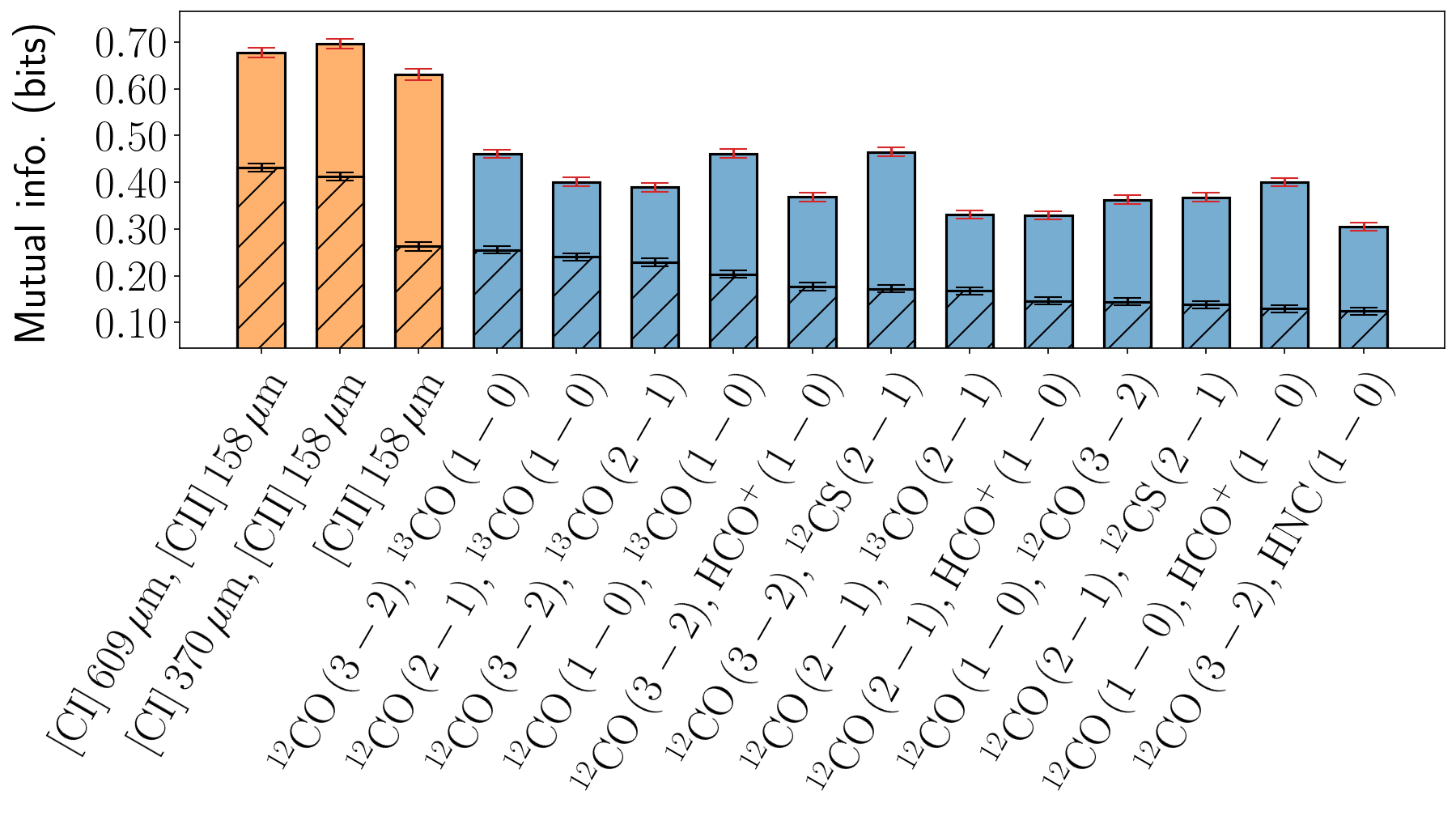}
        \caption{
            All \AV{} environments $(3 \leq \AV \leq 24)$.
        }
        \label{fig:kappa_g0_selection:all}
    \end{subfigure}
    \hspace{5mm}
    \begin{subfigure}{0.46\linewidth}
        \includegraphics[width=\linewidth]{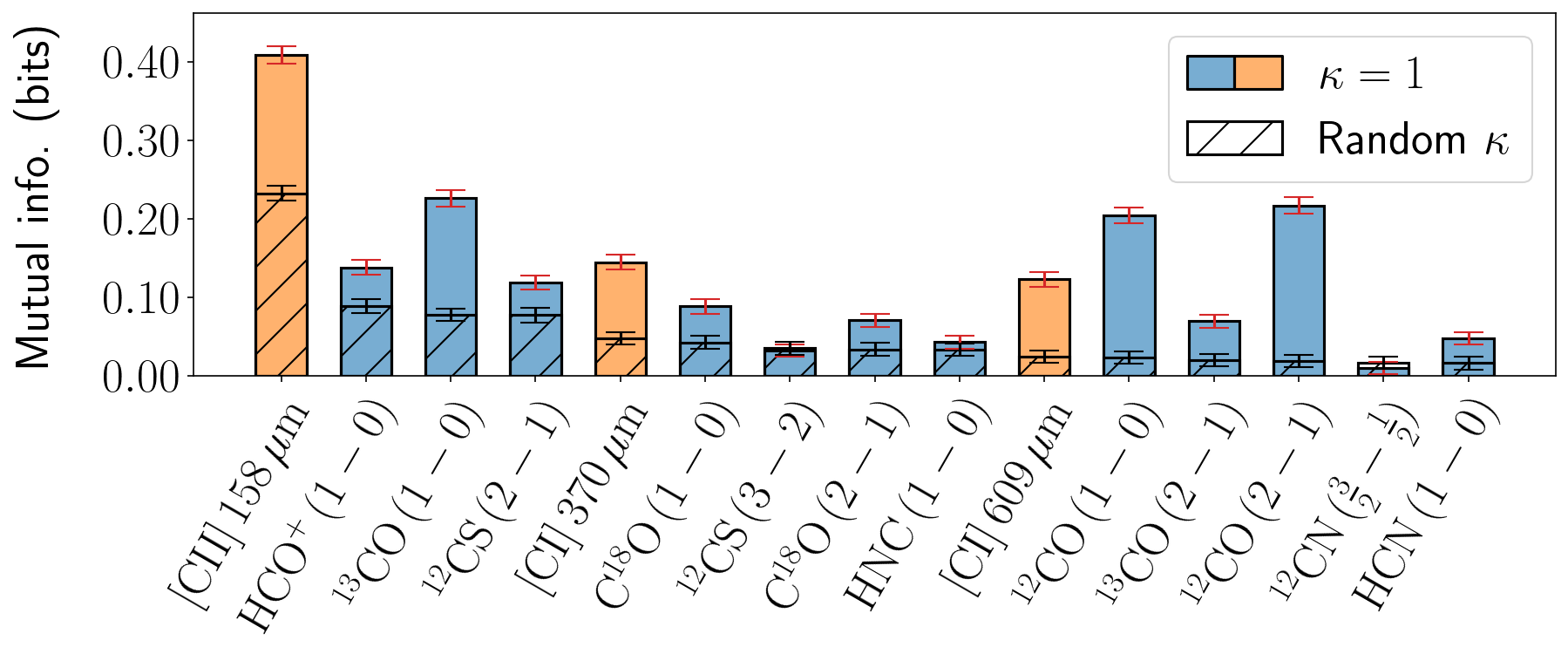}
        \includegraphics[width=\linewidth]{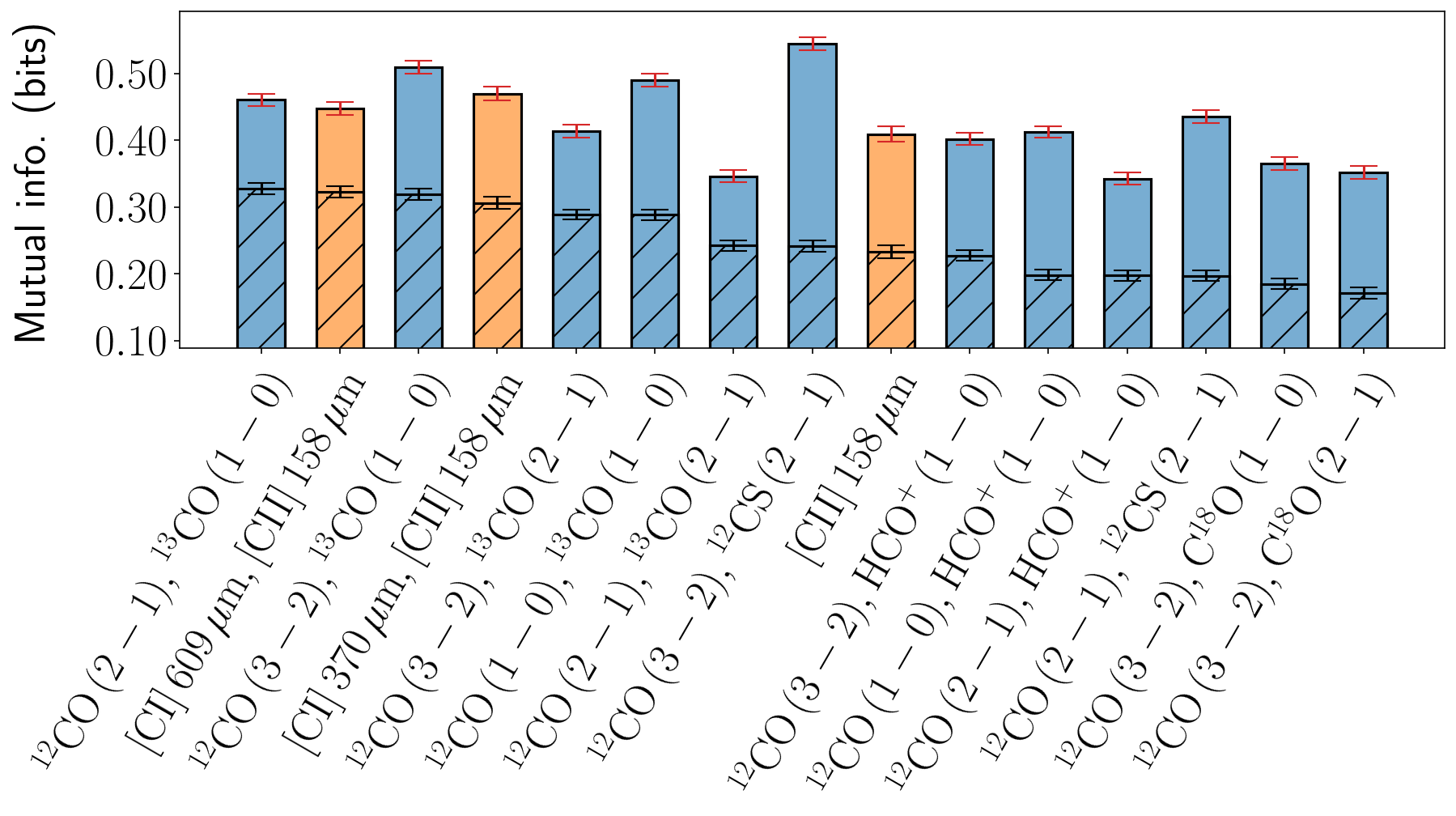}
        \caption{Results for translucent gas $(3 \leq \AV \leq 6)$.}
        \label{fig:kappa_g0_selection:3_6}
    \end{subfigure}
    \par\medskip
    \begin{subfigure}{0.46\linewidth}
        \includegraphics[width=\linewidth]{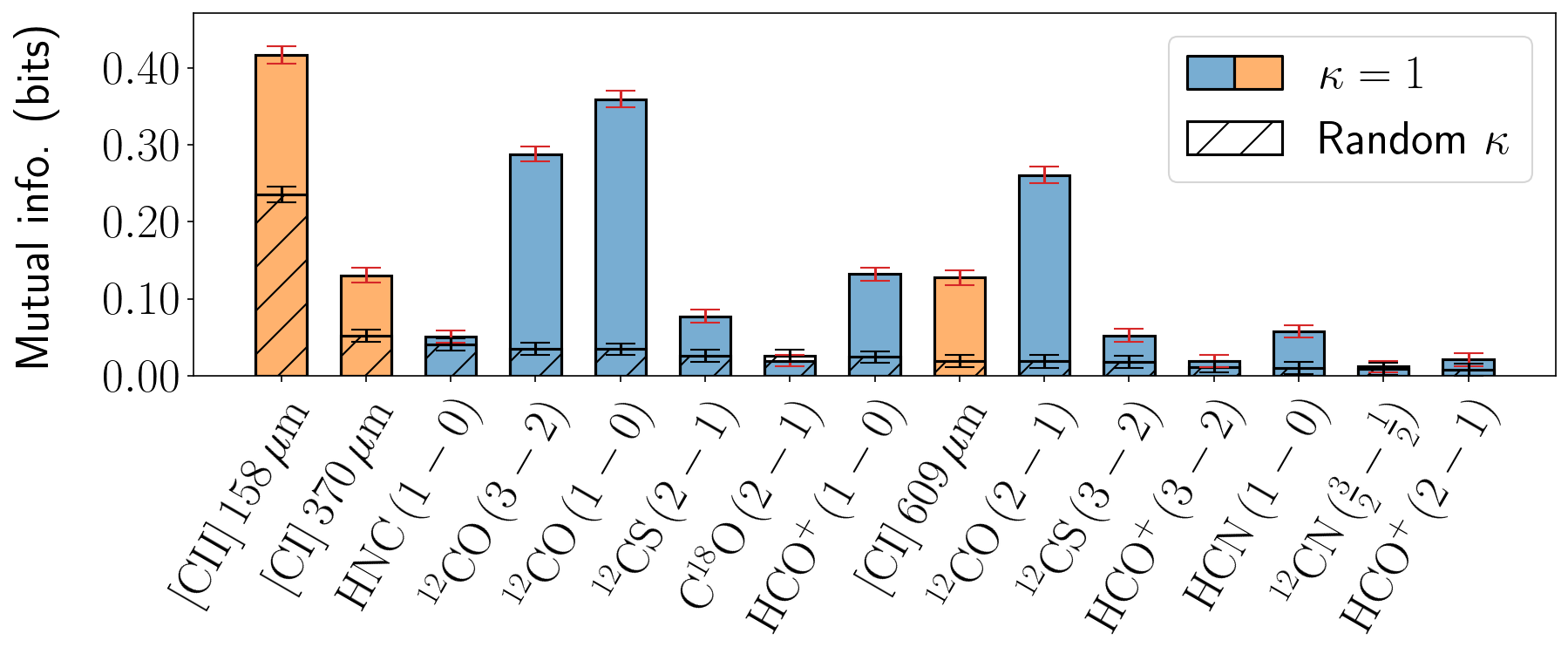}
        \includegraphics[width=\linewidth]{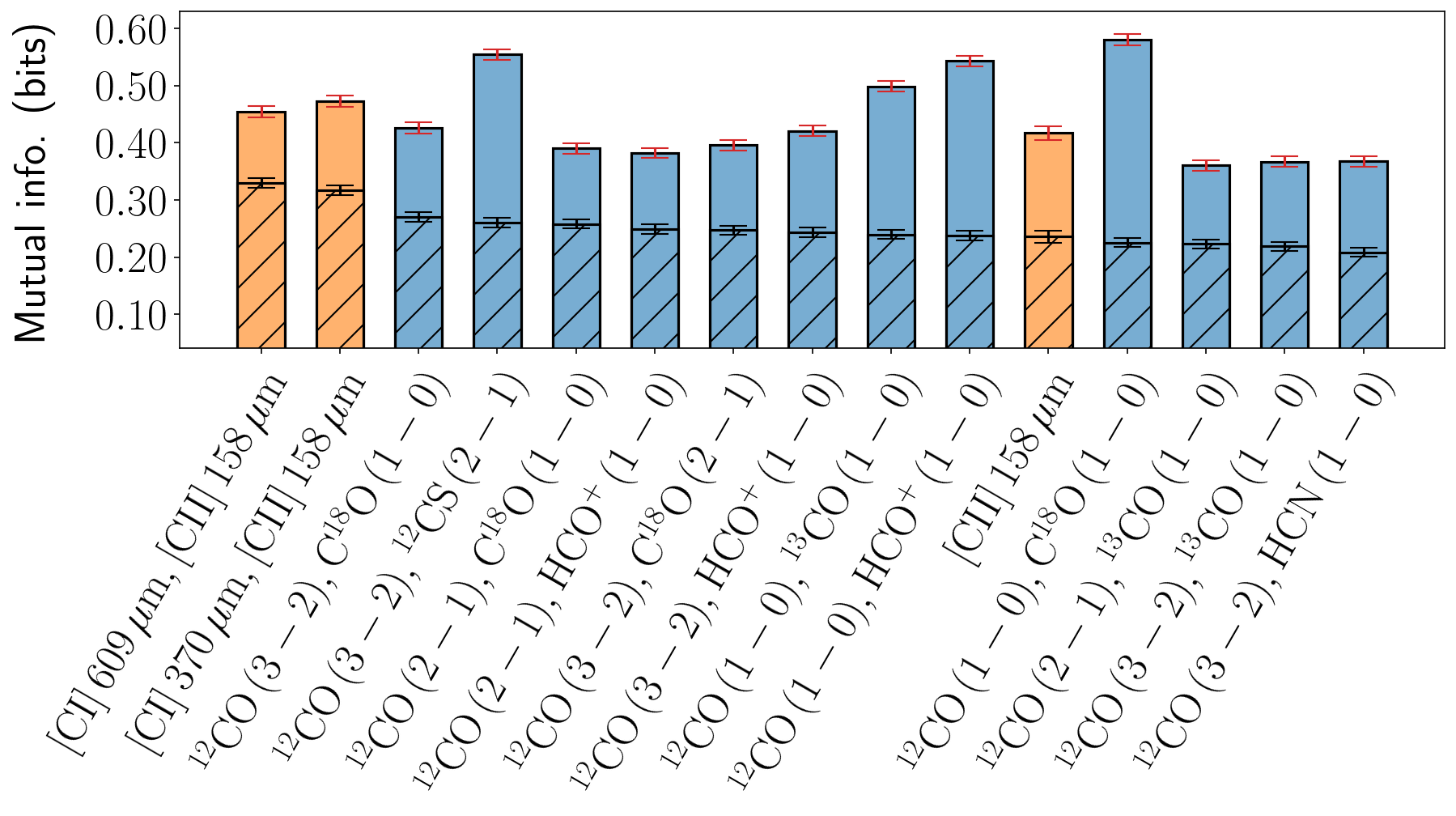}
        \caption{Results for filamentary gas $(6 \leq \AV \leq 12)$.}
        \label{fig:kappa_g0_selection:6_12}
    \end{subfigure}
    \hspace{5mm}
    \begin{subfigure}{0.46\linewidth}
        \includegraphics[width=\linewidth]{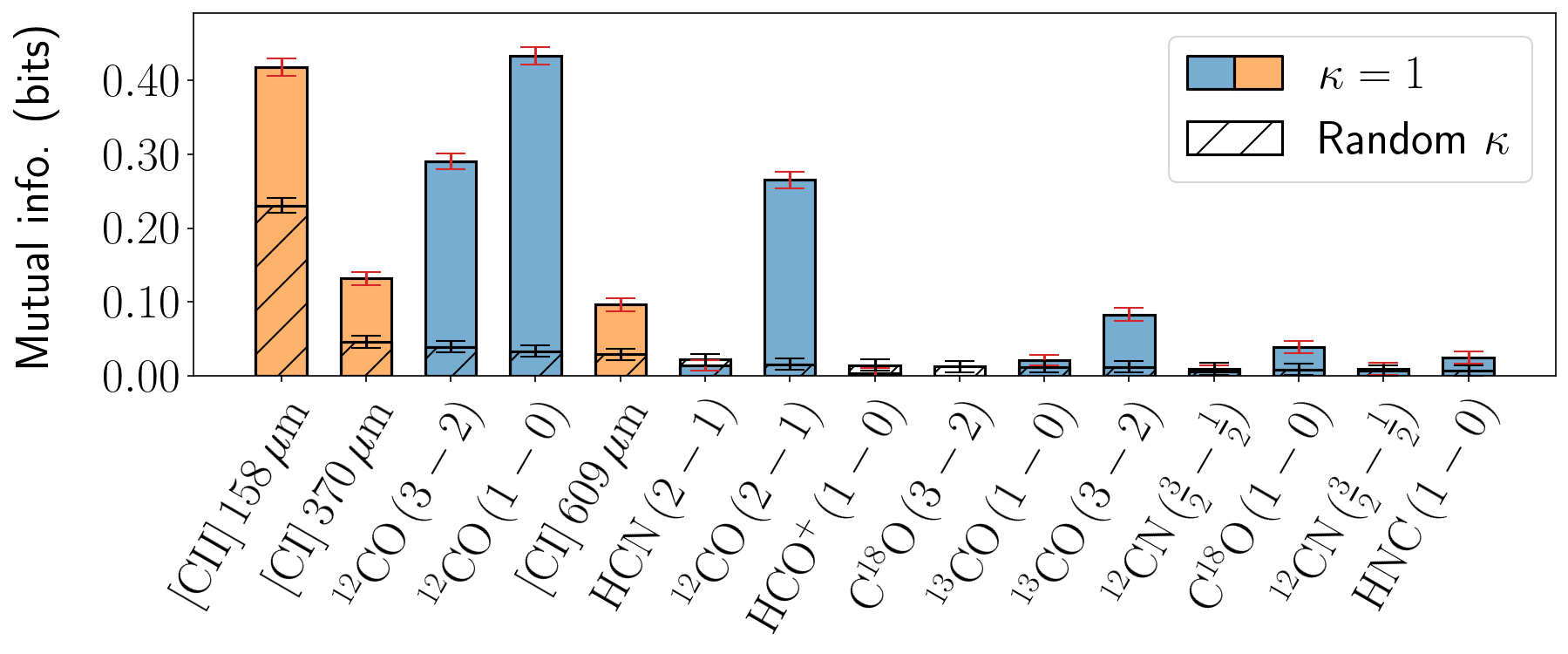}
        \includegraphics[width=\linewidth]{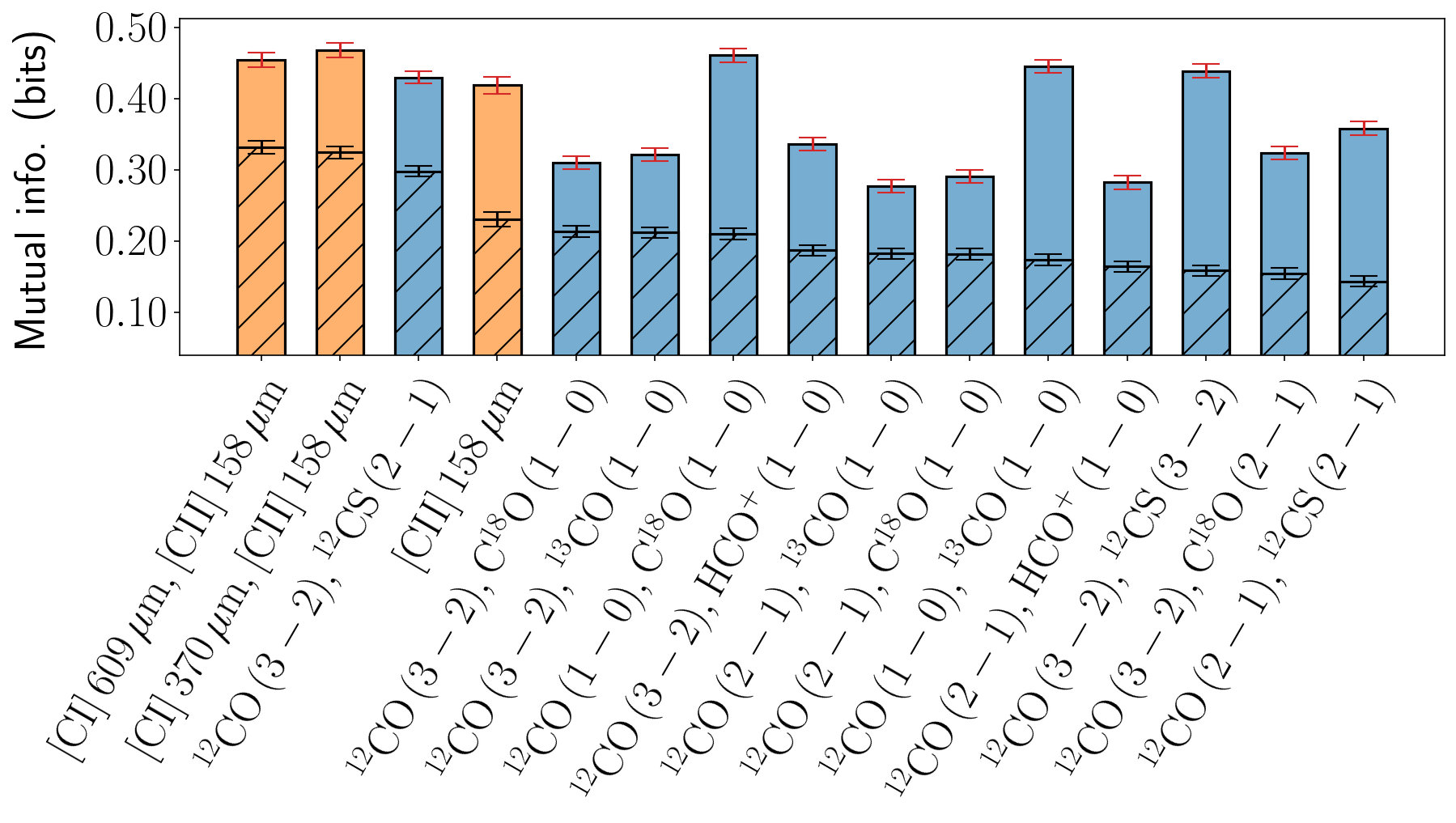}
        \caption{Results for dense cores $(12 \leq \AV \leq 24)$.}
        \label{fig:kappa_g0_selection:12_24}
    \end{subfigure}
    \caption{
        Line selection for \Guv{} for different regimes of \AV{}, in an environment similar to the Horsehead pillar, for the uncertain geometry use case (reference integration time and addition of a scaling factor $\kappa$ in the simulator of observation).
        The blue and orange bars correspond to the results obtained for the reference use case, \ie, without scaling factor (see Fig.~\ref{fig:g0_selection}).
    }%
    \label{fig:kappa_g0_selection}
\end{figure*}

The geometry in ISM clouds is uncertain.
The impact of this uncertainty is more important for physical parameters defined at the surface of the cloud, such as~\Guv{}, than for quantities integrated along the line of sight, such as the visual extinction.
We thus only consider the effect of the uncertain geometry in inferring~\Guv{}.
We simply use a scaling factor (see Eq.~\ref{eq:obs_model_instance_with_scaling}) to take into account the uncertainty about the geometry, such as beam dilution effect and cloud surface orientation.
As a reminder, $\logd\kappa$ is assumed to be uniformly distributed between -0.5 and 0.5.

Figure~\ref{fig:kappa_g0_selection} compares the mutual information between the line intensities and \Guv{} for the reference case and in this uncertain geometry use case.
It shows that the best tracers of \Guv{} remain surface tracers in all \AV{} regimes, \ie, the \cp{} line or the combination of the \cp{} and \ci{} lines.
Note that for translucent gas, the combination of the \latexmol{12co} and \latexmol{13co} molecular lines is formally ranked before the \cp{} and \ci{} lines, but this might be due to estimation error
However, this ranking might be due to estimation error, as the error bars are larger than the difference of estimated mutual information.

While nonzero, the mutual information with \Guv{} is low.
In other words, a precise estimation of \Guv{} is difficult.
It thus is all the more important to select the best tracers.
In this respect, couples of lines overall bring significantly higher information on \Guv{} than single lines.

\subsection{Using line ratios leads to a loss of information}

\begin{figure}[h]
    \centering
    \includegraphics[width=0.9\linewidth]{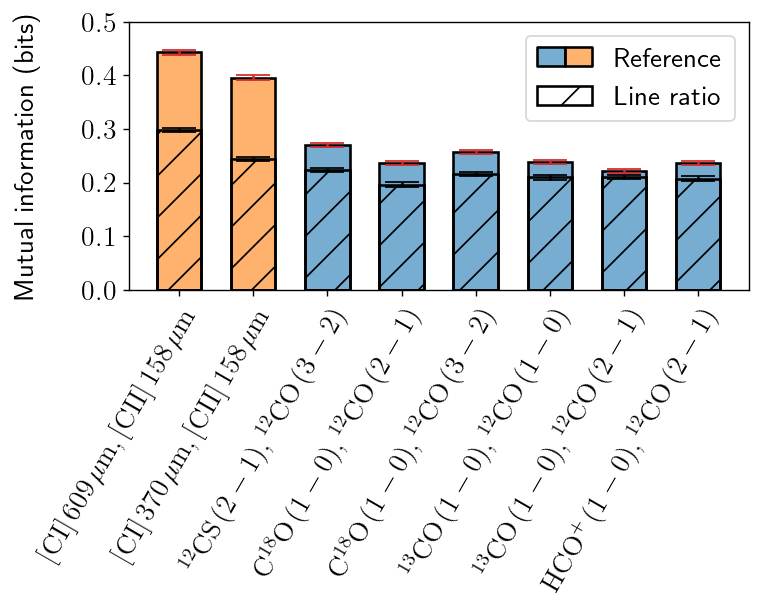}
    \caption{
        Comparison between the amount of information on \Guv{} provided by the five best couples of lines in Fig.~\ref{fig:kappa_g0_selection} (colored bars) and their line ratio (hatched bars).
    }
    \label{fig:line_ratio_mi}
\end{figure}

Using line intensity ratios in the analysis of spectral data of interstellar clouds is common in ISM studies to eliminate observational uncertainties such as the dependency with the cloud geometry -- see, for example,~\citep{cormier2015herschel,kaplan2021near}.
Assuming that the geometry effects impact the line intensities in similar ways, this allows observers to get rid of the scaling factor $\kappa$ from~\eqref{eq:obs_model_instance_with_scaling} for a high enough S/N.
Besides, as line ratios reduce the dimensionality from two or more to one, they allow for simpler visualizations and thus a simpler understanding of ISM properties~\citep{kaufmanFarInfraredSubmillimeter1999}.
Similarly to line selection, assessing the relevance of a large set of line ratios to select the best ones was already done before.
For instance,~\citet{bronTracersIonizationFraction2021} uses random forests to select the line ratio that best traces the ionization fraction.
As mentioned in Sect.~\ref{subsec:obs_model_and_proba}, the line selection method presented in Sect.~\ref{sec:optimization_pb} can be applied with line ratios.
This section illustrates an important specificity of line ratios to have in mind when evaluating their physical relevance. %, and to select those that bring as much information as a joint analysis of the raw line intensities.

Figure~\ref{fig:line_ratio_mi} compares the mutual information between \Guv{} and either a couple of lines or their line ratios.
We perform this comparison for the five most informative line couples for filamentary gas (\mbox{$6 \leq \AV{} \leq 12$}) and with a random scaling factor $\kappa$.
In all cases, the mutual information with the line couple is larger than with the line ratio.

This observation can be explained theoretically.
Computing a line ratio goes from two dimensions or more (the integrated intensities of the two or more lines) to only one (the ratio) and is thus not a bijective operation.
As stated in Sect.~\ref{sec:mi}, a non-bijective transformation results in a loss of information.
However, this loss of information differs from one line couple to another.
On the figure, two classes of line combinations appear.

For couples of [CI] and [CII] lines, the joint analysis yields much larger mutual information values than analyzing the associated line ratio, \ie, using a line intensity ratio instead of the two lines intensities results in a large loss of information.
The [CII] line being a cooling line emitted from the cloud surface, its intensity contains a lot of information on \Guv{}, which is partially lost when using a ratio.
For molecular line combinations, here combining a low-J \latexmol{12co} line and another millimeter line, this loss of information is much smaller, almost negligible.
%
% Since low-J \latexmol{12co} lines quickly become optically thick, they mostly permit
%
In this specific example, studying how \Guv{} depends on a line ratio instead of the original line couple is both simpler and equivalent in terms of informativity.

More generally, line ratios can be valuable tools to inspect ISM properties.
For a given set of lines, noise characteristics and physical regime, mutual information can permit observers to identify lines ratios that are most informative on a physical parameter or combination of physical parameters.
However, working with line ratios instead of the original set of lines can lead to a significant loss of information.
Therefore, for tasks that seek to exploit as much information as possible from a costly dataset such as inference, considering the original set of lines should be more relevant than line ratios.

%%%%%%%%%%%%%%%%%%%%%%%%%%%%%%%%%%%%%%%%%%%%%%%%%%%%%%%%%%%%%%%%%%%%%%%%%%%

\section{Conclusion}%
\label{sec:conclu}

In this work, we showed how information theory concepts such as mutual information~\citep[sect. 8.6]{coverElementsInformationTheory2006} can be used to evaluate quantitatively capability of line observations to constrain physical parameters such the visual extinction $\AV{}$ or the FUV illumination field $\Guv{}$.
Such a quantitative criterion opens a new perspective to visualize and understand the statistical relationships between physical parameters and tracers.
In particular, mutual information relies on few and nonrestrictive assumptions on the considered probability distributions.
Therefore, conclusions drawn from it only depend on the underlying physics and the noise properties of the observations.
In addition, mutual information can also be used to determine the best lines to observe in a future observation campaign given an instrument specifications, and to recommend a target integration time.
To illustrate the potential of the proposed method, we applied it to lines observable with the EMIR instrument at the IRAM~30m radio telescope for physical regimes similar to those found in the Horsehead Nebula.
The results for this case are as follows.
\begin{itemize}
    \item The determination of the optimal combination of lines to estimate a physical parameter depends heavily on the achieved S/N and thus on the integration time for single-dish telescopes.
    For instance, the \latexmol{hcn} and \latexline{hnc}{1}{0} lines achieve their full potential as dense cores tracers only for a S/N $\ga 20$.
    \item The line intensity has to vary significantly as a function of the physical parameters to get a high precision during the inference.
    This implies that the capability of a line to infer, \eg, the visual extinction, depends on the physical regime.
    For instance, the best lines in the Horsehead Nebula -- for an integration time similar to that of the \mbox{ORION-B} Large Program -- are \latexmol{13co} and \latexline{hcop}{1}{0} for translucent gas $(3 \leq \AV{} \leq 6)$, \latexmol{13co}, \latexmol{hcop} and \latexline{c18o}{1}{0} for filamentary gas $(6 \leq \AV{} \leq 12)$, and \latexmol{13co} and \latexline{c18o}{1}{0} for dense cores $(12 \leq \AV{} \leq 24)$.
    \item The low-$J$ lines of CO isotopologues are key tracers of the gas column density for a wide range of the (\AV{}, \Guv{}) space.
    \item Surface tracers such as the \cp{} line, the \ci{} lines, or \latexmol{12co} lines are the most useful tracers of \Guv{}.
    However, \Guv{} is much more difficult to estimate than \AV{}.
    \item The best combination does not always combine the best individual lines.
    Considering the combination of the $K \geq 2$ best individual lines as  the best subset of $K$ lines may thus lead to a suboptimal choice.
\end{itemize}
The proposed methods are general enough to be applicable to any ISM model or even observational dataset.
The latter application will be the subject of the second paper in this set.
The Python software that implements the general method we proposed is available in open access\footnote{\url{https://github.com/einigl/infovar}}.
The simulator of line observations based on Meudon PDR code predictions is also available\footnote{\url{https://github.com/einigl/iram-30m-emir-obs-info}}.
It allows us to simulate observations from the EMIR receiver at the IRAM~30m radio telescope, but can be adapted to any other instruments (including those operating in other frequency ranges).
In this case, the user would only need to specify the noise and calibration properties.
Finally, the scripts that reproduce the exact results presented in this paper are available in another repository\footnote{\url{https://github.com/einigl/informative-obs-paper}}.

To simplify the presentation and interpretations, this work focused on constraining physical parameters individually.
For an observation campaign, the lines to be observed will be used to constrain multiple physical parameters -- such as visual extinction $\AV{}$ and the intensity of the UV radiative field $\Guv{}$ at once.
In this case, mutual information should be used to search for the line combinations that best constrain the combination of these physical parameters.
In particular, this method has the potential to indicate that combinations of physical parameters may be constrained by a given set of lines, even though each individual parameter is not constrained by the same set of lines.

Finally, this work focused line integrated intensities as these are the quantities predicted by the considered ISM model, the Meudon PDR code.
However, the proposed approach could be applied with any observable.
For instance, radio telescopes yield full line profiles.
Since integrating these line profiles is a non-bijective transformation, considering the integrated intensity instead of the line profile results in a loss of information.
Future work could quantify this loss exploiting mutual information.

%%%%%%%%%%%%%%%%%%%%%%%%%%%%%%%%%%%%%%%%%%%%%%%%%%%%%%%%%%%%%%%%%%%%%%%%%%%

\begin{acknowledgements}
    % ANR DAOISM
    This work received support from the French Agence Nationale de la Recherche through the DAOISM grant ANR-21-CE31-0010,
    % PCMI
    and from the Programme National ``Physique et Chimie du Milieu Interstellaire'' (PCMI) of CNRS/INSU with INC/INP, co-funded by CEA and CNES.
    % ANR MIAI
    It also received support through the ANR grant ``MIAI $@$ Grenoble Alpes'' ANR-19-P3IA-0003.
    % 80Prime
    This work was partly supported by the CNRS through 80Prime project OrionStat, a MITI interdisciplinary program,
    % ANR Sherlock
    by the ANR project ``Chaire IA Sherlock'' ANR-20-CHIA-0031-01 held by P. Chainais,
    % ANR ULNE
    and by the national support within the {\em programme d'investissements d'avenir} ANR-16-IDEX-0004 ULNE and Région HDF.
    % Javier & Miriam
    JRG and MGSM thank the Spanish MCINN for funding support under grant PID2019-106110G-100.
    % Miriam
    MSGM acknowledges support from the NSF under grant CAREER 2142300.
    % Darek
    Part of the research was carried out at the Jet Propulsion Laboratory, California Institute of Technology, under a contract with the National Aeronautics and Space Administration (80NM0018D0004). D.C.L. acknowledges financial support from the National Aeronautics and Space Administration (NASA) Astrophysics Data Analysis Program (ADAP).
    Finally, we thank Tommaso Grassi for his constructive comments and feedback that helped us improve this article.
\end{acknowledgements}

% -------------------------------------------------------------------

\bibliographystyle{aa} %
\bibliography{main.bib} %

\begin{appendix}

%%%%%%%%%%%%%%%%%%%%%%%%%%%%%%%%%%%%%%%%%%%%%%%%%%%%%%%%%%%%%%%%%%%%%%%%%%%

\section{Details on the two-dimensional illustrative example}%
\label{subsec:example_detailed}

Section~\ref{subsec:example} introduces a joint distribution on $(\paramfull, \obsfull)$ that follows a two-dimensional lognormal distribution.
Its parameters, $\boldmu$ and $\boldsymbol{\Sigma}$, correspond to the mean vector and covariance matrix in the logarithmic scale, respectively.
They are set to obtain expectations of 1, a standard deviation such that a $1\sigma$ error corresponds to a factor 1.3, and a $\rho = 0.9$ correlation coefficient in linear scale.
One can show that the associated distribution parameters are
\begin{align}
    \boldmu & = - \frac{1}{2}\begin{pmatrix}
        (\ln 1.3)^2 \\
        (\ln 1.3)^2
    \end{pmatrix}
    \simeq
    \begin{pmatrix}
        -0.0344 \\
        -0.0344
    \end{pmatrix}
    \; \text{ and } \; \\
    \boldsymbol{\Sigma} & = \begin{pmatrix}
        (\ln 1.3)^2 \quad \quad \ln \left[ 1 + 0.9 \left(e^{(\ln 1.3)^2} - 1 \right) \right] \\
        \ln \left[ 1 + 0.9 \left(e^{(\ln 1.3)^2} - 1 \right) \right] \quad  \quad(\ln 1.3)^2
    \end{pmatrix}
    \simeq \begin{pmatrix}
        0.0688 & 0.0622 \\
        0.0622 & 0.0688
    \end{pmatrix}
    \nonumber
    .
\end{align}
In this simple case, one can show that $\paramfull \,\vert\, \obsfull \sim \lognormal(\mu,\,\sigma^2)$ with
\begin{align}
    \label{eq:params_lognormal_conditional}
    \begin{dcases}
        \mu = -\frac{1}{2} \boldsymbol{\Sigma}_{1,1}
        + \frac{\boldsymbol{\Sigma}_{1,2}}{\boldsymbol{\Sigma}_{1,1}}
        \left(\ln y + \frac{1}{2} \boldsymbol{\Sigma}_{1,1}\right) \\
        \sigma^2 = \boldsymbol{\Sigma}_{1,1}
        - \frac{\boldsymbol{\Sigma}_{1,2}^2}{\boldsymbol{\Sigma}_{1,1}}
    \end{dcases}
\end{align}

%%%%%%%%%%%%%%%%%%%%%%%%%%%%%%%%%%%%%%%%%%%%%%%%%%%%%%%%%%%%%%%%%%%%%%%%%%%

\section{What is a bit of information?}%
\label{sec:what_is_a_bit}

\begin{figure}
    \centering
    \includegraphics[width=0.48\textwidth]{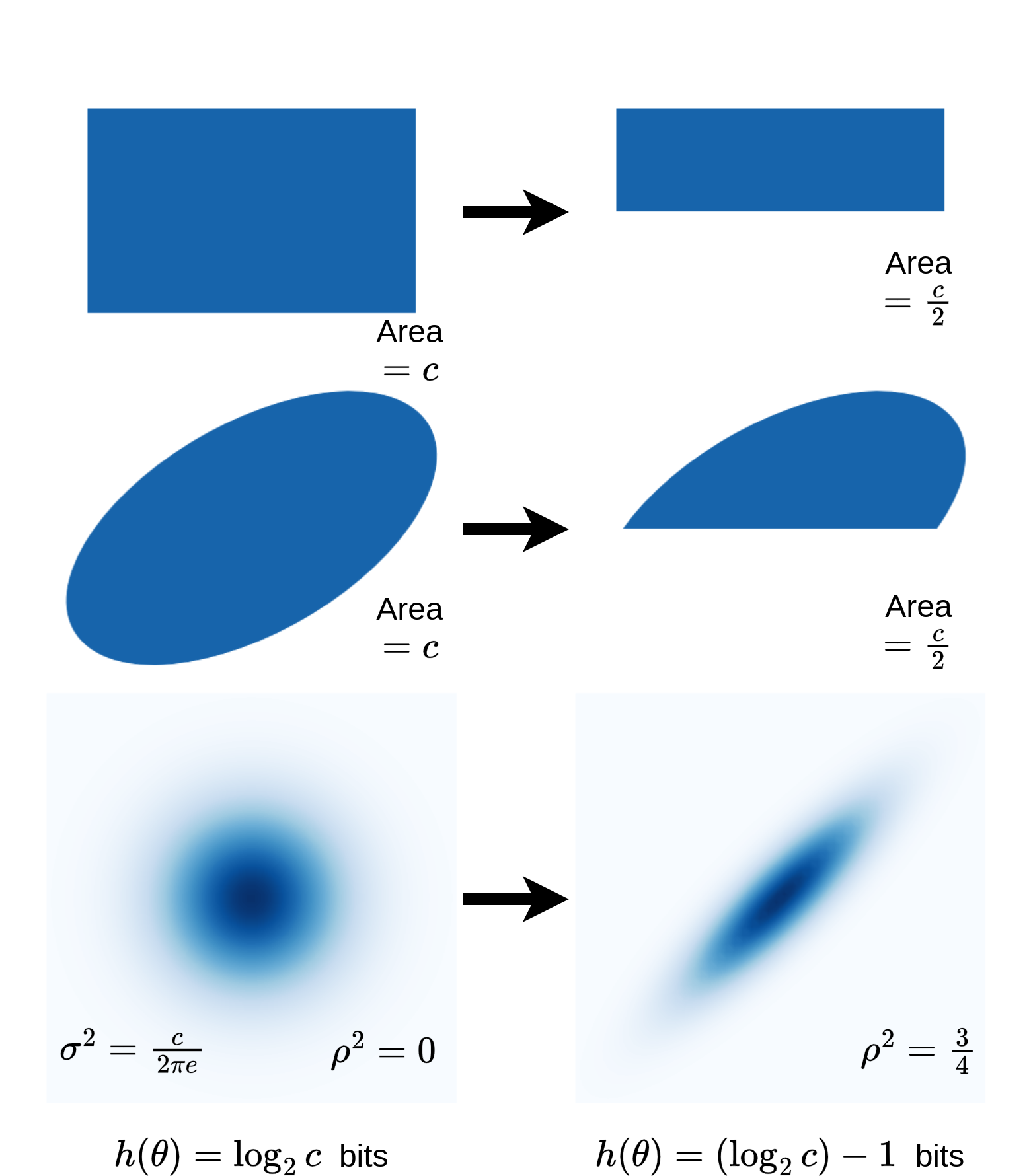}
    \caption{
        Entropy definition illustration for different example distributions on $\paramvect{}$.
        The first two rows show the PDF of uniform distributions on different sets, and the last row of Gaussian distributions.
        The distributions in one column have an equal differential entropy $\h{\paramfull}$ whose value depends on a positive constant $c$.
        Each arrow indicates a gain of 1\,bit of information, \ie, a decrease in the entropy of 1\,bit.
        In the last row, the variance in both horizontal and vertical directions is denoted $\sigma^2$, and the correlation coefficient $\rho$.
    }%
    \label{fig:entropy-def-illustration}
\end{figure}

Figure~\ref{fig:entropy-def-illustration} illustrates six probability distributions on a fictitious two-dimensional physical parameter: four uniform distributions on compact sets and two Gaussian distributions.
The two compact sets on the top left have the same area, denoted $c$.
By construction, the three distributions on the left share the same differential entropy, namely $\h{\paramfull} = \log_2 c$\,bits.
For the two-dimensional normal distribution,
\begin{align*}
    \h{\paramfull} =
    \log_2 \left[ 2 \pi e \sigma^2 \right]
    + \frac{1}{2} \log_2 \left(1 - \rho^2 \right)
    \quad \text{for }
    \Sigma =
    \begin{pmatrix}
        \sigma^2 & \rho \sigma^2 \\
        \rho \sigma^2 & \sigma^2
    \end{pmatrix}
    .
\end{align*}
Using $\sigma^2 = c / (2 \pi e)$, the first term in the sum simplifies to \mbox{$\log_2 c$\,bits}.
The correlation coefficient is $\rho^2 = 0$, so that \mbox{$\frac{1}{2} \log_2 (1 - \rho^2 ) = 0$\,bit} and \mbox{$\h{\paramfull} = \log_2 c$\,bits}.
The three distributions on the right are transformed versions of the left column.
Each transformation results in a decrease in the entropy of 1\,bit.
Indeed, the two compact sets on the top right have a $c/2$ area and the entropy of the associated uniform distributions is thus $\h{\paramfull} = \log_2 (c/2) = \log_2 c - 1$\,bits.
For the two-dimensional Gaussian, the correlation coefficient is $\rho^2 = 3/4$ in the right column, leading to $\frac{1}{2} \log_2 (1 - \rho^2 ) = -1$\,bit.
Note that the use of the binary base to express values in bits provides a simple interpretation when comparing entropy values: a difference of 1 bit of information corresponds to a factor 2 of standard deviation.
Therefore, in an estimation procedure, decreasing the entropy by 1 bit results in improving the estimation precision by a factor 2.

%%%%%%%%%%%%%%%%%%%%%%%%%%%%%%%%%%%

\section{Estimating the mutual information}%
\label{subsec:estimation}

Several Monte Carlo estimators $\miEst{\paramfull}{\obsfullsubset{}}$ of mutual information exist -- see~\citet{walters2009estimation} for a review.
In this section, we compare two such estimators: the nonparametric ``Kraskov estimator''~\citep{kraskovEstimatingMutualInformation2004} (used in this work), and an estimator based on the assumption that the joint PDF of $\left(\paramfull, \obsfullsubset\right)$ is Gaussian.

The Kraskov estimator is based on nearest-neighbors -- see Appendix~\ref{app:knn} for more details on this approach.
It is notably used by the \textsc{SciPy} Python package%
\footnote{\url{https://scipy.org/}}.
It does not make assumptions on the shape of the joint distribution on $\left(\paramfull, \obsfullsubset\right)$.
It can thus capture both linear and nonlinear relationships between lines $\obsfullsubset$ and physical parameters $\paramfull$.
It is asymptotically unbiased, \ie, it converges to the exact mutual information in the large number of observations limit $N \to \infty$.
To reduce the bias that can occur at small $N$, we apply the Gaussian reparametrization strategy from~\citet{holmes2019estimation}, which bijectively transforms each marginal distribution to a Gaussian.
Appendix~\ref{app:bias} provides more details on this bias reduction technique.

Under the assumption that the joint PDF of $\left(\paramfull, \obsfullsubset\right)$ is Gaussian, the mutual information is simply a function of the canonical correlations~\citep[CC,][]{schreierUnifyingDiscussionCorrelation2008}.
Since canonical correlation can be estimated based on second order empirical moment, our second mutual information estimator is obtained by injecting the estimated canonical correlation coefficient in the analytical entropy formula for a Gaussian distribution after application of the Gaussian reparametrization strategy~\citep{holmes2019estimation}.
The ``CC estimator'' has shorter computation time than the Kraskov estimator, because it only requires evaluations of second order moments.
However, as imposing the Gaussianity of marginal is generally not sufficient to match the multivariate Gaussian assumption, the ``CC estimator'' is only asymptotically unbiased in the general case.
Appendix~\ref{app:cca} provides more details on this estimator.

For both estimators, the variance evolution with different sample sizes $N$ allows us to assess their accuracy and to estimate error bars.
To do this, we follow a method introduced in~\citet{holmes2019estimation}, and summarized in Appendix~\ref{app:variance}.

\begin{figure}[t]
    \centering
    \includegraphics[width=\linewidth]{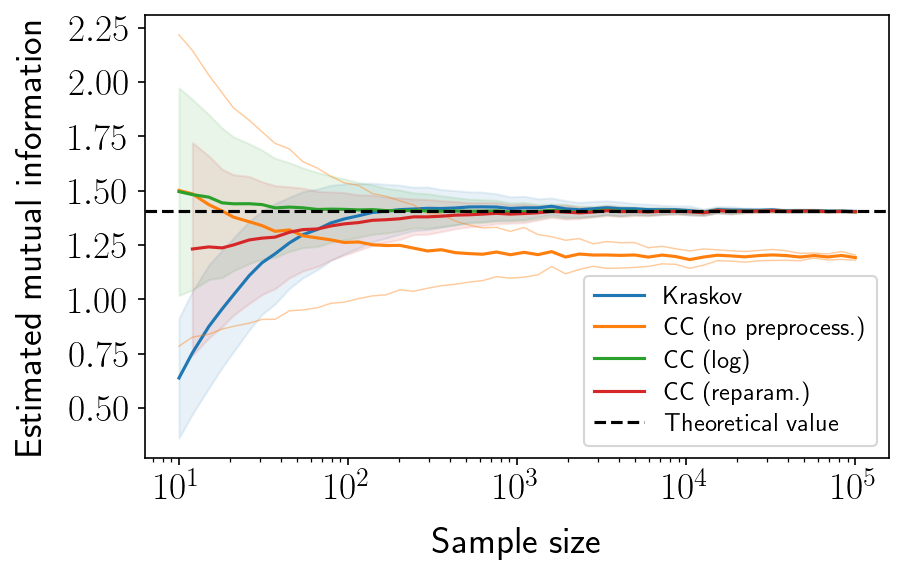}
    \caption{
        Comparison of four mutual information estimators applied on the simple lognormal bivariate distribution presented in Sect.~\ref{subsec:obs_model_and_proba}.
        The black dashed line corresponds to the theoretical value.
        The 1$\sigma$ interval is not shown for the simple canonical correlation-based estimator as it appears to be asymptotically biased.
        The Kraskov estimator (in blue) converges to the correct value for a large number of simulated observations~$N$.
        The CC estimator is used with three preprocessing: no preprocessing, log and Gaussian reparametrization.
        As the joint distribution is non-Gaussian, the no-preprocessing case does not converge to the theoretical value.
        However, the other two preprocessing transform it to a Gaussian, the associated CC estimators converge to the theoretical value for large values of~$N$.
        %
        %\PC{reparam = on se perd un peu en technique}
    }%
    \label{fig:mi-estimators-lognormal}
\end{figure}

Figure~\ref{fig:mi-estimators-lognormal} quantitatively shows the behavior of both estimators as a function of the number of  $N$ for the bivariate lognormal case introduced in Sect.~\ref{subsec:example}.
The Kraskov estimator is biased for a low number of observations $N$ but is very close to the theoretical value for $N \geq 10^3$.
The canonical estimator is combined with three different transformations of the marginal distributions of $\obsfullsubset$ and $\paramfull$:
1)~no preprocessing,
2)~taking the logarithm of the random variables,
and 3)~the Gaussian reparametrization described above.
In the no preprocessing case, the CC estimator does not converge to the true value, because the samples are log-normally distributed instead of being normally distributed as required by the estimator.
For instance, for \mbox{$N=10^3$}, the mean error on the estimation in the no preprocessing case is about twice its standard deviation, while it is 3 and 5 times lower than its standard deviation for the Kraskov estimator, and the CC estimator with Gaussian reparameterization, respectively.

Astrophysical models produce complex and nonlinear relationships between lines $\obsfullsubset$ and physical parameters $\paramfull$.
The previous discussion shows that the canonical estimator is potentially useful when the sample size is small.
Applying a marginal Gaussian reparametrization is a simple solution to reduce the bias, even though this transformation does not always yield normal joint distributions on $\left(\paramfull,\obsfullsubset \right)$.
Using this strategy, the Kraskov estimator seems to give adequate results for $N \geq 10^4$, and does not require any Gaussianity assumption.

In the remainder of this work, we use the Kraskov estimator to evaluate the mutual information.
This estimator is evaluated with the \textsc{NPEET} Python package\footnote{\url{https://github.com/gregversteeg/NPEET}}.
This package handles many-to-many relationships, \ie, it permits the evaluation of the mutual information between combinations of lines and combinations of physical parameters.
Conversely, as of today, the more common implementation from \textsc{SciPy} only handles one-to-one relationships.

%%%%%%%%%%%%%%%%%

\section{Nearest neighbors-based estimators}%
\label{app:knn}

\subsection{Naive estimation of entropy}

Calculating entropy involves estimating the variable's PDF.
Traditionally, this is done using a histogram~\citep{beirlant1997nonparametric}.
However, this approach creates widely skewed PDF, leading to biases in the entropy estimator.
Moreover, this approach suffers in high dimensions~\citep{miller1955note}, due to the so-called curse of dimensionality~\citep{kouiroukidis2011effects}.

\begin{figure}
    \centering
    \includegraphics[width=0.7\linewidth]{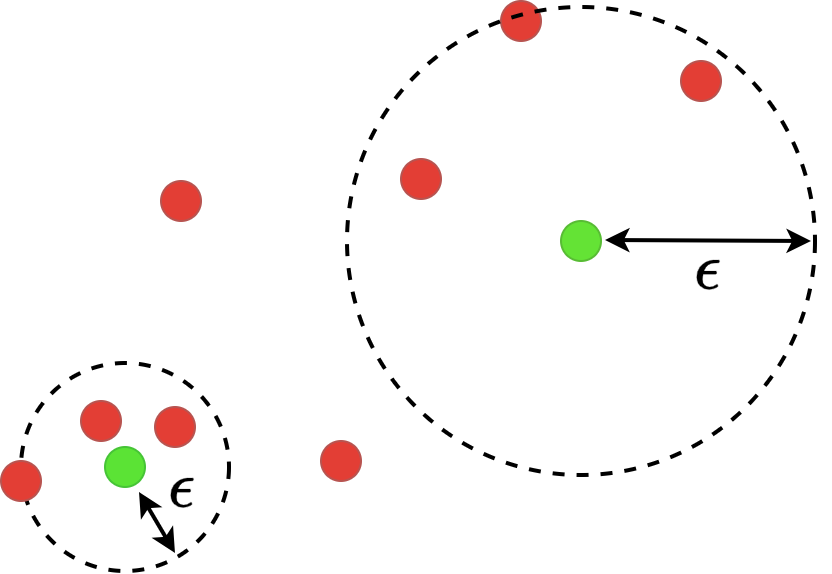}
    \caption{Illustration of the $k$-NN estimators for $k=3$. For each green point, the distance to its third nearest-neighbor, denoted $\epsilon_d$, is represented. A low distance implies a locally high density.}
    \label{fig:knn}
\end{figure}

A popular alternative is to estimate the PDF using the nearest-neighbors method~\citep{beirlant1997nonparametric}.
Indeed, intuitively, if the $k$-th nearest neighbor of a point is close to it, then the PDF of the random variable in its neighborhood is high (see Fig.~\ref{fig:knn}).
The PDF of the variable $X$ in the neighborhood of $X_i$ is then approximated by the expression
\begin{equation}
    \widehat{\pi}_X(x_i) = \frac{k/N}{\mathcal{V}_d\left(\epsilon_k^{(i)}\right)}
    \label{eq:knn-density}
\end{equation}
where $N$ is the total number of samples, $\epsilon_k^{(i)}$ the distance from $x_i$ to its $k$-th nearest neighbor and $\mathcal{V}_d(r)$ the volume of a ball of radius $r$ in $\R^d$.
This then allows the entropy to be estimated by the following Monte Carlo estimator,
\begin{equation}
    \widehat{h}(X) = -\frac{1}{N} \sum_{i=1}^N \log \widehat{\pi}_X(x_i).
    \label{eq:mc-entropy}
\end{equation}

Combining equations 1 and 2, and using the fact that the volume of a $d$-ball of radius $r$ is $\mathcal{V}_d(r) = r^d\,V_d$, where $V_d$ is the volume of the unit $d$-ball, \ie, the $d$-ball of radius 1, we obtain the following expression of the estimator,
\begin{equation}
    \widehat{h}(X) = \log N - \log k + \log V_d + \frac{d}{N}\sum_{i=1}^N \log \epsilon_k^{(i)}.
    \label{eq:knn-entropy}
\end{equation}

\subsection{Kozachenko-Leonenko estimator of entropy}

\begin{figure}
    \centering
    \includegraphics[width=0.9\linewidth]{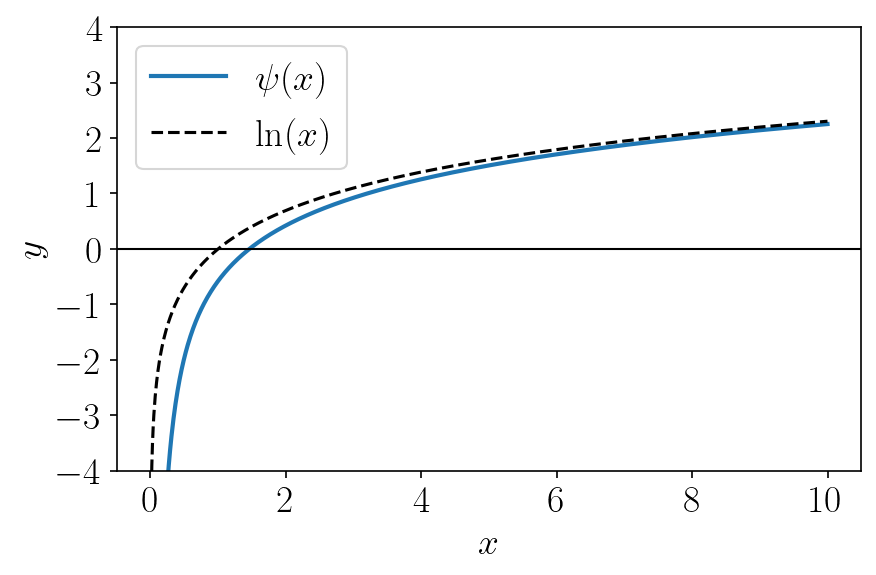}
    \caption{Graph of the digamma function $\psi$ on $\R_+^*$, and comparison with the natural logarithm. The digamma function is equivalent to the latter for $x\to\infty$.}
    \label{fig:digamma}
\end{figure}

The previous estimator is prone to high bias, especially when the number of neighbors $k$ or the number of samples $N$ are small.
To address this issue, \citet{kozachenko1987sample} proposed the following estimator,
\begin{equation}
    h_{\text{KL}}(X) = \psi(N) -\psi(k) + \log V_d + \frac{d}{N}\sum_{i=1}^N \log \epsilon_k^{(i)}
    ,
    \label{eq:kozachenko}
\end{equation}
where $\psi$ is the digamma function.
The digamma function behaves similarly to the logarithm for high values.
On the other hand, it differs for small values (see Fig.~\ref{fig:digamma}).

The digamma function acts as a correction term and ensures that this estimator remains asymptotically unbiased, which is only the case for the naive one if $k$ and $N$ are high.
More details about how the digamma function appears in the Kozachenko-Leonenko estimator are provided in~\citet{kraskovEstimatingMutualInformation2004}.

\subsection{Kraskov estimator of mutual information}

\begin{figure}
    \centering
    \includegraphics[width=\linewidth]{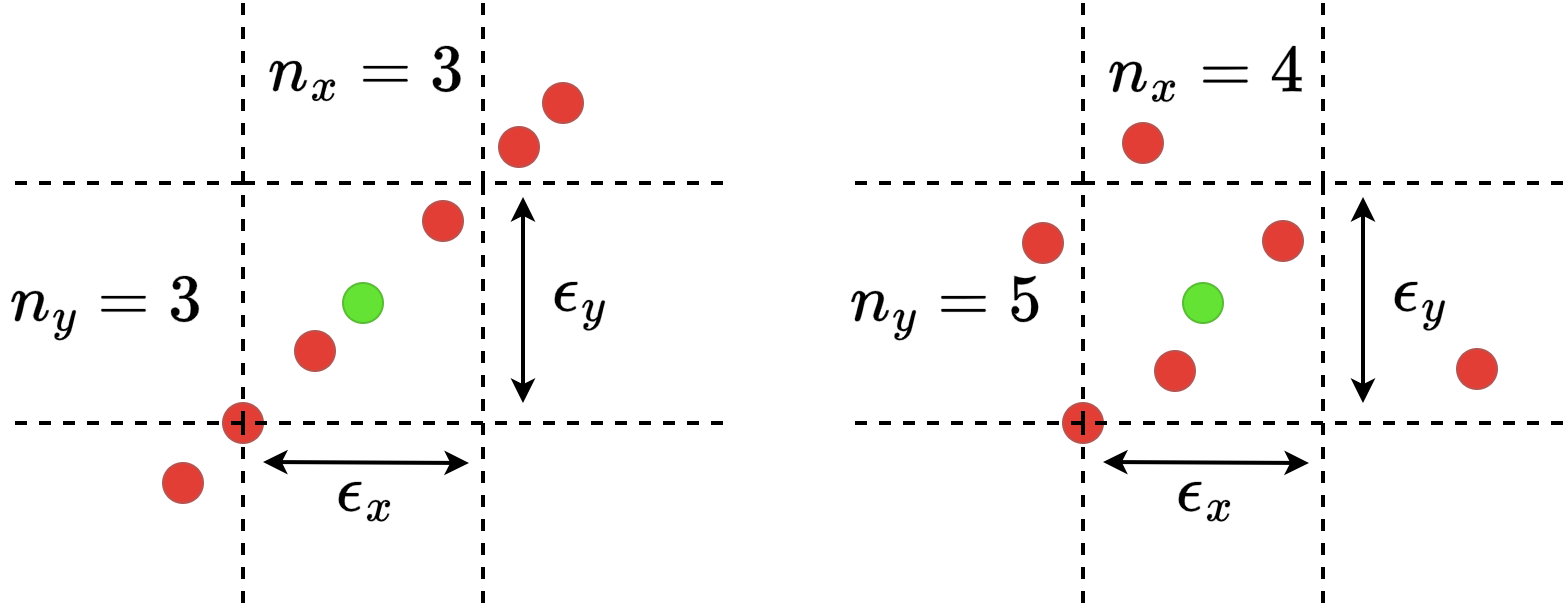}
    \caption{
        Illustration of the $k$-NN estimator of mutual information for $k=3$.
        In the first panel, the mutual information is high so $n_x$ and $n_y$ are close to $k$.
        In the second panel, the mutual information is low so $n_x$ and $n_y$ are much higher than $k$.
    }
    \label{fig:knn-mi}
\end{figure}

A naive mutual information estimator could be based directly on the Kozachenko-Leonenko estimator from the relationship \mbox{$\mi{X}{Y} = \h{X} + \h{Y} - \h{X, Y}$}.
\citet{kraskovEstimatingMutualInformation2004} argued that this solution would be highly biased.
Instead, they proposed the following estimator,
\begin{equation}
    I_{\text{KSG}}(X, Y) = \psi(k) - 1/k - \left\langle \psi(n_x(k)) + \psi(n_y(k))\right\rangle + \psi (N ),
\end{equation}
where $n_x(k)$ is the number of points $j$ such that $\left\lVert x_j - x_i \right\rVert \leq \epsilon_x^{(i)}/2$, $n_y(k)$ is the number of points $j$ such that $\left\lVert y_j - y_i \right\rVert \leq \epsilon_y^{(i)}/2$ and $\left\langle\cdot\right\rangle$ denotes the average value over all points $i$.
This approach to calculating mutual information is illustrated in Fig.~\ref{fig:knn-mi}.

Nearest-neighbors entropy estimates, and in a lower extent Kraskov's estimate of mutual information, are sensitive to duplicates in the data.
In fact, it means that $\epsilon_k^{(i)}=0$ for at least one $i$, which leads to an infinitely negative entropy.
This result is not absurd: it is the theoretical value we would expect to obtain for a distribution containing one or more diracs.
If duplicates are not handled properly, for example by adding noise or reparameterizing, they can lead to a significant bias in estimates.

%%%%%%%%%%%%%%%%%%%%%%%%%%%%%%%%%%%

\section{Bias and variance of the estimator}
\label{app:bias-variance}

\subsection{Bias of the estimator}
\label{app:bias}

The bias of an estimator quantifies the systematic error in the estimation, that is the difference between the true value and the average estimated value over many datasets drawn from the same distribution.
\citet{kraskovEstimatingMutualInformation2004} identify that non-skewed distribution, in particular Gaussian distribution, led to a lower bias and suggests that reparameterizing the marginal distributions into Gaussians could be a way of controlling the bias.
\citet{holmes2019estimation} proposed the following formula to transform any univariate distribution into a Gaussian one
\begin{equation}
    x_i' = \sqrt{2}\,\mathrm{erf}^{-1}\left(\frac{2r_i - 1}{N} - 1\right).
\end{equation}
where $1 \leq r_i \leq N$ is the rank of the sample $x_i$ in a sorted array (regardless of whether it is in ascending or descending order).
This formula consists of two parts. First, the $\frac{2r_i - 1}{N}$ transformation is used to transform any distribution into a uniform distribution over the $[0, 1]$ segment.
Secondly, the Gaussian cumulative distribution function (CDF) $\Phi$,
\begin{equation}
    \Phi(x) = \frac12 \left[ 1 + \mathrm{erf}^{-1}\left(\frac{x}{\sqrt2}\right) \right],
\end{equation}
is used to transform the uniform distribution into a reduced-centered normal distribution
Note that, by changing the CDF $\Phi$, we could reparametrize the data in any distribution which has an analytic CDF.
We emphasize that even though this reparametrization transforms all the marginal distributions to Gaussians, the obtained joint distribution is not a multivariate normal in general.

It appears that the bias becomes substantial when calculating the mutual information between several lines and physical parameters.
The intuitive reason is that it then becomes more difficult to identify the statistical relationships, which can be arbitrarily complex.
If the number of observations is small, these can be missed, resulting in a significant underestimation of the mutual information.

\subsection{Variance of the estimator}%
\label{app:variance}

The variance of an estimator quantifies the dispersion of estimations for different numbers of samples from the same population.
Knowing this dispersion is important in determining how reliable a single estimate is.
However, the variance of the nonparametric mutual information estimator does not have a closed-form formula.
Usually, this problem is solved by bootstrapping, which is a method of resampling with replacement~\citep{johnson2001introduction}.
However, this is not possible here because the estimate is not linear in the probability distribution (\eg, duplicate data does not count twice).
As proposed by~\citet{holmes2019estimation}, estimating the variance of the Kraskov estimator can be achieved by considering that the variance is inversely proportional to the sample size. % (see Fig.~\ref{fig:var-hist}).
This is a property shared by many estimators, such as the mean estimator.
In the case of the Kraskov estimator, the variance can then be expressed as
\begin{equation}
    \mathrm{Var}\left(\widehat{I}_N\right) = \frac{B}{N},
    \label{eq:var-formula}
\end{equation}
where $B$ is a model fitting parameter to the empirical variances that remains to be estimated and depends on the data distribution.

\begin{figure}
    \centering
    \includegraphics[width=0.9\linewidth]{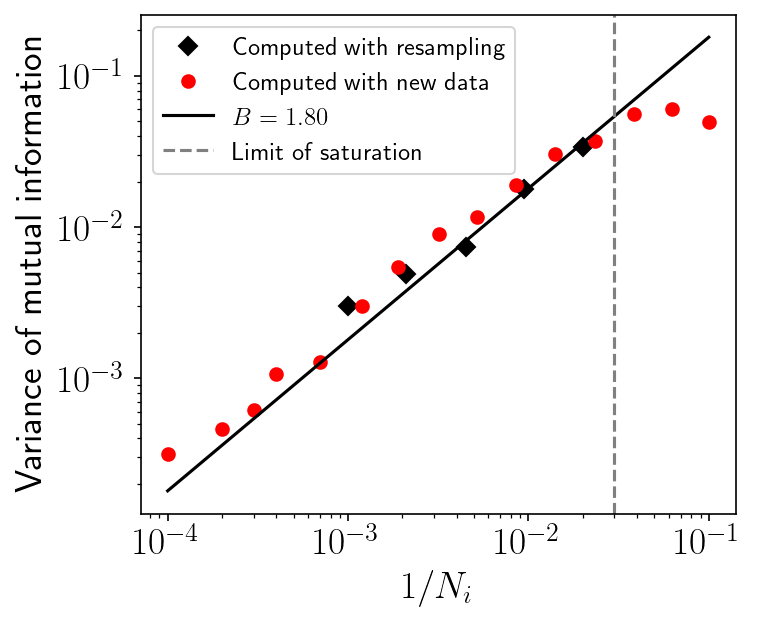}
    \caption{
        Comparison between the estimation of variances for different numbers of samples for a bivariate normal distribution with $\rho=0.8$.
        Red markers: variances estimated with several different datasets.
        Black markers: variances estimated with subsampling of a single dataset.
        Black line: regression line to predict the variance for any number of samples.
        The limit of saturation indicated by a dashed line corresponds to the number of samples for which the relationship of~\eqref{eq:var-formula} no longer holds.}
    \label{fig:var-slope}
\end{figure}

To estimate the value of $B$, we calculate the variance for different numbers of samples.
To do this, we separate the data into several subsets of equal size.
For example, for a total number of 1\,000 samples, it is possible to create 10 subsets of 100 samples, or 20 subsets of 50 samples.
Once the variance is computed for several numbers of samples, the $B$ value can be estimated by fitting a line curve.
More precisely, \citet{holmes2019estimation} proposed to estimate $B$ as
\begin{equation}
    \widehat{B} = \frac{N\sum_i \frac{n_i-1}{n_i}\widehat{\sigma}^2(N_i)}{\sum_i n_i-1}.
\end{equation}
This method is illustrated in Fig.~\ref{fig:var-slope}.
Empirically, the value of $B$ is usually between 1 and 3.

%%%%%%%%%%%%%%%%%%%%%%%%%%%%%%%%%%%

\section{Canonical correlations-based estimation of mutual information}
\label{app:cca}

Under the assumption that the joint distribution of observations and parameters $(X, \theta)$ is a multivariate normal distribution, the mutual information between observations and parameters can be expressed as follows
\begin{equation}
    I^{\text{CC}} = -\frac12 \sum_i \log\left(1-\lambda_i^2\right)
    \label{eq:cca-to-info}
\end{equation}
where the $\left\{\lambda_i\right\}_i$ are called the canonical correlations of $(X, \theta)$.
They satisfy the constraint $\forall i,\,0 \leq \lambda_i \leq 1$ and are the singular values of the normalized correlation matrix $M_{X\theta}$ defined as
\begin{equation}
    M_{X\theta} = C_{XX}^{-\frac12}\, C_{X\theta}\, C_{\theta\theta}^{-\frac12}
    \label{eq:cca}
\end{equation}
where $C$ denotes the empirical correlation matrix and $C^{-\frac12}$ the inverse of the matrix square root.
These coefficients are the basis of the method known as ``canonical correlation analysis''.
Notably, the coefficient $\lambda_1$ can be interpreted as the highest possible correlation coefficient between any linear combination of observables and any linear combination of lines.

Compared with the Kraskov estimator, the estimator in~\eqref{eq:cca-to-info} is much faster to compute.
However, when the joint distribution is different from a multivariate normal distribution, the mutual information estimate may be asymptotically biased.
A critical case occurs when the data are decorrelated yet statistically dependent (\eg, $\theta = X^2 + \epsilon$).
The correlation coefficient is then zero, resulting in a zero CC estimate of the mutual information while in the limit $\epsilon \to 0$ the analytical mutual information tends towards infinity.

%%%%%%%%%%%%%%%%%%%%%%%%%%%%%%%%%%%

\section{Considered lines}
\label{app:line-selection}

\begin{table}
    \centering
    \caption{Description of retained EMIR lines}
    \renewcommand{\arraystretch}{1.6}
    \addtolength{\tabcolsep}{-1.5mm}
    \footnotesize
    \begin{tabular}{ccccc}
        \hline
        \multirow{2}{*}{Species} & \multirow{2}{*}{Transition} & Frac. > 3\,$\sigma_a$ & Cal. err. & Integ. noise $\sigma_a$ \\
        & & (\%) & (\%) & (erg cm$^{-2}$ s$^{-1}$ sr$^{-1}$) \\
        \hline
        \multirow{3}{*}{\latexmol{12co}}%
        & $J=1-0$ & 90.6 & 5.0 & $1.22\times10^{-9}$\\
        & $J=2-1$ & 92.8 & 10.0 & $6.57\times10^{-9}$\\
        & $J=3-2$ & 91.5 & 10.0 & $6.92\times10^{-8}$\\
        \hline
        \multirow{3}{*}{\latexmol{13co}}%
        & $J=1-0$ & 81.4 & 5.0 & $5.27\times10^{-10}$\\
        & $J=2-1$ & 84.5 & 10.0 & $5.59\times10^{-9}$\\
        & $J=3-2$ & 57.0 & 10.0 & $1.42\times10^{-7}$\\
        \hline
        \multirow{3}{*}{\latexmol{c18o}}%
        & $J=1-0$ & 58.3 & 5.0 & $5.12\times10^{-10}$\\
        & $J=2-1$ & 61.7 & 10.0 & $5.53\times10^{-9}$\\
        & $J=3-2$ & 3.3 & 10.0 & $2.69\times10^{-7}$\\
        \hline
        \multirow{4}{*}{\latexmol{hcop}}%
        & $J=1-0$ & 61.9 & 5.0 & $2.52\times10^{-10}$\\
        & $J=2-1$ & 42.6 & 7.5 & $1.36\times10^{-8}$\\
        & $J=3-2$ & 45.8 & 10.0 & $1.20\times10^{-8}$\\
        & $J=4-3$ & 15.6 & 10.0 & $1.12\times10^{-7}$\\
        \hline
        \multirow{5}{*}{\latexmol{12cs}}%
        & $J=2-1$ & 70.8 & 5.0 & $3.25\times10^{-10}$\\
        & $J=3-2$ & 67.6 & 7.5 & $1.17\times10^{-9}$\\
        & $J=5-4$ & 53.4 & 10.0 & $8.22\times10^{-9}$\\
        & $J=6-5$ & 37.8 & 10.0 & $2.14\times10^{-8}$\\
        & $J=7-6$ & 22.3 & 10.0 & $6.71\times10^{-8}$\\
        \hline
        \multirow{3}{*}{\latexmol{hcn}}%
        & $J=1-0$ & 46.5 & 5.0 & $2.49\times10^{-10}$\\ % $F=2-1$
        & $J=2-1$ & 10.6 & 7.5 & $8.28\times10^{-9}$\\ % $F=3-2$
        & $J=3-2$ & 26.3 & 10.0 & $1.15\times10^{-8}$\\ % $F=3-3$
        \hline
        \multirow{2}{*}{\latexmol{hnc}}%
        & $J=1-0$ & 39.8 & 5.0 & $2.62\times10^{-10}$\\
        & $J=3-2$ & 18.0 & 10.0 & $1.27\times10^{-8}$\\
        \hline
        \multirow{6}{*}{\latexmol{12cn}}%
        & $n=1-0$, $J=\frac{1}{2}-\frac{1}{2}$ & 10.1 & 5.0 & $7.11\times10^{-10}$\\
        & $n=1-0$, $J=\frac{3}{2}-\frac{1}{2}$ & 17.5 & 5.0 & $7.51\times10^{-10}$\\
        & $n=2-1$, $J=\frac{3}{2}-\frac{1}{2}$ & 8.6 & 10.0 & $6.17\times10^{-9}$\\
        & $n=2-1$, $J=\frac{5}{2}-\frac{3}{2}$ & 15.3 & 10.0 & $6.18\times10^{-9}$\\
        & $n=3-2$, $J=\frac{5}{2}-\frac{3}{2}$ & 2.6 & 10.0 & $6.31\times10^{-8}$\\
        & $n=3-2$, $J=\frac{7}{2}-\frac{5}{2}$ & 6.3 & 10.0 & $6.31\times10^{-8}$\\
        \hline
        \multirow{4}{*}{\latexmol{c2h}}%
        & $n=2-1$, $J=\frac{5}{2}-\frac{3}{2}$ & 1.0 & 7.5 & $4.33\times10^{-9}$\\
        & $n=3-2$, $J=\frac{5}{2}-\frac{3}{2}$ & 6.3 & 10.0 & $1.08\times10^{-8}$\\
        & $n=3-2$, $J=\frac{7}{2}-\frac{5}{2}$ & 9.6 & 10.0 & $1.08\times10^{-8}$\\
        & $n=4-3$, $J=\frac{9}{2}-\frac{7}{2}$ & 4.0 & 10.0 & $7.69\times10^{-8}$\\
        \hline
    \end{tabular}
    \label{tab:emir-lines}
\end{table}

\begin{table}
    \centering
    \caption{Description of FIR lines}
    \renewcommand{\arraystretch}{1.6}
    \addtolength{\tabcolsep}{0mm}
    \footnotesize
    \begin{tabular}{ccccc}
        \hline
        Species & $\lambda$ & Frac. > 3\,$\sigma_a$ & Cal. err. & Integ. noise $\sigma_a$ \\
        & (\microm) & (\%) & (\%) & (erg cm$^{-2}$ s$^{-1}$ sr$^{-1}$) \\
        \hline
        \multirow{2}{*}{\ci}%
        & 158 & 77.2 & 83.0 & $1.93\times10^{-7}$\\
        & 609 & 66.3 & 73.8 & $8.59\times10^{-7}$\\
        \cp{} & 370 & 50.2 & 5.0 & $2.20\times10^{-5}$\\
        \hline
    \end{tabular}
    \label{tab:fir-lines}
\end{table}

\begin{figure}
    \centering
    \includegraphics[width=\linewidth]{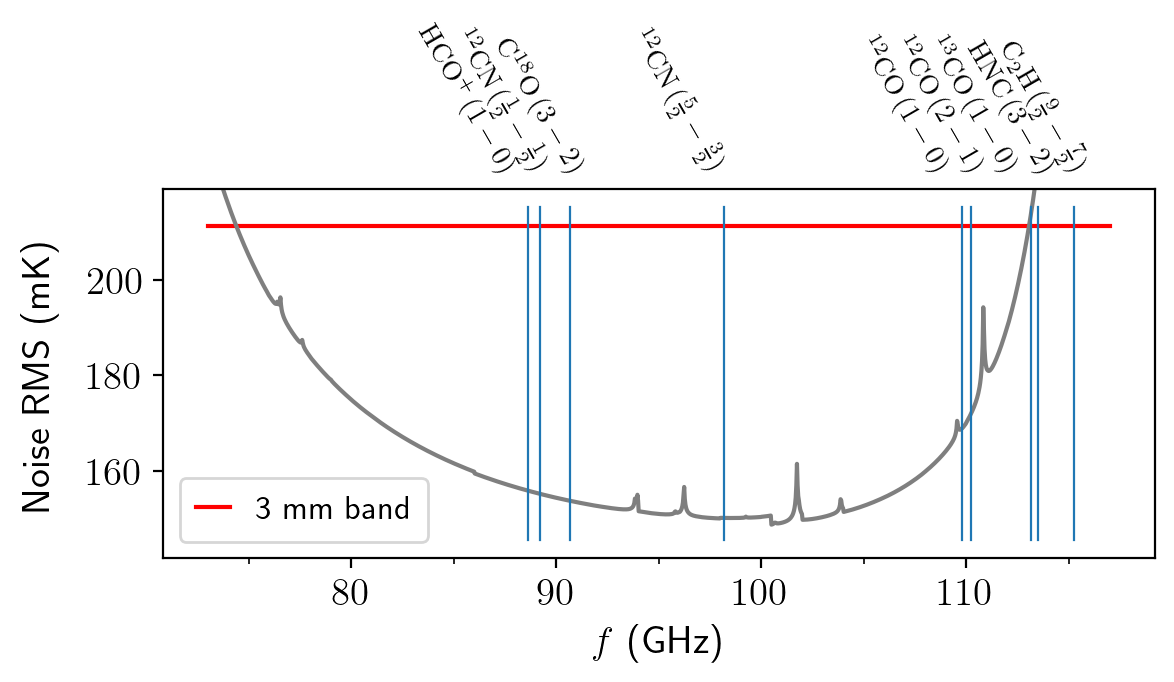}\vspace{2mm}
    \includegraphics[width=\linewidth]{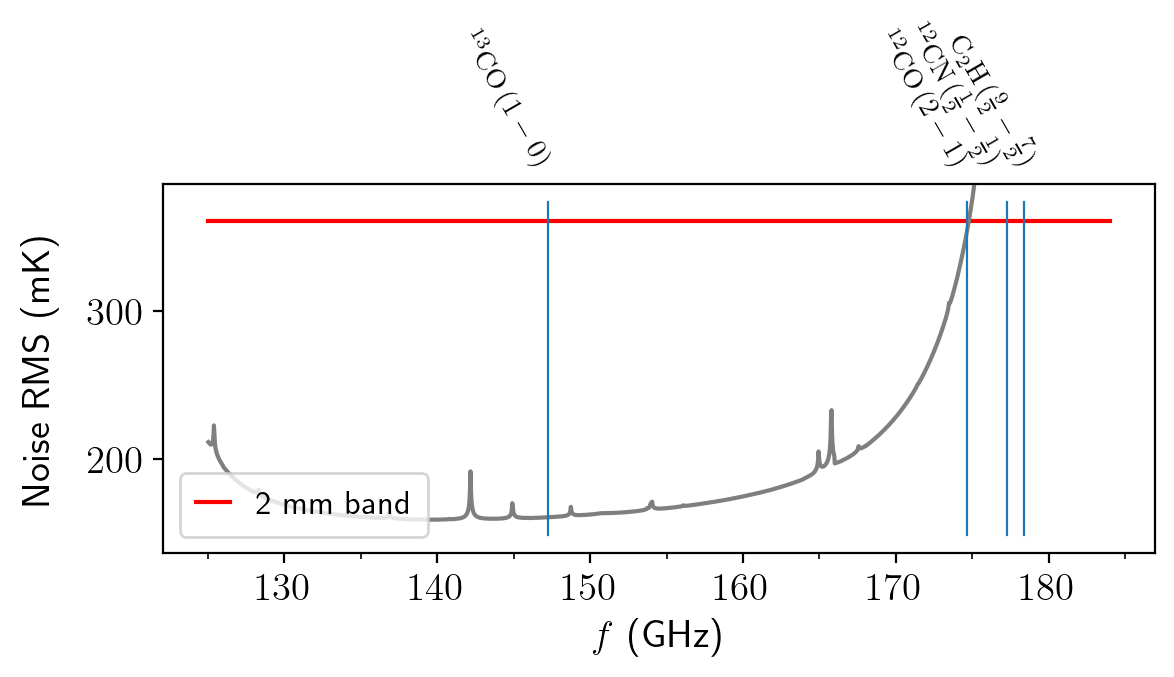}\vspace{2mm}
    \includegraphics[width=\linewidth]{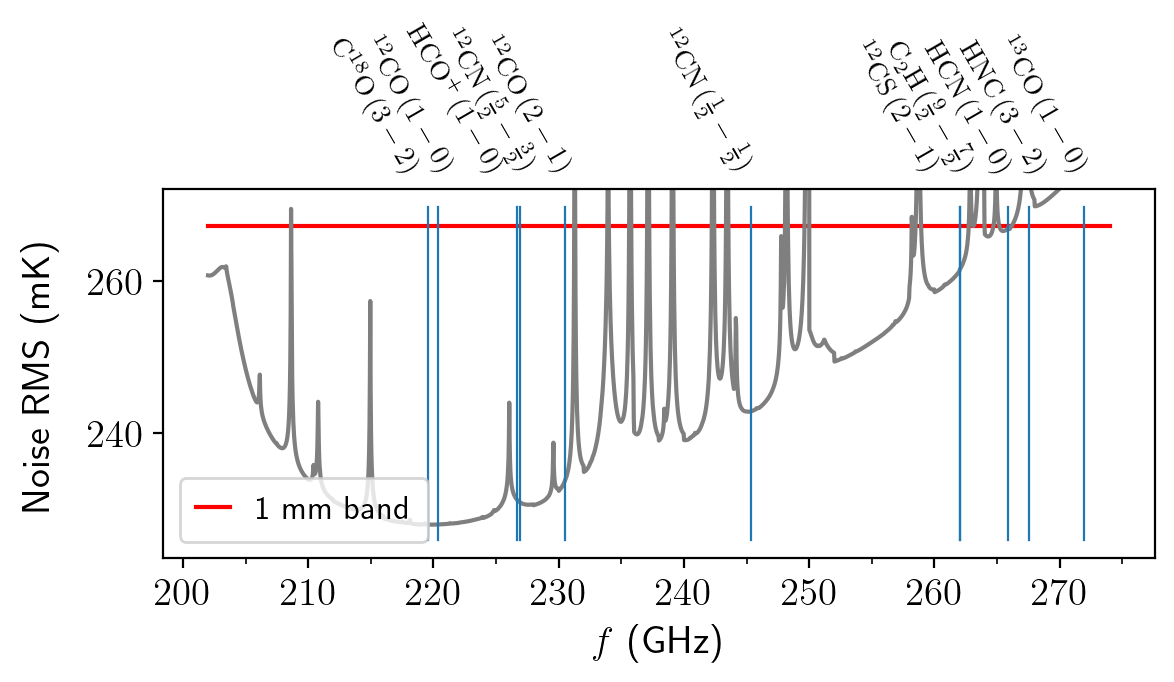}\vspace{2mm}
    \includegraphics[width=\linewidth]{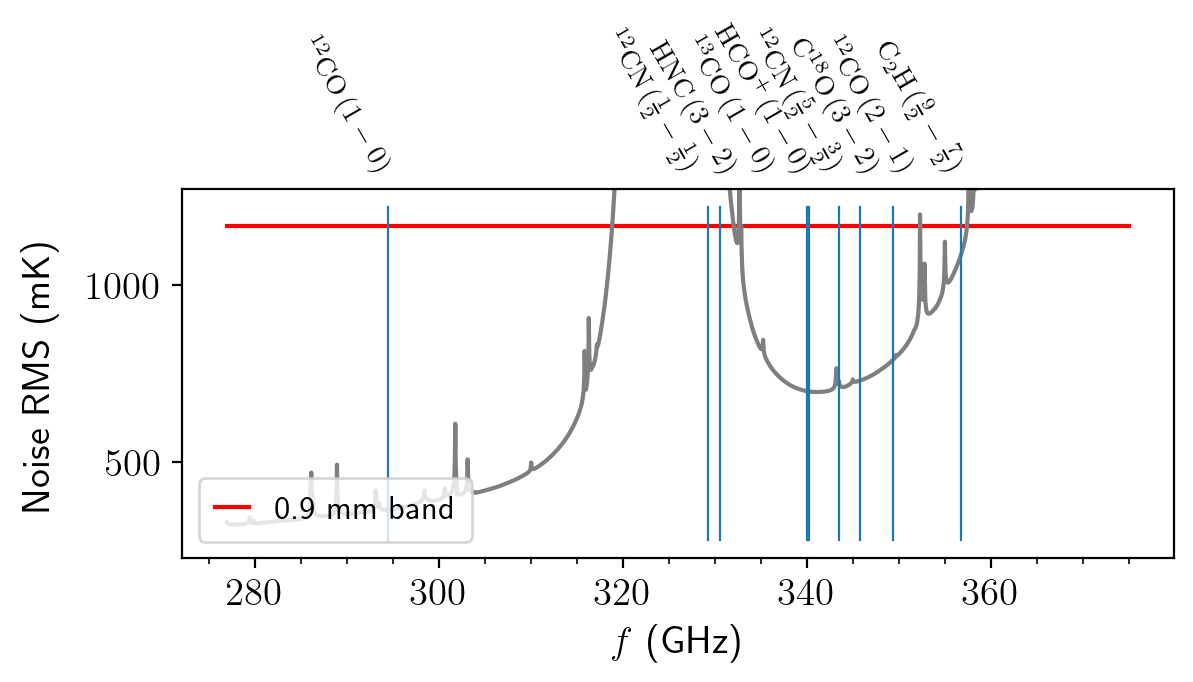}
    \caption{
        33 selected molecular lines by EMIR band.
    }
    \label{fig:emir-bands}
\end{figure}

In this section, we describe in more detail the 36 lines retained in Sect.~\ref{subsec:dataset_generation}.
They are used for mutual information maps (Sect.~\ref{subsec:maps_mi}) and line selection (Sect.~\ref{sec:results_line_selection}).
Table~\ref{tab:emir-lines} gathers the additive and multiplicative noise levels for each of the 33 millimeter lines, and provide the fraction of the full parameter space for which the S/N is greater than 3.
The millimeter lines selected are those for which this fraction is greater than 1\%.
Table~\ref{tab:fir-lines} gathers the same information for the three \ci{} and \cp{} lines.

Figure~\ref{fig:emir-bands} displays the considered lines in each of the four frequency bands of the EMIR receiver, namely the 3\,mm, 2\,mm, 1\,mm and 0.9\,mm bands.
It also shows the additive noise for the reference integration time.

%%%%%%%%%%%%%%%%%%%%%%%%%%%%%%%%%%%

\section{Gradients of log intensities}%
\label{app:gradients}

\begin{figure*}[h]
    \centering
    \includegraphics[width=\linewidth]{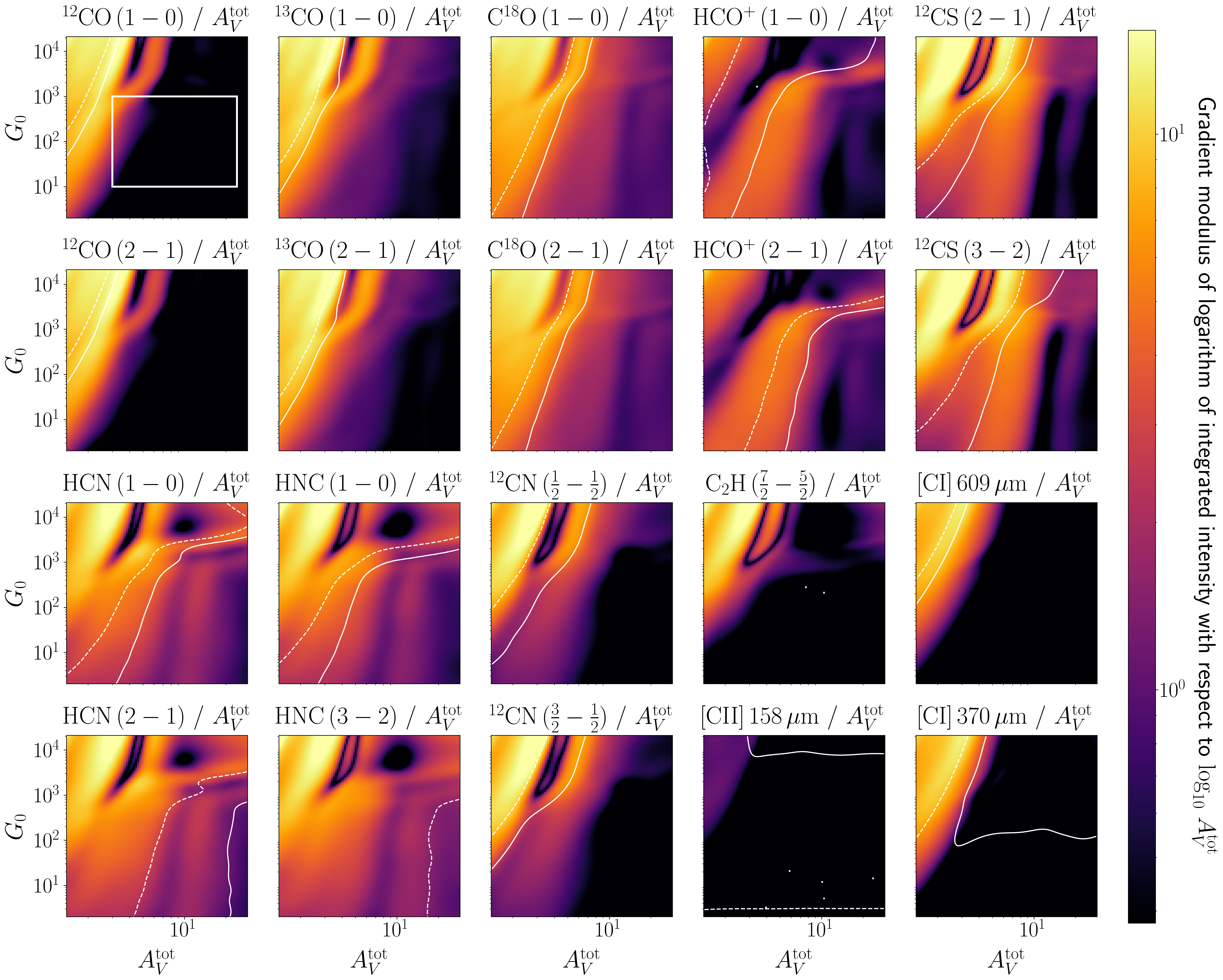}
    \caption{
        Partial derivative of the log of the predicted intensity with respect to $\logd\AV$ for multiple lines.
        The white full line represents the standard deviation $\sigma_{a,\ell}$ of the additive noise from~\eqref{eq:obs_model_instance} for the \mbox{ORION-B} observations~\citep{petyAnatomyOrionGiant2017}.
        The white dashed line indicates the standard deviation with a ten times longer integration time (deeper integration use case).
        The white rectangle on the first panel delimits the parameter space characterizing the Horsehead Nebula.
        Black and thin contours corresponds to critical points, \ie, points where the gradient cancels out (typically local extrema or inflection points).
    }%
    \label{fig:grad_av}
\end{figure*}

\begin{figure*}[h]
    \centering
    \includegraphics[width=\linewidth]{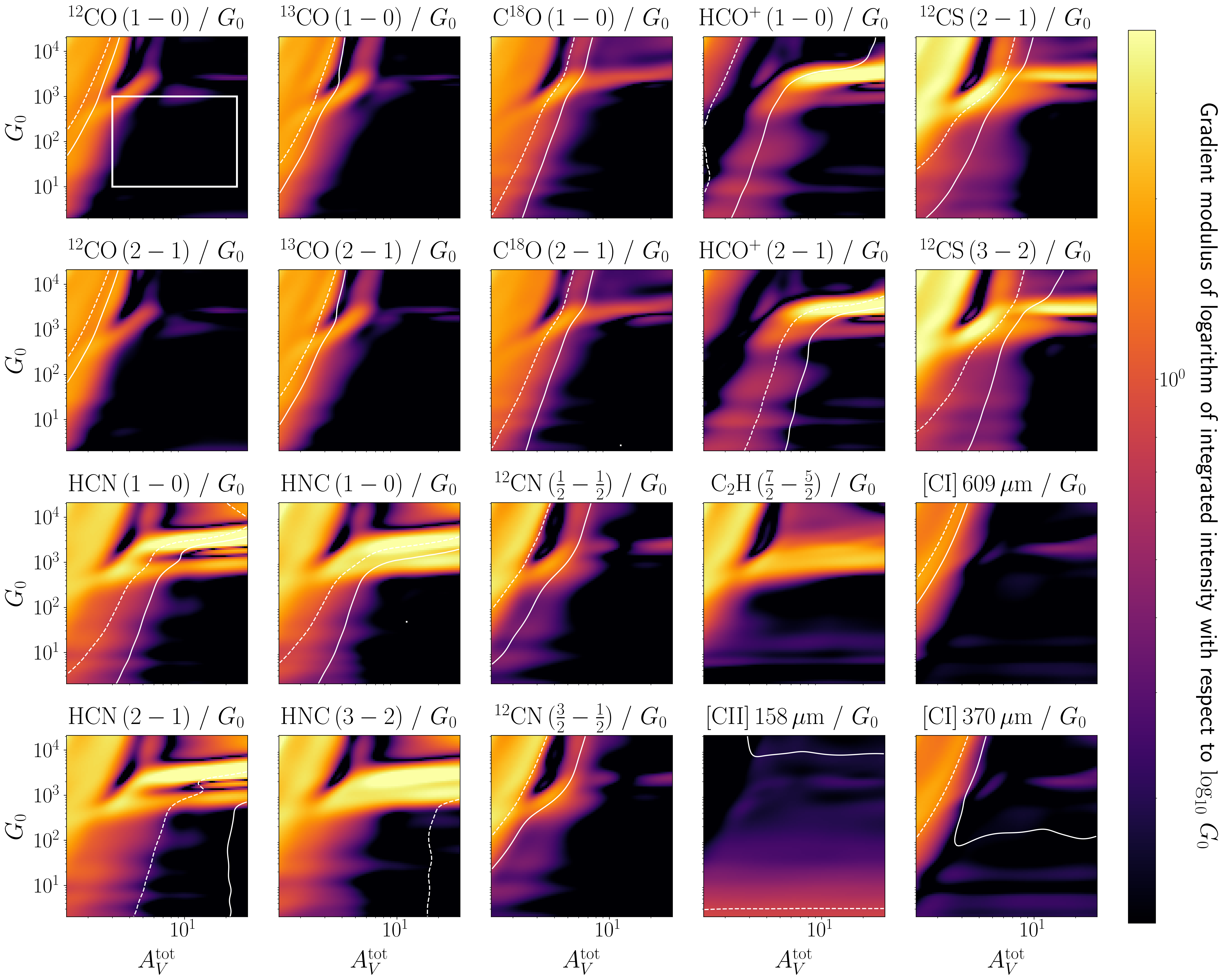}
    \caption{
        Partial derivative of the log of the predicted intensity with respect to $\log\Gnaught$ for multiple lines.
        The white full line represents the standard deviation $\sigma_{a,\ell}$ of the additive noise from~\eqref{eq:obs_model_instance} for the \mbox{ORION-B} observations~\citep{petyAnatomyOrionGiant2017}.
        The white dashed line indicates the standard deviation with a ten times longer integration time (deeper integration use case).
        The white rectangle on the first panel delimits the parameter space characterizing the Horsehead Nebula.
        Black and thin contours corresponds to critical points, \ie, points where the gradient cancels out (typically local extrema or inflection points).
    }%
    \label{fig:grad_g0}
\end{figure*}

Figures~\ref{fig:grad_av} and~\ref{fig:grad_g0} show the absolute value of the partial derivative of the predicted log integrated intensities with respect to $\AV$ and $\Gnaught$, respectively.
In other words:
\begin{align}
    \text{Fig.~\ref{fig:grad_av} displays}
    \left|
        \frac{\partial \log \truefell}{\partial \AV}
    \right|,
    \;
    \text{Fig.~\ref{fig:grad_g0} displays}
    \left|
        \frac{\partial \log \truefell}{\partial \Gnaught}
    \right|.
\end{align}
These figures permit one to better assess the variations of $\log \truefell$, which are hard to read from Fig.~\ref{fig:predicted_maps}.
When compared with the mutual information maps in Fig.~\ref{fig:maps-grid-av} and~\ref{fig:maps-grid-g0}, they highlight the fact that a high mutual information requires both a large S/N and a large gradient.
This is easy to see for $\AV$ and the first two transitions of $^{12}$CO, $^{13}$CO and C$^{18}$O, for instance.
For $^{12}$CO, the gradient quickly goes to zero, as the two lines become optically thick and saturate.
To achieve a S/N > 1, $^{13}$CO lines require larger values of $\AV$.
They eventually also become optically thick and saturate, but for much larger values of $\AV$ than $^{12}$CO.
Lines of C$^{18}$O never saturate: their partial derivative is always $\gtrsim 10^{-2}$.

%%%%%%%%%%%%%%%%%%%%%%%%%%%%%%%%%%%

\section{Influence of integration time}%
\label{app:additional-results}

\begin{figure*}
    \centering
    \includegraphics[width=\linewidth, trim={0 1.3cm 0 0.5cm}, clip]{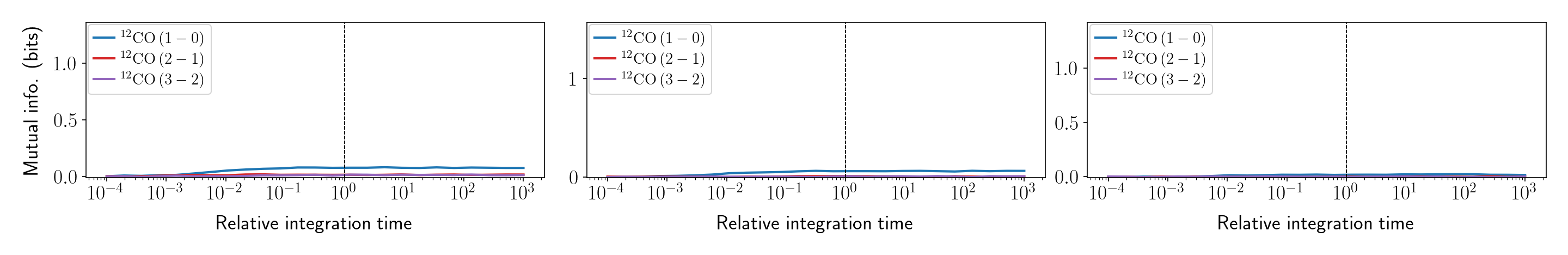}
    \includegraphics[width=\linewidth, trim={0 1.3cm 0 0.5cm}, clip]{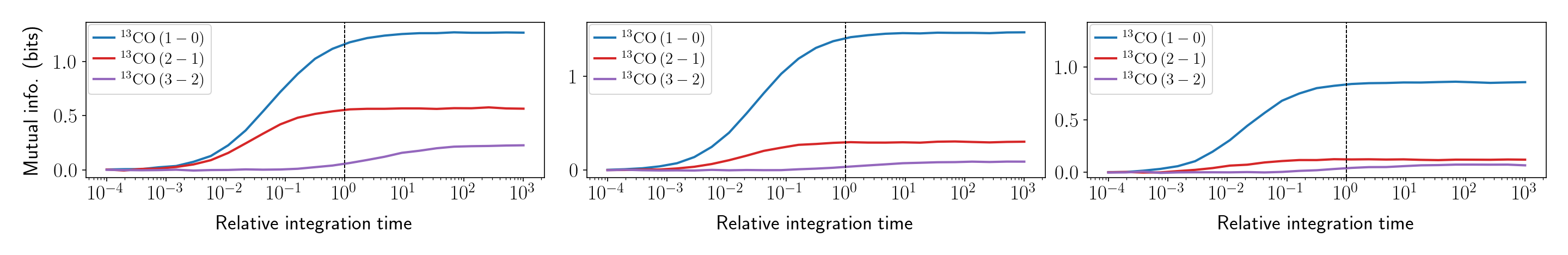}
    \includegraphics[width=\linewidth, trim={0 1.3cm 0 0.5cm}, clip]{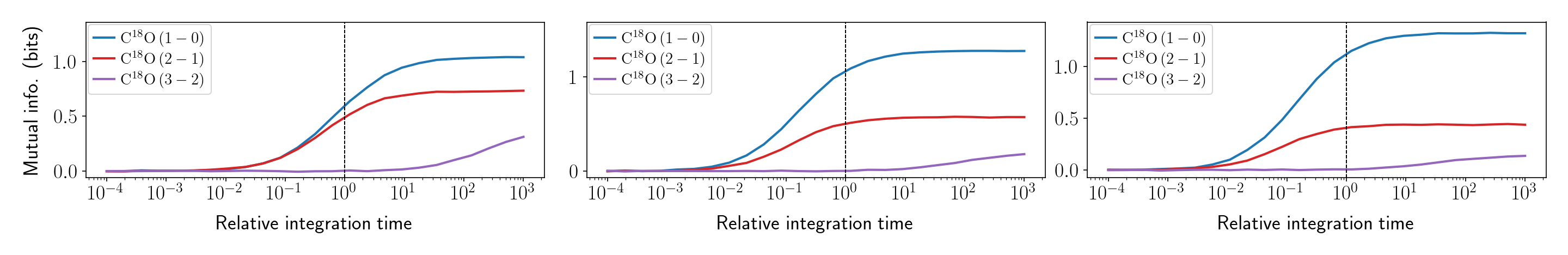}
    \includegraphics[width=\linewidth, trim={0 1.3cm 0 0.5cm}, clip]{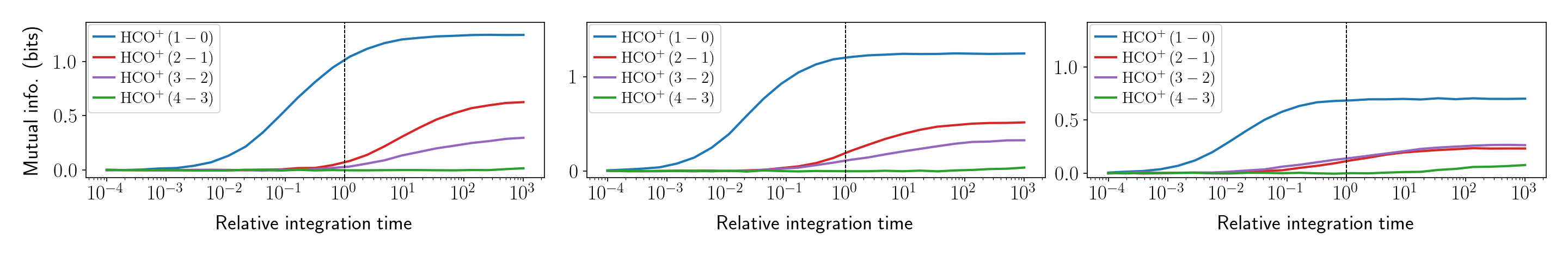}
    \includegraphics[width=\linewidth, trim={0 1.3cm 0 0.5cm}, clip]{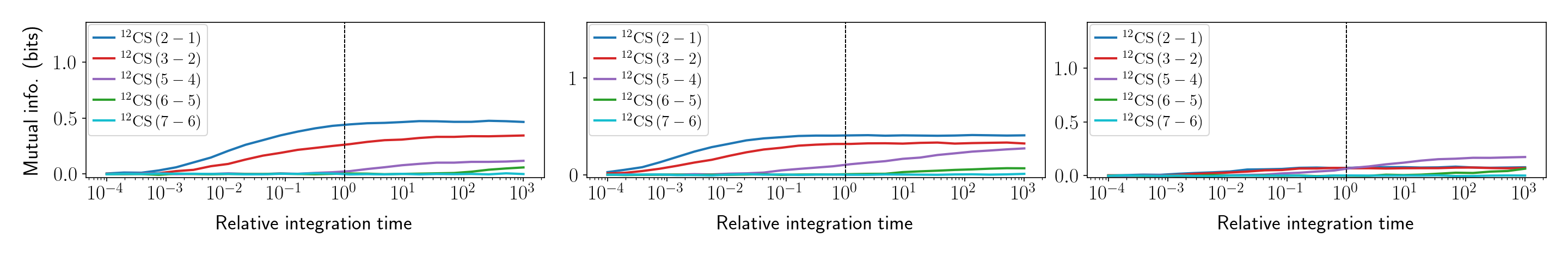}
    \includegraphics[width=\linewidth, trim={0 1.3cm 0 0.5cm}, clip]{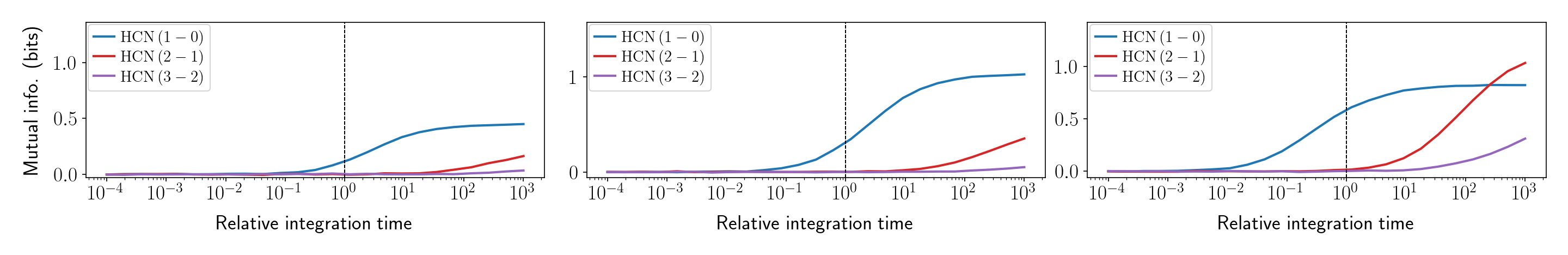}
    \includegraphics[width=\linewidth, trim={0 1.3cm 0 0.5cm}, clip]{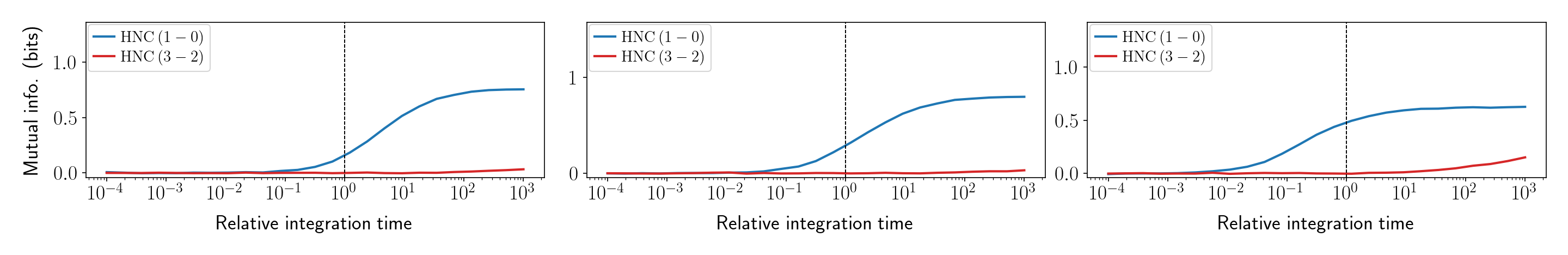}
    \includegraphics[width=\linewidth, trim={0 1.3cm 0 0.5cm}, clip]{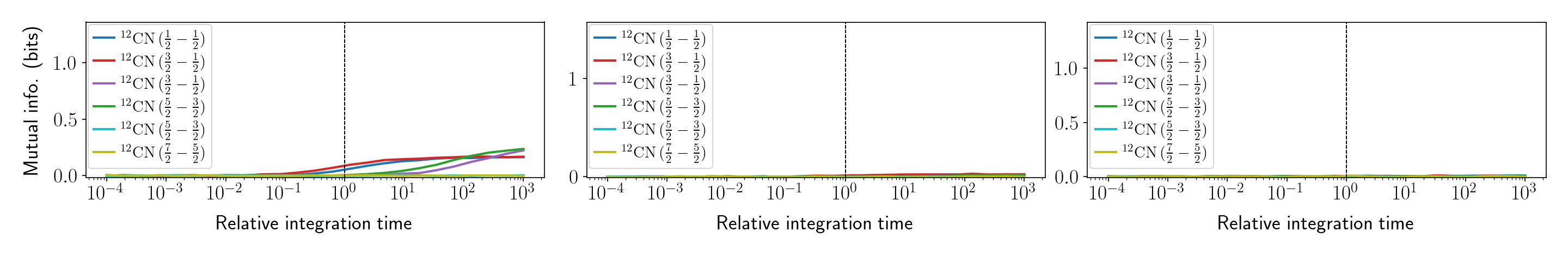}
    \includegraphics[width=\linewidth, trim={0 1.3cm 0 0.5cm}, clip]{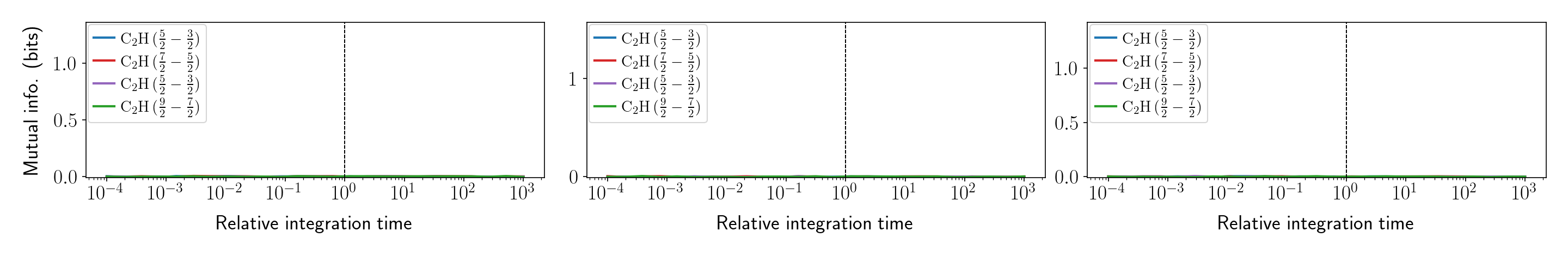}
    \includegraphics[width=\linewidth, trim={0 0.3cm 0 0.5cm}, clip]{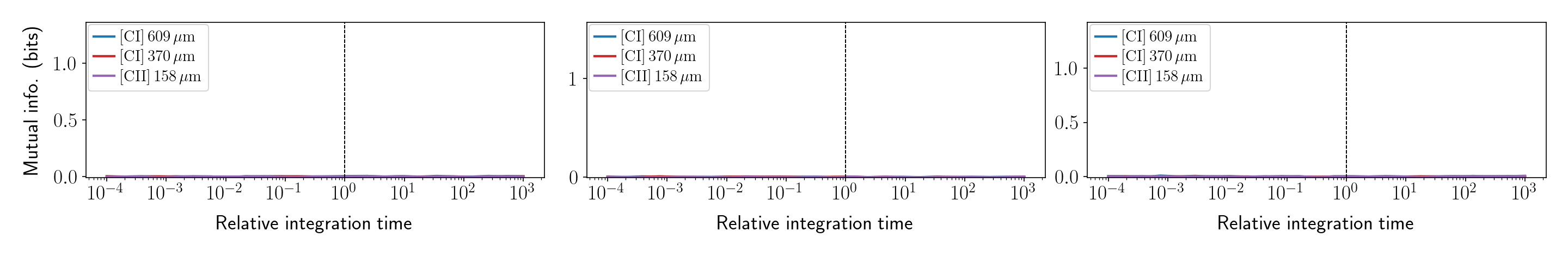}
    \caption{
        Evolution of mutual information between $\AV{}$ and integrated line intensities as a function of integration time.
        Results are shown for translucent gas ($3 \leq \AV{} \leq 6$, left), filamentary gas ($6 \leq \AV{} \leq 12$, middle) and dense cores ($12 \leq \AV{} \leq 24$, right).
    }%
    \label{fig:obs_time_av_grid}
\end{figure*}

\begin{figure*}
    \centering
    \includegraphics[width=\linewidth, trim={0 1.3cm 0 0.5cm}, clip]{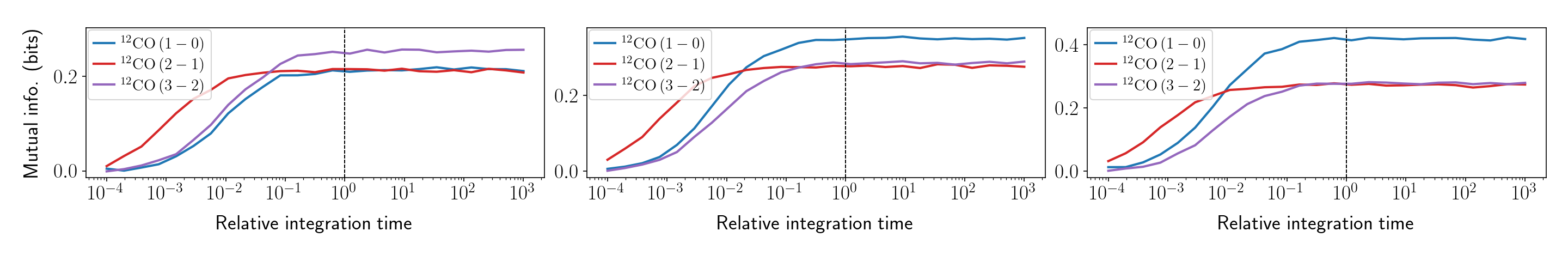}
    \includegraphics[width=\linewidth, trim={0 1.3cm 0 0.5cm}, clip]{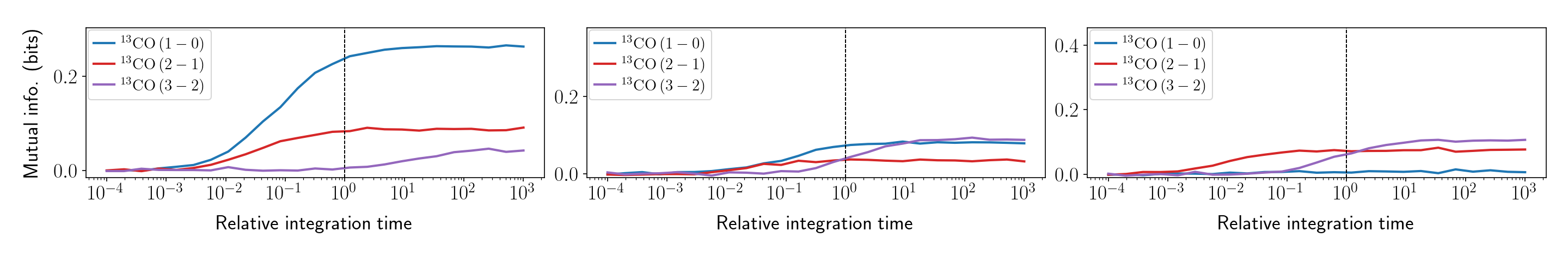}
    \includegraphics[width=\linewidth, trim={0 1.3cm 0 0.5cm}, clip]{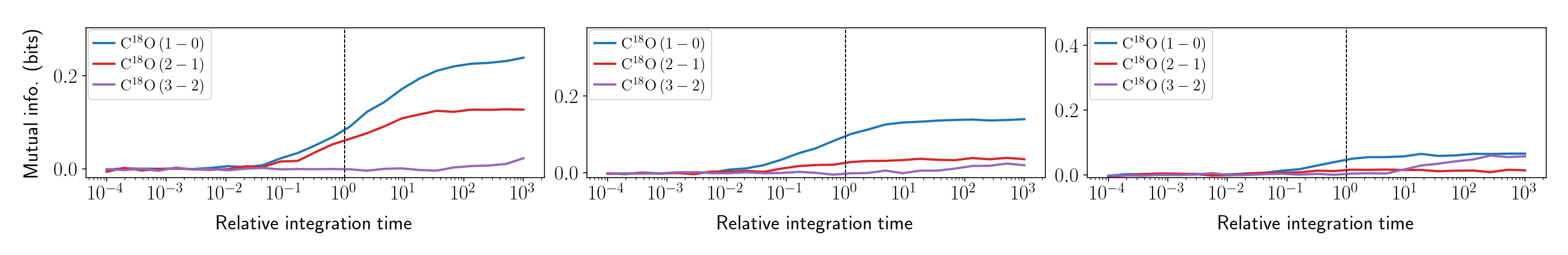}
    \includegraphics[width=\linewidth, trim={0 1.3cm 0 0.5cm}, clip]{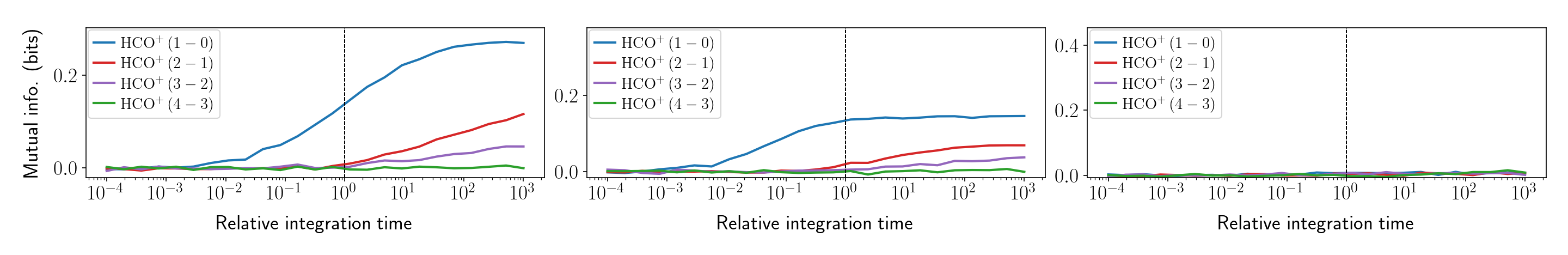}
    \includegraphics[width=\linewidth, trim={0 1.3cm 0 0.5cm}, clip]{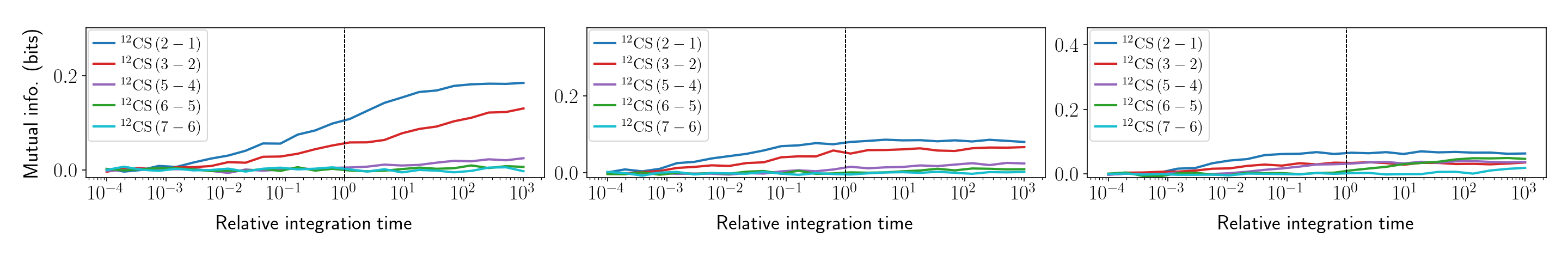}
    \includegraphics[width=\linewidth, trim={0 1.3cm 0 0.5cm}, clip]{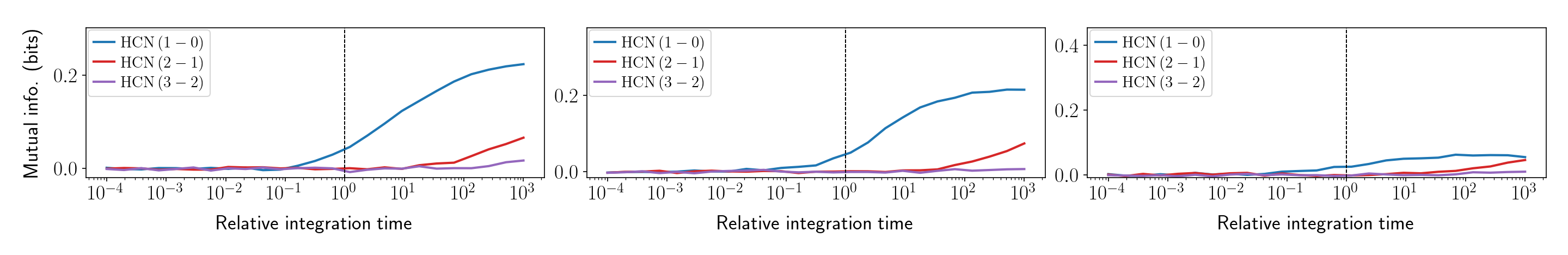}
    \includegraphics[width=\linewidth, trim={0 1.3cm 0 0.5cm}, clip]{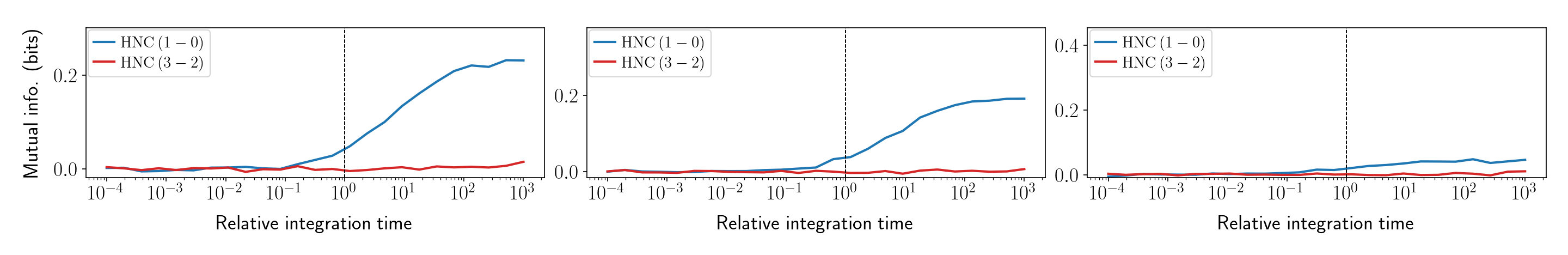}
    \includegraphics[width=\linewidth, trim={0 1.3cm 0 0.5cm}, clip]{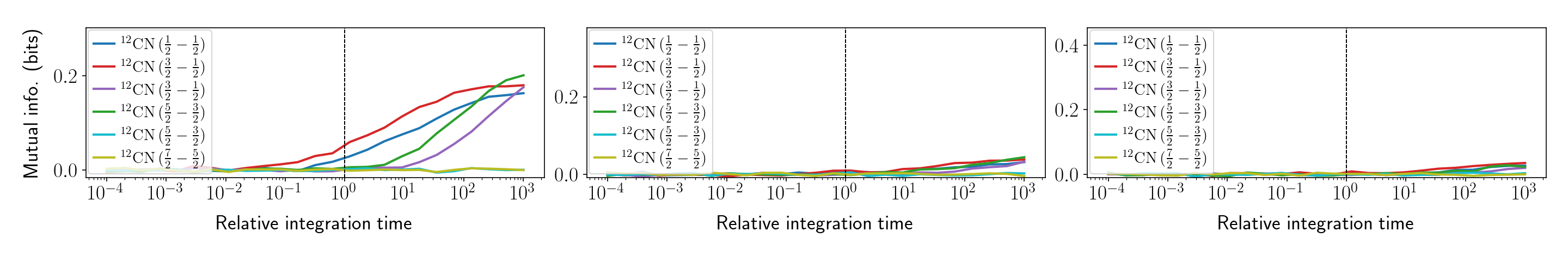}
    \includegraphics[width=\linewidth, trim={0 1.3cm 0 0.5cm}, clip]{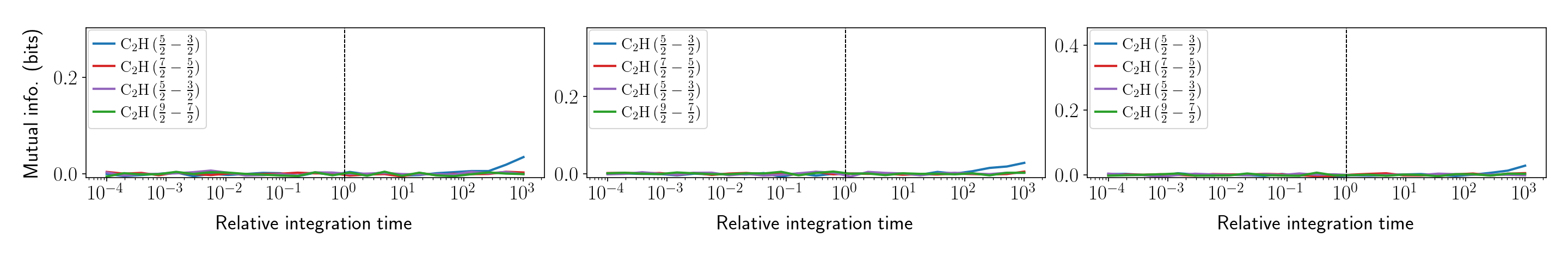}
    \includegraphics[width=\linewidth, trim={0 0.3cm 0 0.5cm}, clip]{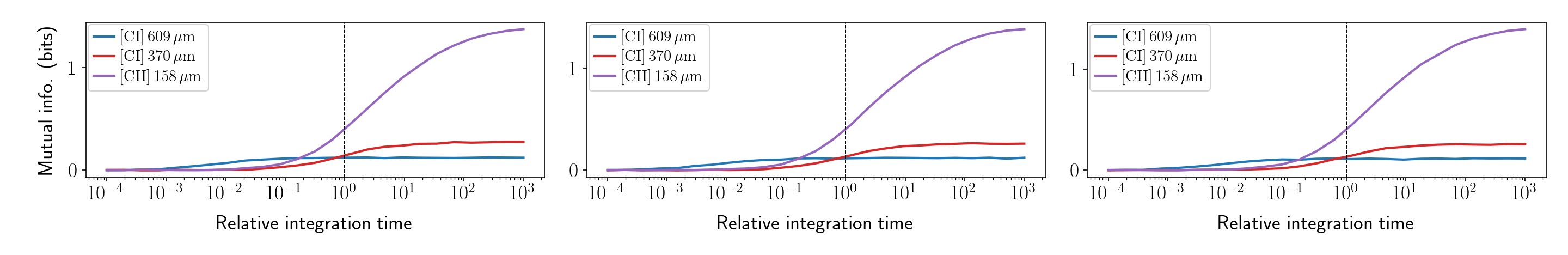}
    \caption{
        Evolution of mutual information between $\Guv{}$ and integrated line intensities as a function of integration time.
        Results are shown for
        translucent gas ($3 \leq \AV{} \leq 6$, left),
        filamentary gas ($6 \leq \AV{} \leq 12$, middle)
        and dense cores ($12 \leq \AV{} \leq 24$, right).
    }%
    \label{fig:obs_time_g0_grid}
\end{figure*}

Figure~\ref{fig:obs_time_av_grid} and~\ref{fig:obs_time_g0_grid} show how the integration time influences the mutual information of individual lines with \AV{} and \Guv{}, respectively.
Note that some lines, such as the C$_2$H lines, always yield a near-zero mutual information for both \AV{} and \Guv{}.

\end{appendix}

\end{document}

%%%%% End of main.tex